\documentclass[english,10pt,journal,twocolumn]{IEEEtran}
\usepackage[T1]{fontenc}
\usepackage[latin9]{inputenc}
\usepackage{float}
\usepackage{calc}
\usepackage{amsmath}
\usepackage{amssymb}
\usepackage{stackrel}
\usepackage{graphicx}

\makeatletter

\floatstyle{ruled}
\newfloat{algorithm}{tbp}{loa}
\providecommand{\algorithmname}{Algorithm}
\floatname{algorithm}{\protect\algorithmname}

\@ifundefined{date}{}{\date{}}
\usepackage{babel}

\@ifundefined{showcaptionsetup}{}{%
 \PassOptionsToPackage{caption=false}{subfig}}
\usepackage{subfig}
\makeatother

\usepackage{babel}
\begin{document}
\title{Designing sequence set with minimal peak side-lobe level for applications
in high resolution RADAR imaging}
\author{Surya Prakash Sankuru, R Jyothi, Prabhu Babu, Mohammad Alaee-Kerahroodi}
\maketitle
\begin{abstract}
Constant modulus sequence set with low peak side-lobe level is a necessity
for enhancing the performance of modern active sensing systems like
Multiple Input Multiple Output (MIMO) RADARs. In this paper, we consider
the problem of designing a constant modulus sequence set by minimizing
the peak side-lobe level, which can be cast as a non-convex minimax
problem, and propose a Majorization-Minimization technique based iterative
monotonic algorithm. The iterative steps of our algorithm are computationally
not very demanding and they can be efficiently implemented via Fast
Fourier Transform (FFT) operations. We also establish the convergence
of our proposed algorithm and discuss the computational and space
complexities of the algorithm. Finally, through numerical simulations,
we illustrate the performance of our method with the state-of-the-art
methods. To highlight the potential of our approach, we evaluate the
performance of the sequence set designed via our approach in the context
of probing sequence set design for MIMO RADAR angle-range imaging
application and show results exhibiting good performance of our method
when compared with other commonly used sequence set design approaches.

\end{abstract}

\section{{Introduction and literature review }}

In recent years, Multiple Input Multiple Output (MIMO) RADAR has become
a trending technology and plays a key role in modern warfare systems.
Unlike the phased array RADAR system \cite{Phas_mimo} which transmits
the scaled versions of single sequence, the MIMO RADAR system will
take advantage of the waveform diversity \cite{calder_mag} and transmits
more number of sequences simultaneously which results in creating
a very large virtual aperture that inturn gives enhanced resolution
images \cite{R_Imaging_Li_PS} and paves way for better target detection
\cite{Localization,RADARSimi,EffRad,Betterperf}. But to reap all
the benefits of the MIMO RADAR, the probing sequence set employed
should have good correlation side-lobe levels \cite{spsp,4_signal_App}.
In any practical RADAR system, there are many challenges like limited
energy budget, high cost of the practical hardware components, and
the necessity to work on the linear range of power amplifiers, which
force one to consider using either unimodular (or) low Peak to the
Average Power Ratio (PAPR) constrained probing set of sequences \cite{9_waveform_textbook,S8}.
Besides MIMO RADAR, some notable applications where constant modulus
sequence set with better correlation side-lobe levels play a prominent
role are wireless communication systems \cite{4_signal_App,9_waveform_textbook},
MIMO SONAR \cite{sd7,1_DSP_SONAR,8_Probing_waveform,Activesonar},
Cryptography \cite{4_signal_App}, channel estimation \cite{S8,PM31},
CDMA and spread spectrum applications \cite{PM33,MC20,12binary,CDMA}.
Hence, designing constant modulus sequence set with better auto-correlation
and cross-correlation side-lobe levels is always desired.

Designing sequences with good auto-correlation properties for applications
in active sensing, wireless communication is an active area of research
and countless number of researchers have contributed to it. In the
following, we will briefly discuss some key contributions. The foundation
for research on the probing signal design is done by notable researchers
like Nyquist, Shannon, Tesla, and was continued by the Barker, Golomb,
Frank, Woodward, etc. In the early years, researchers studied the
single sequence design problem via analytical approaches and proposed
the maximal length, Gold, Kasami sequences which are known to possess
better periodic correlation properties, and later the binary Barker
\cite{Barkerbook}, Golomb \cite{15_polyphaseseq_Zhang}, Frank \cite{13_Polyphasecode_frank},
polyphase \cite{14_polyphaseseq_borwein} sequences have been developed
which have better aperiodic correlation properties. A major drawback
of the analytical approaches is the sequences obtained by these approaches
are known to exist only for the limited lengths and have lesser degrees
of freedom. To overcome such issues, in the recent decade (or) so,
researchers have used numerical optimization methods and used different
metrics (depending upon the applications) and designed algorithms
to generate larger length sequences with good correlation properties.
The authors in \cite{17_MISL,21_MM_PSL,CPM_kehrodi} proposed methods
to design sequence with good correlation properties by optimizing
the correlation related metrics like Integrated Side-lobe Level (ISL)
and approximated Peak Side-lobe level (PSL), works done in \cite{minmax_logexp,fan_psl,SpectralShaping}
studied the sequence design along with correlation and spectral constraints.
The authors in \cite{aubry} studied the problem of designing sequences
with better ambiguity function - which ensures sequences with good
autocorrelation and as well as immune to Doppler ambiguities. All
the above mentioned works studied only the single sequence design
problem and in the following, we discuss the literature on the sequence
set design which is the main focus of this paper.

The authors in \cite{Multi_CAN} minimized the approximated ISL metric
using the alternating minimization method and proposed the Multi-CAN
algorithm that can design very large length sequence sets. In \cite{MMCORR,20_fast_alg_waveform_LI},
the authors have proposed an optimization technique that minimizes
the original ISL metric and the resultant algorithm was capable of
designing large length sequence sets with better correlation side-lobe
levels than the one generated via the Multi-CAN approach. Slightly
different from the ISL based approaches, the authors in \cite{sd8}
tried to minimize the PSL metric by approximating it with a Chebyshev
metric and designed sequence set for applications in MIMO RADAR. An
iterative direct search algorithm that updates each element of the
sequence set sequentially while considering the remaining elements
fixed using exhaustive pattern search method was proposed in \cite{sd15}.
In \cite{PSL_Kehroodi}, the authors considered the Pareto function
(weighted combination) of PSL, ISL metrics and optimized using the
Block Coordinate Descent (BCD) method and proposed an algorithm BiST
to design set of sequences with discrete phase constraint. The authors
in \cite{Bound_Kehroodi} optimized the ISL metric via Coordinate
Descent (CD) framework to design set of sequences that meets Welch
bound on the ISL value. It is very clear from the literature survey,
that none of the researchers have designed sequence set by directly
minimizing the peak side-lobe level metric - a major reason being
the challenging nature of the design metric as it is not differentiable
and the associated optimization problem is a saddle point problem.
A brief summary of the literature survey on the sequence set design
problem is as follows: 
\begin{itemize}
\item There are analytical as well as computational approaches minimizing
the ISL metric to design large length single sequences with better
periodic, aperiodic correlation properties. 
\item To design sequence set, there are ISL minimization based algorithms
(Multi-CAN, MM-Corr, ISL-NEW, Iteration Direct Search Algorithm) but
none that can generate sequence set by optimizing the exact PSL metric. 
\item There is PSL (or) ISL (or) both minimization based algorithm named
as BiST, which can able to design set of sequences but only under
the discrete phase constraint.
\end{itemize}
So to the best of our knowledge, nobody has approached the problem
of designing sequence set by directly minimizing (without any approximation)
the peak side-lobe level metric. One may ask, what can one achieve
special by minimizing the PSL when compared to the ISL metric? The
answer can be explained as follows - In a RADAR system with a matched
filter receiver, the peak side-lobe level of the transmit sequence
will dictate the false alarm probability, so the lower the PSL level,
the lower the probability of false alarm. Thus, on top of the computational
challenge, designing sequence set by minimizing the PSL metric has
also pratical significance.

The major contributions of this paper are as follows: 
\begin{itemize}
\item We have developed a monotonic algorithm which is based on the technique
of Majorization-Minimization to design a constant modulus sequence
set by minimizing the PSL metric. 
\item We show a computationally efficient way of implementing our proposed
algorithm using FFT and IFFT operations. 
\item We prove the convergence of our algorithm to a stationary point of
the PSL metric. 
\item Finally, we evaluate the performance of the proposed algorithm through
numerical simulations for different problem dimensions and compare
them with the state-of-the-art techniques. We also evaluate the proposed
algorithm in the context of the MIMO RADAR imaging application. 
\end{itemize}
The rest of the paper is organized as follows. Formulation of the
sequence set design problem and a review of the Majorization-Minimization
technique is discussed in section II. The proposed algorithm, its
proof of monotonic convergence to the stationary point, computational
and space complexities of the proposed algorithm are discussed in
section III. The numerical simulations of the proposed algorithm and
the MIMO RADAR angle-range imaging experimental results of the generated
sequence set are given in section IV and finally section V concludes
the paper.

Throughout the paper the following mathematical notations are used
hereafter: Matrices are indicated by the boldface upper case letters,
column vectors are indicated by the boldface lowercase letters and
the scalars are indicated by the italics. Complex conjugate, conjugate
transpose, and the transpose are indicated by the superscripts $()^{*},()^{H},()^{T}$.
Trace of a matrix is denoted by $\text{Tr}()$ . $\boldsymbol{z}_{i}(m)$
denote the $m^{th}$ element of a vector $\boldsymbol{\boldsymbol{z}}_{i}$.
$\boldsymbol{\boldsymbol{I}}_{w}$ denote the $w\times w$ identity
matrix. $\boldsymbol{\boldsymbol{O}}_{w}$ denote the $w\times w$
matrix full of zeros. $||.||_{2}$ denote the $l_{2}$ norm. $\text{vec}(\boldsymbol{\boldsymbol{A}})$
is a column vector stacked with all the columns of matrix-$\boldsymbol{\boldsymbol{A}}$.
$\left|.\right|^{2}$ denote the absolute squared value. The maximum
eigenvalue of $\boldsymbol{\boldsymbol{\boldsymbol{A}}}$ is denoted
by $\lambda_{\text{max}}(\boldsymbol{\boldsymbol{\boldsymbol{A}}})$.
$\mathbb{R}$ and $\mathbb{C}$ represent the real and complex fields.
The real and imaginary parts are denoted by $\text{Re}(.)$ and $\text{Im}(.)$
respectively. $\nabla g(.)$ (or) $\partial g(.)$ denote the gradient
of a function $g(.).$ $\boldsymbol{\boldsymbol{z}}^{t}$ denote the
value of $\boldsymbol{\boldsymbol{z}}$ at $t^{th}$ iteration. $\bigodot$
represents the Hadamard product.

\section{{PROBLEM FORMULATION AND MAJORIZATION-MINIMIZATION METHOD}}

\subsection{PROBLEM FORMULATION }

Let $\left\{ \boldsymbol{\boldsymbol{s}}_{1},\boldsymbol{\boldsymbol{s}}_{2},...,\boldsymbol{\boldsymbol{s}}_{L}\right\} $
denote the $L$ number of $M$ length probing phase sequences. The
Peak Side-lobe Level (PSL) metric which is our design criterion to
evaluate the quality of the sequence-set is defined as:

\begin{equation}
\text{PSL}=\text{\text{\ensuremath{\underset{k\in a_{k}}{\text{\text{max}}}}}}\left\{ \Bigl|r_{i,j}(k)\Bigr|\right\} ,\,i,j=1,2,..,L,\label{eq:PSL-1}
\end{equation}

where

\[
a_{k}=\begin{cases}
1,2,...,(M-1) & \text{if}\,i=j\\
0,1,2,..,(M-1) & \text{if}\,i\neq j
\end{cases}
\]

and $r_{i,j}(k)$ is the aperiodic cross-correlation of the sequences
$\boldsymbol{\boldsymbol{s}}_{i},\boldsymbol{\boldsymbol{s}}_{j}$
at lag $k$, which is defined as:

\begin{equation}
r_{i,j}(k)=\sum_{m=1}^{M-k}s_{i}^{*}(m)s_{j}(m+k)=r_{j,i}^{*}(-k)\label{eq:acor}
\end{equation}

which will be the auto-correlation function when $i=j$. Since we
are interested in sequence set design with each element of any sequence
to have constant modulus, the problem of interest can be formulated
as:

\begin{equation}
\begin{aligned} & \underset{\left\{ \boldsymbol{\boldsymbol{s}}_{1},\boldsymbol{\boldsymbol{s}}_{2},.,\boldsymbol{\boldsymbol{s}}_{L}\right\} }{\text{\text{min}}} &  & \text{\ensuremath{\underset{k\in a_{k}}{\text{\text{max}}}}}\hphantom{nn}\Bigl|r_{i,j}(k)\Bigr|,\,i,j=1,..,L,\\
 & \text{ subject to} &  & \bigl|\boldsymbol{\boldsymbol{s}}_{i}(m)\bigr|=1,\,i=1,..,L;m=1,..,M.
\end{aligned}
\label{eq:PSLprob}
\end{equation}
The problem in (\ref{eq:PSLprob}) is a minimax problem and is non-convex
in nature. Moreover, the objective function (PSL metric) is non-differentiable
(due to the maximum), which makes the problem even more challenging.
Before we move on to the presentation of our algorithm, we will present
a brief summary of the Majorization-Minimization (MM) framework, which
would be central in the development of our algorithm.

\subsection{Majorization-Minimization Method }

In this sub-section, we will briefly discuss the general Majorization-Minimization
method (which is defined for minimization problems) and its extension
to the minimax optimization problems.

\subsubsection{Majorization-Minimization for minimization problems}

Majorization-Minimization (MM) method was first developed by De Leeuw
while solving the multidimensional scaling problem \cite{Leeuw}.
Later, after realizing its strength to solve non-convex (or) even
convex problems efficiently, many researchers adopted the technique
to address problems in various fields like signal processing, communications,
machine learning, etc. Consider a minimization problem as follows:

\begin{equation}
\min_{\boldsymbol{\boldsymbol{x}\in\boldsymbol{\boldsymbol{\boldsymbol{\chi}}}}}f(\boldsymbol{\boldsymbol{x}})\label{eq:linfun}
\end{equation}
where $f(\boldsymbol{\boldsymbol{x}})$ is some non-linear function
and $\boldsymbol{\boldsymbol{\boldsymbol{\chi}}}$ denotes the constraint
set. The MM approach is two-step method \cite{23_Tutorial_MM} in
which at the first step, at any given point $\boldsymbol{\boldsymbol{\boldsymbol{x}}}^{t}$
($\boldsymbol{\boldsymbol{\boldsymbol{x}}}$ at $t^{th}$ iteration),
an upper bound (majorization) function $u(\boldsymbol{\boldsymbol{\boldsymbol{x}}}|\boldsymbol{\boldsymbol{\boldsymbol{x}}}^{t})$
to the original objective function $f(\boldsymbol{\boldsymbol{x}})$
is constructed, and in the second step the upper bound function $u(\boldsymbol{\boldsymbol{\boldsymbol{x}}}|\boldsymbol{\boldsymbol{\boldsymbol{x}}}^{t})$
is minimized to obtain the next iterate point $\boldsymbol{\boldsymbol{\boldsymbol{x}}}^{t+1}$.
Again at $\boldsymbol{\boldsymbol{\boldsymbol{x}}}^{t+1}$, the above
mentioned two steps are implemented and the series of these steps
will continue until the optimum stationary point of an original function
$f(\boldsymbol{\boldsymbol{x}})$ is reached. The majorization function
$u(\boldsymbol{\boldsymbol{\boldsymbol{x}}}|\boldsymbol{\boldsymbol{\boldsymbol{x}}}^{t})$
(which is constructed in the first step) has to satisfy the following
properties:

\begin{equation}
u(\boldsymbol{\boldsymbol{\boldsymbol{x}}}^{t}|\boldsymbol{\boldsymbol{\boldsymbol{x}}}^{t})=f(\boldsymbol{\boldsymbol{\boldsymbol{x}}}^{t}),\quad\forall\boldsymbol{\boldsymbol{x}}\in\boldsymbol{\boldsymbol{\boldsymbol{\chi}}}.\label{eq:5-1}
\end{equation}

\begin{equation}
u(\boldsymbol{\boldsymbol{x}}|\boldsymbol{\boldsymbol{\boldsymbol{x}}}^{t})\geq f(\boldsymbol{\boldsymbol{x}}),\quad\forall\boldsymbol{\boldsymbol{x}}\in\boldsymbol{\boldsymbol{\boldsymbol{\chi}}}.\label{eq:6-1}
\end{equation}

For a given problem, more than one possible majorization function
will exist and this gives a possibility to develop two different iterative
algorithms for the same problem. A survey to construct the majorizing
function is given in \cite{22_MM_prabhubabu}. The objective function
value evaluated at every iteration generated by MM will satisfy the
descent property, i.e. 
\begin{equation}
f(\boldsymbol{\boldsymbol{\boldsymbol{x}}}^{t+1})\leq u(\boldsymbol{\boldsymbol{\boldsymbol{x}}}^{t+1}|\boldsymbol{\boldsymbol{\boldsymbol{x}}}^{t})\leq u(\boldsymbol{\boldsymbol{\boldsymbol{x}}}^{t}|\boldsymbol{\boldsymbol{\boldsymbol{x}}}^{t})=f(\boldsymbol{\boldsymbol{\boldsymbol{x}}}^{t}).\label{eq:MM-1}
\end{equation}

\subsubsection{Majorization-Minimization for minimax problems}

Consider the minimax problem as follows:

\begin{equation}
\min_{\boldsymbol{\boldsymbol{y}\in\boldsymbol{\boldsymbol{\boldsymbol{\chi}}}}}g(\boldsymbol{\boldsymbol{y}}),\label{minmax}
\end{equation}

where $g(\boldsymbol{\boldsymbol{y}})=\underset{k=1,..,N}{\text{max}}g_{k}(\boldsymbol{\boldsymbol{y}})$.
Similar to the case of minimization problems, the majorization function
for the objective in (\ref{minmax}) can be constructed as follows:

\begin{equation}
u(\boldsymbol{\boldsymbol{y}}|\boldsymbol{\boldsymbol{y}}^{t})=\underset{k=1,..,N}{\text{max}}\tilde{u}_{k}(\boldsymbol{\boldsymbol{y}}|\boldsymbol{\boldsymbol{y}}^{t}),\label{eq:ubd2}
\end{equation}

where each $\tilde{u}_{k}(\boldsymbol{\boldsymbol{y}}|\boldsymbol{\boldsymbol{y}}^{t})$
is an upper bound for the respective $g_{k}(\boldsymbol{\boldsymbol{y}})$
at any given $\boldsymbol{\boldsymbol{y}}^{t},\,\forall k$. Here,
every majorization function $\tilde{u}_{k}(\boldsymbol{\boldsymbol{y}}|\boldsymbol{\boldsymbol{y}}^{t}),\,\forall k$
will also satisfy the conditions mentioned in (\ref{eq:5-1}), (\ref{eq:6-1})
i.e.,

\begin{equation}
\tilde{u}_{k}(\boldsymbol{\boldsymbol{y}}^{t}|\boldsymbol{\boldsymbol{y}}^{t})=g_{k}(\boldsymbol{\boldsymbol{y}}^{t}),\,\forall k,\boldsymbol{y}\in\boldsymbol{\boldsymbol{\boldsymbol{\chi}}}\label{eq:mimaeq}
\end{equation}

\begin{equation}
\tilde{u}_{k}(\boldsymbol{\boldsymbol{y}}|\boldsymbol{\boldsymbol{y}}^{t})\geq g_{k}(\boldsymbol{\boldsymbol{y}}),\,\forall k,\boldsymbol{y}\in\boldsymbol{\boldsymbol{\boldsymbol{\chi}}}\label{eq:mimaubnd}
\end{equation}

It can be easily shown that the choice of $u(\boldsymbol{\boldsymbol{y}}|\boldsymbol{\boldsymbol{y}}^{t})$
in (\ref{eq:ubd2}) is a global upper bound function of $g(\boldsymbol{\boldsymbol{y}})$,
i.e.,

\begin{equation}
u(\boldsymbol{\boldsymbol{y}}^{t}|\boldsymbol{\boldsymbol{y}}^{t})=\underset{k=1,..,N}{\text{max}}\tilde{u}_{k}(\boldsymbol{\boldsymbol{y}}^{t}|\boldsymbol{\boldsymbol{y}}^{t})=\underset{k=1,..,N}{\text{max}}g_{k}(\boldsymbol{\boldsymbol{y}}^{t})=g(\boldsymbol{\boldsymbol{y}}^{t})\label{eq:finsreq}
\end{equation}

\begin{equation}
\begin{array}{cc}
\tilde{u}_{k}(\boldsymbol{\boldsymbol{y}}|\boldsymbol{\boldsymbol{y}}^{t})\geq g_{k}(\boldsymbol{\boldsymbol{y}}) & \Leftrightarrow\underset{k=1,..,N}{\text{max}}\tilde{u}_{k}(\boldsymbol{\boldsymbol{y}}|\boldsymbol{\boldsymbol{y}}^{t})\geq\underset{k=1,..,N}{\text{max}}g_{k}(\boldsymbol{\boldsymbol{y}})\\
\Leftrightarrow u(\boldsymbol{\boldsymbol{y}}|\boldsymbol{\boldsymbol{y}}^{t})\geq g(\boldsymbol{\boldsymbol{y}})
\end{array}\label{eq:finubsgf}
\end{equation}

Similar to the MM for minimization problems, here too the sequence
of points $\left\{ \boldsymbol{\boldsymbol{y}}\right\} =\boldsymbol{\boldsymbol{y}}^{1},\boldsymbol{\boldsymbol{\boldsymbol{y}}}^{2},\boldsymbol{\boldsymbol{\boldsymbol{y}}}^{3},.....,\boldsymbol{\boldsymbol{y}}^{m}$
obtained via the MM update rule will monotonically decrease the objective
function.

\section{{PROPOSED ALGORITHM}}

\subsection{PROPOSED ALGORITHM }

In this section, we present our algorithm which monotonically minimizes
the PSL metric. So we start with the PSL minimization problem (\ref{eq:PSLprob}):
\begin{equation}
\begin{aligned} & \underset{\left\{ \boldsymbol{\boldsymbol{s}}_{1},\boldsymbol{\boldsymbol{s}}_{2},.,\boldsymbol{\boldsymbol{s}}_{L}\right\} }{\text{\text{min}}} &  & \text{\ensuremath{\underset{k\in a_{k}}{\text{\text{max}}}}}\;\;2\Bigl|r_{i,j}(k)\Bigr|^{2},\,i,j=1,..,L,\\
 & \text{ subject to} &  & \bigl|\boldsymbol{\boldsymbol{s}}_{i}(m)\bigr|=1,\,i=1,..,L;m=1,..,M.
\end{aligned}
\label{eq:Pslprob2}
\end{equation}
Please note that we have squared the objective function (as squaring
the absolute valued objective will not change the optimum) and have
also scaled the objective by a factor $2$, we have done these things
for future convenience. Let us define

\begin{equation}
\begin{aligned}\boldsymbol{\boldsymbol{s}} & =[\boldsymbol{\boldsymbol{s}}_{1}^{T},\boldsymbol{\boldsymbol{s}}_{2}^{T},...,\boldsymbol{\boldsymbol{s}}_{L}^{T}]^{T},\\
\boldsymbol{\boldsymbol{\boldsymbol{B}}}_{s} & =\Bigl[\boldsymbol{\boldsymbol{\boldsymbol{0}}}_{M\times(s-1)M},\boldsymbol{\boldsymbol{\boldsymbol{I}}}_{M},\boldsymbol{\boldsymbol{\boldsymbol{0}}}_{M\times(L-s)M}\Bigr],\,s=1,..,L,\\
\boldsymbol{\boldsymbol{\boldsymbol{A}}}_{k} & =\begin{cases}
1 & ;b-a=k\\
0 & ;else
\end{cases},\,k=0,..,(M-1),
\end{aligned}
\label{eq:matrices}
\end{equation}
where $\boldsymbol{\boldsymbol{s}}$ is a $ML\times1$ dimensional
sequence vector consisting of all the $L$ number of $M$ length sequences
stacked one above the other, $\boldsymbol{\boldsymbol{\boldsymbol{B}}}_{s}$
is the $M\times ML$ dimensional block selection matrix and $\boldsymbol{\boldsymbol{\boldsymbol{A}}}_{k}$
is the $M\times M$ dimensional Toeplitz shift matrix with $a,b$
denoting its row and column indexes respectively. By using (\ref{eq:matrices}),
we can write $r_{i,j}(k)=\boldsymbol{\boldsymbol{\boldsymbol{s}}}_{i}^{H}\boldsymbol{\boldsymbol{\boldsymbol{A}}}_{k}\boldsymbol{\boldsymbol{\boldsymbol{s}}}_{j}=\bigl(\boldsymbol{\boldsymbol{\boldsymbol{B}}}_{i}\boldsymbol{\boldsymbol{\boldsymbol{s}}}\bigr)^{H}\boldsymbol{\boldsymbol{\boldsymbol{A}}}_{k}\bigl(\boldsymbol{\boldsymbol{\boldsymbol{B}}}_{j}\boldsymbol{\boldsymbol{\boldsymbol{s}}}\bigr)$.
Then the cost function in problem (\ref{eq:Pslprob2}) can be rewritten
as:

\begin{equation}
\begin{aligned}2\Bigl|r_{i,j}(k)\Bigr|^{2} & =\Bigl|\bigl(\boldsymbol{\boldsymbol{\boldsymbol{B}}}_{i}\boldsymbol{\boldsymbol{\boldsymbol{s}}}\bigr)^{H}\boldsymbol{\boldsymbol{\boldsymbol{A}}}_{k}\bigl(\boldsymbol{\boldsymbol{\boldsymbol{B}}}_{j}\boldsymbol{\boldsymbol{\boldsymbol{s}}}\bigr)\Bigr|^{2}+\Bigl|\bigl(\boldsymbol{\boldsymbol{\boldsymbol{B}}}_{i}\boldsymbol{\boldsymbol{\boldsymbol{s}}}\bigr)^{H}\boldsymbol{\boldsymbol{\boldsymbol{A}}}_{k}^{H}\bigl(\boldsymbol{\boldsymbol{\boldsymbol{B}}}_{j}\boldsymbol{\boldsymbol{\boldsymbol{s}}}\bigr)\Bigr|^{2}\\
 & =\Bigl|\boldsymbol{\boldsymbol{\boldsymbol{s}}}^{H}\Bigl(\bar{\boldsymbol{\boldsymbol{\boldsymbol{A}}}}_{i,j}(k)\Bigr)\boldsymbol{\boldsymbol{\boldsymbol{s}}}\Bigr|^{2}+\Bigl|\boldsymbol{\boldsymbol{\boldsymbol{s}}}^{H}\Bigl(\bar{\boldsymbol{\boldsymbol{\boldsymbol{A}}}}_{i,j}(k)\Bigr)^{H}\boldsymbol{\boldsymbol{\boldsymbol{s}}}\Bigr|^{2}
\end{aligned}
\label{eq:rk}
\end{equation}

where $\bar{\boldsymbol{\boldsymbol{\boldsymbol{A}}}}_{i,j}(k)=\bigl(\boldsymbol{\boldsymbol{\boldsymbol{B}}}_{i}\bigr)^{H}\boldsymbol{\boldsymbol{\boldsymbol{A}}}_{k}\bigl(\boldsymbol{\boldsymbol{\boldsymbol{B}}}_{j}\bigr).$

By defining $\boldsymbol{\boldsymbol{\boldsymbol{S}}}=\boldsymbol{\boldsymbol{s}}\boldsymbol{\boldsymbol{s}}^{H}$,
(\ref{eq:rk}) can be further rewritten as:

\begin{equation}
\begin{array}{ccc}
 & \Bigl|\boldsymbol{\boldsymbol{\boldsymbol{s}}}^{H}\Bigl(\bar{\boldsymbol{\boldsymbol{\boldsymbol{A}}}}_{i,j}(k)\Bigr)\boldsymbol{\boldsymbol{\boldsymbol{s}}}\Bigr|^{2}+\Bigl|\boldsymbol{\boldsymbol{\boldsymbol{s}}}^{H}\Bigl(\bar{\boldsymbol{\boldsymbol{\boldsymbol{A}}}}_{i,j}(k)\Bigr)^{H}\boldsymbol{\boldsymbol{\boldsymbol{s}}}\Bigr|^{2}\\
 & =\text{Tr}\Bigl(\Bigl(\bar{\boldsymbol{\boldsymbol{\boldsymbol{A}}}}_{i,j}(k)\Bigr)\boldsymbol{\boldsymbol{\boldsymbol{S}}}\Bigr)\text{Tr}\Bigl(\Bigl(\bar{\boldsymbol{\boldsymbol{\boldsymbol{A}}}}_{i,j}(k)\Bigr)^{H}\boldsymbol{\boldsymbol{\boldsymbol{S}}}\Bigr)\\
 & \hphantom{nnnnnn}+\text{Tr}\Bigl(\boldsymbol{\boldsymbol{\boldsymbol{S}}}\Bigl(\bar{\boldsymbol{\boldsymbol{\boldsymbol{A}}}}_{i,j}(k)\Bigr)^{H}\Bigr)\text{Tr}\Bigl(\Bigl(\bar{\boldsymbol{\boldsymbol{\boldsymbol{A}}}}_{i,j}(k)\Bigr)\boldsymbol{\boldsymbol{\boldsymbol{S}}}\Bigr).
\end{array}\label{eq:treq}
\end{equation}

By using (\ref{eq:treq}) and the relation $\text{Tr}\Bigl(\Bigl(\bar{\boldsymbol{\boldsymbol{\boldsymbol{A}}}}_{i,j}(k)\Bigr)\boldsymbol{\boldsymbol{\boldsymbol{S}}}\Bigr)=\text{vec}^{H}\Bigl(\boldsymbol{\boldsymbol{\boldsymbol{S}}}\Bigr)\text{vec}\Bigl(\bar{\boldsymbol{\boldsymbol{\boldsymbol{A}}}}_{i,j}(k)\Bigr)$,
problem in (\ref{eq:Pslprob2}) can be rewritten as:

\begin{equation}
\begin{aligned} & \underset{\boldsymbol{\boldsymbol{\boldsymbol{s}}},\boldsymbol{\boldsymbol{\boldsymbol{\boldsymbol{S}}}}}{\text{\text{min}}} &  & \text{\ensuremath{\underset{k\in a_{k}}{\text{\text{max}}}}}\;\text{vec}^{H}\Bigl(\boldsymbol{\boldsymbol{\boldsymbol{S}}}\Bigr)\Bigl(\boldsymbol{\boldsymbol{\boldsymbol{\Phi}}}_{i,j}(k)\Bigr)\text{vec}\Bigl(\boldsymbol{\boldsymbol{\boldsymbol{S}}}\Bigr),\,i,j=1,.,L,\\
 & \text{ s.t} &  & \boldsymbol{\boldsymbol{\boldsymbol{S}}}=\boldsymbol{\boldsymbol{s}}\boldsymbol{\boldsymbol{s}}^{H},\\
 &  &  & \bigl|\boldsymbol{\boldsymbol{s}}(m)\bigr|=1,\,m=1,..,ML,
\end{aligned}
\label{eq:vecprob}
\end{equation}
where

$\begin{array}{c}
\boldsymbol{\boldsymbol{\boldsymbol{\Phi}}}_{i,j}(k)=\text{vec}\Bigl(\bar{\boldsymbol{\boldsymbol{\boldsymbol{A}}}}_{i,j}(k)\Bigr)\text{vec}^{H}\Bigl(\Bigl(\bar{\boldsymbol{\boldsymbol{\boldsymbol{A}}}}_{i,j}(k)\Bigr)^{H}\Bigr)\\
\hphantom{nnnnnnnnnn}+\text{vec}\Bigl(\Bigl(\bar{\boldsymbol{\boldsymbol{\boldsymbol{A}}}}_{i,j}(k)\Bigr)^{H}\Bigr)\text{vec}^{H}\Bigl(\bar{\boldsymbol{\boldsymbol{\boldsymbol{A}}}}_{i,j}(k)\Bigr)
\end{array}$ 

is a matrix of dimension $(ML)^{2}\times(ML)^{2}$ and here s.t stands
for subject to.

It's worth noting that the objective function in (\ref{eq:vecprob})
is a quadratic function in the auxiliary variable $\boldsymbol{S}$.
In the following we will present a lemma using which we can find a
tighter upper bound for the objective in (\ref{eq:vecprob}) at any
give $\boldsymbol{S}^{t}$ (which of course can be obtained from any
$\boldsymbol{s}^{t}$).

\textbf{Lemma-1}: Let $g:\mathbb{\mathbb{C}}^{N}\rightarrow\mathbb{R}$
be a continuously twice differentiable function with a bounded curvature,
then there exists a matrix $\boldsymbol{\boldsymbol{C}}\succeq\nabla^{2}g(\boldsymbol{\boldsymbol{\boldsymbol{z}}})$
such that at any fixed point $\boldsymbol{\boldsymbol{z}}^{t}$, $g(\boldsymbol{\boldsymbol{\boldsymbol{z}}})$
can be majorized as:

\begin{equation}
g(\boldsymbol{\boldsymbol{\boldsymbol{z}}})\leq g(\boldsymbol{\boldsymbol{\boldsymbol{z}}}^{t})+\textrm{Re}(\nabla g(\boldsymbol{\boldsymbol{\boldsymbol{z}}}^{t})^{H}(\boldsymbol{\boldsymbol{\boldsymbol{z}}}-\boldsymbol{\boldsymbol{\boldsymbol{z}}}^{t}))+\frac{1}{2}(\boldsymbol{\boldsymbol{\boldsymbol{z}}}-\boldsymbol{\boldsymbol{\boldsymbol{z}}}^{t})^{H}\boldsymbol{\boldsymbol{C}}(\boldsymbol{\boldsymbol{\boldsymbol{z}}}-\boldsymbol{\boldsymbol{\boldsymbol{z}}}^{t}).\label{eq:surg}
\end{equation}

\textbf{Proof }: By using second order Taylor series expansion, any
quadratic function can be written as follows:

\begin{equation}
\begin{array}{cc}
g(\boldsymbol{\boldsymbol{\boldsymbol{z}}})= & g(\boldsymbol{\boldsymbol{\boldsymbol{z}}}^{t})+\textrm{Re}(\nabla g(\boldsymbol{\boldsymbol{\boldsymbol{z}}}^{t})^{H}(\boldsymbol{\boldsymbol{\boldsymbol{z}}}-\boldsymbol{\boldsymbol{\boldsymbol{z}}}^{t}))\\
 & +\frac{1}{2}(\boldsymbol{\boldsymbol{\boldsymbol{z}}}-\boldsymbol{\boldsymbol{\boldsymbol{z}}}^{t})^{H}\nabla^{2}g(\boldsymbol{\boldsymbol{\boldsymbol{z}}}^{t})(\boldsymbol{\boldsymbol{\boldsymbol{z}}}-\boldsymbol{\boldsymbol{z}}^{t}).
\end{array}\label{eq:equalty}
\end{equation}

If such a quadratic function has bounded curvature, then there exist
a matrix $\boldsymbol{\boldsymbol{C}}\succeq\nabla^{2}g(\boldsymbol{\boldsymbol{\boldsymbol{z}}})$,
such that $g(z)$ is upper bounded as follows:

\[
g(\boldsymbol{\boldsymbol{\boldsymbol{z}}})\leq g(\boldsymbol{\boldsymbol{\boldsymbol{z}}}^{t})+\textrm{Re}(\nabla g(\boldsymbol{\boldsymbol{\boldsymbol{z}}}^{t})^{H}(\boldsymbol{\boldsymbol{\boldsymbol{z}}}-\boldsymbol{\boldsymbol{\boldsymbol{z}}}^{t}))+\frac{1}{2}(\boldsymbol{\boldsymbol{\boldsymbol{z}}}-\boldsymbol{\boldsymbol{\boldsymbol{z}}}^{t})^{H}\boldsymbol{\boldsymbol{C}}(\boldsymbol{\boldsymbol{\boldsymbol{z}}}-\boldsymbol{\boldsymbol{\boldsymbol{z}}}^{t}).
\]

Hence, it concludes the proof. $\hphantom{nnnnnnnnnnnnnnnnnn}$$\blacksquare$

Example: Let $g(\boldsymbol{\boldsymbol{\boldsymbol{z}}})=\boldsymbol{\boldsymbol{z}}^{H}\boldsymbol{\boldsymbol{G}}\boldsymbol{\boldsymbol{z}}$
be any quadratic function, then by using the Lemma-1 we can majorize
it as:

\[
g(\boldsymbol{\boldsymbol{\boldsymbol{z}}})\leq(\boldsymbol{\boldsymbol{z}}^{t})^{H}\boldsymbol{\boldsymbol{G}}\boldsymbol{\boldsymbol{z}}^{t}+\textrm{Re}((2\boldsymbol{\boldsymbol{G}}\boldsymbol{\boldsymbol{z}}^{t})^{H}(\boldsymbol{\boldsymbol{\boldsymbol{z}}}-\boldsymbol{\boldsymbol{z}}^{t}))+(\boldsymbol{\boldsymbol{\boldsymbol{z}}}-\boldsymbol{\boldsymbol{\boldsymbol{z}}}^{t})^{H}\boldsymbol{\boldsymbol{C}}(\boldsymbol{\boldsymbol{\boldsymbol{z}}}-\boldsymbol{\boldsymbol{\boldsymbol{z}}}^{t}),
\]

where $\boldsymbol{\boldsymbol{C}}=\lambda_{\text{max}}(\boldsymbol{\boldsymbol{G}})\boldsymbol{\boldsymbol{I}}$
and the above majorized function can be rearranged as:

\begin{equation}
g(\boldsymbol{\boldsymbol{\boldsymbol{z}}})\leq-(\boldsymbol{\boldsymbol{\boldsymbol{z}}}^{t})^{H}(\boldsymbol{\boldsymbol{G}}-\boldsymbol{\boldsymbol{C}})\boldsymbol{\boldsymbol{\boldsymbol{z}}}^{t}+2\text{Re}((\boldsymbol{\boldsymbol{z}}^{t})^{H}(\boldsymbol{\boldsymbol{G}}-\boldsymbol{\boldsymbol{C}})\boldsymbol{\boldsymbol{\boldsymbol{z}}})+\boldsymbol{\boldsymbol{z}}{}^{H}\boldsymbol{\boldsymbol{C}}\boldsymbol{\boldsymbol{z}}.\label{eq:Maj}
\end{equation}

The problem in (\ref{eq:vecprob}) is quadratic in $\textrm{vec}(\boldsymbol{\boldsymbol{\boldsymbol{S}}})$
and by using the Lemma-1, at any given point $\boldsymbol{\boldsymbol{\boldsymbol{S}}}^{t}$
we can majorize the objective as follows:

\begin{equation}
\begin{array}{c}
\!\!\!\!\!\!\!\!\!\!\!\!\!\!\!\!\!\!\!\!\!\!\!\!\!\!\!\!\!\!\!\!\!\!\!\!\!\!\!\!\!\!\!\!\!\!\!\!\text{vec}^{H}\Bigl(\boldsymbol{\boldsymbol{\boldsymbol{S}}}\Bigr)\Bigl(\boldsymbol{\boldsymbol{\boldsymbol{\Phi}}}_{i,j}(k)\Bigr)\text{vec}\Bigl(\boldsymbol{\boldsymbol{\boldsymbol{S}}}\Bigr)\leq\\
\hphantom{nn}\Biggl(-\text{vec}^{H}\Bigl(\boldsymbol{\boldsymbol{\boldsymbol{S}}}^{t}\Bigr)\Bigl(\boldsymbol{\boldsymbol{\boldsymbol{\Phi}}}_{i,j}(k)-\boldsymbol{\boldsymbol{\boldsymbol{C}}}_{i,j}(k)\Bigr)\text{vec}\Bigl(\boldsymbol{\boldsymbol{\boldsymbol{S}}}^{t}\Bigr)\\
\hphantom{nnn}+2\text{Re}\Bigl(\text{vec}^{H}\Bigl(\boldsymbol{\boldsymbol{\boldsymbol{S}}}^{t}\Bigr)\Bigl(\boldsymbol{\boldsymbol{\boldsymbol{\Phi}}}_{i,j}(k)-\boldsymbol{\boldsymbol{\boldsymbol{C}}}_{i,j}(k)\Bigr)\text{vec}\Bigl(\boldsymbol{\boldsymbol{\boldsymbol{S}}}\Bigr)\Bigr)\\
\hphantom{nn}+\text{vec}^{H}\Bigl(\boldsymbol{\boldsymbol{\boldsymbol{S}}}\Bigr)\Bigl(\boldsymbol{\boldsymbol{\boldsymbol{C}}}_{i,j}(k)\Bigr)\text{vec}\Bigl(\boldsymbol{\boldsymbol{\boldsymbol{S}}}\Bigr)\Biggr),
\end{array}\label{eq:MajOrg}
\end{equation}

where $\boldsymbol{\boldsymbol{\boldsymbol{C}}}_{i,j}(k)=\lambda_{\text{max}}\Bigl(\boldsymbol{\boldsymbol{\boldsymbol{\Phi}}}_{i,j}(k)\Bigr)\boldsymbol{\boldsymbol{I}}_{(ML)^{2}}.$

It can be noted that for obtaining the upper bound function, one has
to calculate the maximum eigenvalue of $\boldsymbol{\Phi}_{i,j}(k)$.
In the following lemma, we prove that the maximum eigenvalue of $\boldsymbol{\Phi}_{i,j}(k)$
can be obtained in closed form.

\textbf{Lemma-}2: The maximum eigenvalue of the $(ML)^{2}\times(ML)^{2}$
dimension sparse matrix $\boldsymbol{\boldsymbol{\boldsymbol{\Phi}}}_{i,j}(k)$
is equal to $(M-k),\,\forall k\in a_{k}.$

\textbf{Proof }: Let $a=\bar{\boldsymbol{\boldsymbol{\boldsymbol{A}}}}_{i,j}(k)$,
$a_{k}=\text{vec}(a)$, $b_{k}=\text{vec}(a^{H})$, then $\boldsymbol{\boldsymbol{\boldsymbol{\Phi}}}_{i,j}(k)=a_{k}b_{k}^{H}+b_{k}a_{k}^{H}$,
which is an aggregation of two rank-1 matrices and its maximum possible
rank is $2.$

Let $\mu_{1},\mu_{2}$ are the two different eigen values of $\boldsymbol{\boldsymbol{\boldsymbol{\Phi}}}_{i,j}(k)$
and its corresponding characteristic equation is given by: 
\begin{equation}
x^{2}-(\mu_{1}+\mu_{2})x+(\mu_{1}\mu_{2})=0.\label{eq:Cheqn}
\end{equation}

We know that $a_{k}b_{k}^{H}$ is a $(ML)^{2}\times(ML)^{2}$ dimensional
sparse matrix filled with zeros along the diagonal. Hence,

\begin{equation}
\mu_{1}+\mu_{2}=\text{Tr}\bigl(a_{k}b_{k}^{H}+b_{k}a_{k}^{H}\bigr)=0\label{eq:trc}
\end{equation}

We have the relation $\mu_{1}\mu_{2}=\frac{1}{2}((\mu_{1}+\mu_{2})^{2}-(\mu_{1}^{2}+\mu_{2}^{2}))$
and by using (\ref{eq:trc}), it becomes as $\mu_{1}\mu_{2}=-\frac{1}{2}(\mu_{1}^{2}+\mu_{2}^{2})$.

We know that 
\[
\mu_{1}^{2}+\mu_{2}^{2}=2\text{Tr}\bigl(\bigl(a_{k}b_{k}^{H}\bigr)\bigl(b_{k}a_{k}^{H}\bigr)\bigr)=2\bigl\Vert a_{k}\bigr\Vert_{2}^{2}\bigl\Vert b_{k}\bigr\Vert_{2}^{2}
\]

Since the vectors $a_{k}$ and $b_{k}$ have only $(M-k)$ number
of ones and remaining elements as zeros, we get $\bigl\Vert a_{k}\bigr\Vert_{2}^{2}=M-k$
and $\bigl\Vert b_{k}\bigr\Vert_{2}^{2}=M-k$, then

\begin{equation}
\mu_{1}^{2}+\mu_{2}^{2}=2\bigl\Vert a_{k}\bigr\Vert_{2}^{2}\bigl\Vert b_{k}\bigr\Vert_{2}^{2}=2\bigl(M-k\bigr)^{2}\label{eq:prod}
\end{equation}

By using (\ref{eq:trc}) and (\ref{eq:prod}), the characteristic
equation (\ref{eq:Cheqn}) becomes as $x^{2}-\bigl(M-k\bigr)^{2}=0$,
which implies $x=\pm(M-k)$. Among the two possibilities, the maximum
will be $(M-k)$ and this concludes the proof.$\hphantom{nnnnnnnnnnnnnnnnnnnnnnnnnnnnnnnnnnnn}\blacksquare$

So, according to the Lemma-2 the maximum eigenvalue of $\boldsymbol{\boldsymbol{\boldsymbol{\Phi}}}_{i,j}(k)$
is taken as $M-k\,\Bigl(i.e.\,\lambda_{\text{max}}\Bigl(\boldsymbol{\boldsymbol{\boldsymbol{\Phi}}}_{i,j}(k)\Bigr)=(M-k),\,\forall k\in a_{k}\Bigr)$.
Since $\text{vec}^{H}(\boldsymbol{\boldsymbol{\boldsymbol{S}}})\text{vec}(\boldsymbol{\boldsymbol{\boldsymbol{S}}})=(\boldsymbol{\boldsymbol{s}}^{H}\boldsymbol{\boldsymbol{s}})^{2}=ML$,
the surrogate function in (\ref{eq:MajOrg}) can be rewritten as:

\begin{equation}
\begin{aligned}u_{i,j,k}\Bigl(\boldsymbol{\boldsymbol{\boldsymbol{S}}}|\boldsymbol{\boldsymbol{\boldsymbol{S}}}^{t}\Bigr)= & -\text{vec}^{H}\Bigl(\boldsymbol{\boldsymbol{\boldsymbol{S}}}^{t}\Bigr)\Bigl(\boldsymbol{\boldsymbol{\boldsymbol{\Phi}}}_{i,j}(k)\Bigr)\text{vec}\Bigl(\boldsymbol{\boldsymbol{\boldsymbol{S}}}^{t}\Bigr)\\
 & +2\text{Re}\Bigl(\text{vec}^{H}\Bigl(\boldsymbol{\boldsymbol{\boldsymbol{S}}}^{t}\Bigr)\Bigl(\boldsymbol{\boldsymbol{\boldsymbol{\Phi}}}_{i,j}(k)\Bigr)\text{vec}\Bigl(\boldsymbol{\boldsymbol{\boldsymbol{S}}}\Bigr)\Bigr)\\
 & -2\Bigl(M-k\Bigr)\text{Re}\Bigl(\text{vec}^{H}\Bigl(\boldsymbol{\boldsymbol{\boldsymbol{S}}}^{t}\Bigr)\text{vec}\Bigl(\boldsymbol{\boldsymbol{\boldsymbol{S}}}\Bigr)\Bigr)\\
 & +2\Bigl(M-k\Bigr)ML.
\end{aligned}
\label{eq:vecfinal}
\end{equation}

By substituting back $\boldsymbol{\boldsymbol{\boldsymbol{S}}}=\boldsymbol{\boldsymbol{s}}\boldsymbol{\boldsymbol{s}}^{H}$,
the surrogate function in (\ref{eq:vecfinal}) can be expressed in
the original variable $\boldsymbol{s}$ as follows:

\begin{equation}
\begin{aligned}u_{i,j,k}\Bigl(\boldsymbol{\boldsymbol{\boldsymbol{s}}}|\boldsymbol{\boldsymbol{\boldsymbol{s}}}^{t}\Bigr)= & -2\left|a\right|^{2}+2\left(ab^{H}+ba^{H}\right)\\
 & -2\Bigl(M-k\Bigr)\Bigl(\boldsymbol{\boldsymbol{\boldsymbol{s}}}^{H}\boldsymbol{\boldsymbol{\boldsymbol{s}}}^{t}\Bigl(\boldsymbol{\boldsymbol{\boldsymbol{s}}}^{t}\Bigr)^{H}\boldsymbol{\boldsymbol{\boldsymbol{s}}}\Bigr)\\
 & +2\Bigl(M-k\Bigr)ML.
\end{aligned}
\label{eq:backx}
\end{equation}

where $a=\Bigl(\boldsymbol{\boldsymbol{\boldsymbol{s}}}^{t}\Bigr)^{H}\Bigl(\bar{\boldsymbol{\boldsymbol{\boldsymbol{A}}}}_{i,j}(k)\Bigr)\boldsymbol{\boldsymbol{\boldsymbol{s}}}^{t}$
and $b=\Bigl(\boldsymbol{\boldsymbol{\boldsymbol{s}}}\Bigr)^{H}\Bigl(\bar{\boldsymbol{\boldsymbol{\boldsymbol{A}}}}_{i,j}(k)\Bigr)\boldsymbol{\boldsymbol{\boldsymbol{s}}}$.

The above surrogate function (\ref{eq:backx}) can be rewritten more
compactly as:

\begin{equation}
\begin{aligned}u_{i,j,k}\Bigl(\boldsymbol{\boldsymbol{\boldsymbol{s}}}|\boldsymbol{\boldsymbol{\boldsymbol{s}}}^{t}\Bigr)= & -2\left|a\right|^{2}+2\Bigl(\boldsymbol{\boldsymbol{\boldsymbol{s}}}^{H}\Bigl(\boldsymbol{\boldsymbol{\boldsymbol{D}}}_{i,j}(k)\Bigr)\boldsymbol{\boldsymbol{\boldsymbol{s}}}\Bigr)\\
 & -2\Bigl(M-k\Bigr)\Bigl(\boldsymbol{\boldsymbol{\boldsymbol{s}}}^{H}\boldsymbol{\boldsymbol{\boldsymbol{s}}}^{t}\Bigl(\boldsymbol{\boldsymbol{\boldsymbol{s}}}^{t}\Bigr)^{H}\boldsymbol{\boldsymbol{\boldsymbol{s}}}\Bigr)\\
 & +2\Bigl(M-k\Bigr)ML,
\end{aligned}
\label{eq:backx2}
\end{equation}

where $\begin{array}{cc}
\boldsymbol{\boldsymbol{\boldsymbol{D}}}_{i,j}(k)= & \Bigl(\bar{\boldsymbol{\boldsymbol{\boldsymbol{A}}}}_{i,j}(k)\Bigr)a^{H}+\Bigl(\bar{\boldsymbol{\boldsymbol{\boldsymbol{A}}}}_{i,j}(k)\Bigr)^{H}a\end{array}$.

The surrogate function in (\ref{eq:backx2}) is a quadratic function
in the variable $\boldsymbol{s}$ which would be difficult to minimize,
so in the following, we again use the following lemma to further majorize
the surrogate (we find a tighter surrogate to the surrogate function).

\textbf{Lemma-}3: Let $g:\mathbb{\mathbb{C}}^{N}\rightarrow\mathbb{R}$
be any differentiable concave function, then at any fixed point $\boldsymbol{\boldsymbol{z}}^{t}$,
$g(\boldsymbol{\boldsymbol{\boldsymbol{z}}})$ can be upper bounded
(majorized) as,

\begin{equation}
g(\boldsymbol{\boldsymbol{\boldsymbol{z}}})\leq g(\boldsymbol{\boldsymbol{\boldsymbol{z}}}^{t})+\textrm{Re}(\nabla g(\boldsymbol{\boldsymbol{\boldsymbol{z}}}^{t})^{H}(\boldsymbol{\boldsymbol{\boldsymbol{z}}}-\boldsymbol{\boldsymbol{\boldsymbol{z}}}^{t}))\label{eq:orgeq-1}
\end{equation}

\textbf{Proof }: For any bounded concave function $g(\boldsymbol{\boldsymbol{\boldsymbol{z}}})$,
linearizing at a point $\boldsymbol{\boldsymbol{\boldsymbol{z}}}^{t}$
using first order Taylor series expansion will result in the above
mentioned upper bounded function and it concludes the proof. $\hphantom{nnnnnnnnnnnnnnnnnnnnnnnnn}$$\blacksquare$

Let $\boldsymbol{\bar{\boldsymbol{\boldsymbol{D}}}}_{i,j}(k)=\Bigl(\Bigl(\boldsymbol{\boldsymbol{\boldsymbol{D}}}_{i,j}(k)\Bigr)-\Bigl(\lambda_{\text{max}}\Bigl(\boldsymbol{\boldsymbol{\boldsymbol{D}}}_{i,j}(k)\Bigr)\boldsymbol{\boldsymbol{I}}_{ML}\Bigr)\Bigr)$,
then (\ref{eq:backx2}) can be rewritten as:

\begin{equation}
\begin{aligned}u_{i,j,k}\Bigl(\boldsymbol{\boldsymbol{\boldsymbol{s}}}|\boldsymbol{\boldsymbol{\boldsymbol{s}}}^{t}\Bigr)= & -2\left|a\right|^{2}+2\Bigl(\boldsymbol{\boldsymbol{\boldsymbol{s}}}^{H}\boldsymbol{\bar{\boldsymbol{\boldsymbol{D}}}}_{i,j}(k)\boldsymbol{\boldsymbol{\boldsymbol{s}}}\Bigr)+2\lambda_{\text{max}}\Bigl(\boldsymbol{\boldsymbol{\boldsymbol{D}}}_{i,j}(k)\Bigr)\\
 & -2\Bigl(M-k\Bigr)\Bigl(\Bigl(\boldsymbol{\boldsymbol{\boldsymbol{s}}}^{H}\boldsymbol{\boldsymbol{\boldsymbol{s}}}^{t}\Bigl(\boldsymbol{\boldsymbol{\boldsymbol{s}}}^{t}\Bigr)^{H}\boldsymbol{\boldsymbol{\boldsymbol{s}}}\Bigr)-ML.\Bigr)
\end{aligned}
\label{eq:backx3}
\end{equation}

The surrogate function in (\ref{eq:backx3}) is a quadratic concave
function and can be further majorized by lemma-3. So, by majorizing
(\ref{eq:backx3}) as in lemma-3, we obtain the surrogate to the surrogate
function as below $\Bigl($we substituted back $a=\Bigl(\boldsymbol{\boldsymbol{\boldsymbol{s}}}^{t}\Bigr)^{H}\Bigl(\bar{\boldsymbol{\boldsymbol{\boldsymbol{A}}}}_{i,j}(k)\Bigr)\boldsymbol{\boldsymbol{\boldsymbol{s}}}^{t}$$\Bigr)$
:

\begin{equation}
\begin{aligned}\tilde{u}_{i,j,k}\Bigl(\boldsymbol{\boldsymbol{\boldsymbol{s}}}|\boldsymbol{\boldsymbol{\boldsymbol{s}}}^{t}\Bigr) & =-2\left|\Bigl(\boldsymbol{\boldsymbol{\boldsymbol{s}}}^{t}\Bigr)^{H}\Bigl(\bar{\boldsymbol{\boldsymbol{\boldsymbol{A}}}}_{i,j}(k)\Bigr)\boldsymbol{\boldsymbol{\boldsymbol{s}}}^{t}\right|^{2}+2\lambda_{\text{max}}\Bigl(\boldsymbol{\boldsymbol{\boldsymbol{D}}}_{i,j}(k)\Bigr)\\
 & +2\Bigl(-\Bigl(\boldsymbol{\boldsymbol{\boldsymbol{s}}}^{t}\Bigr)^{H}\boldsymbol{\bar{\boldsymbol{\boldsymbol{D}}}}_{i,j}(k)\boldsymbol{\boldsymbol{\boldsymbol{s}}}^{t}\\
 & \hphantom{nnnnnn}+2\text{Re}\Bigl(\Bigl(\boldsymbol{\boldsymbol{\boldsymbol{s}}}^{t}\Bigr)^{H}\boldsymbol{\bar{\boldsymbol{\boldsymbol{D}}}}_{i,j}(k)\boldsymbol{\boldsymbol{\boldsymbol{s}}}\Bigr)\Bigr)\\
 & -2\Bigl(M-k\Bigr)\Bigl(\Bigl(-ML+2\text{Re}\Bigl(\boldsymbol{\boldsymbol{\boldsymbol{s}}}^{H}\boldsymbol{\boldsymbol{\boldsymbol{s}}}^{t}\Bigr)\Bigr)-ML\Bigr).
\end{aligned}
\label{eq:backxfinal}
\end{equation}

We would like to note that the surrogate in (\ref{eq:backxfinal})
is a tighter upper bound for the surrogate in (\ref{eq:backx3}),
which is again a tighter upper bound for the PSL metric, we can directly
formulate (\ref{eq:backxfinal}) as a tighter surrogate for the PSL
metric. Thus using (\ref{eq:backxfinal}), the surrogate minimization
problem is given as:

\begin{equation}
\begin{aligned} & \underset{\boldsymbol{\boldsymbol{\boldsymbol{s}}}}{\text{\text{min}}} &  & \text{\ensuremath{\underset{k\in a_{k}}{\text{\text{max}}}}}\hphantom{nn}4\text{Re}\Bigl(\boldsymbol{\boldsymbol{s}}{}^{H}\boldsymbol{\boldsymbol{\boldsymbol{d}}}_{i,j}(k)\Bigr)+p_{i,j}(k),\,i,j=1,.,L,\\
 & \text{ s.t} &  & \bigl|\boldsymbol{\boldsymbol{s}}(m)\bigr|=1,\,m=1,..,ML,
\end{aligned}
\label{eq:Majfinal}
\end{equation}

where 
\begin{equation}
\begin{aligned}\boldsymbol{\boldsymbol{\boldsymbol{d}}}_{i,j}(k) & =\Bigl(\boldsymbol{\bar{\boldsymbol{\boldsymbol{D}}}}_{i,j}(k)\Bigr)\boldsymbol{\boldsymbol{\boldsymbol{s}}}^{t}-\Bigl(M-k\Bigr)\boldsymbol{\boldsymbol{s}}^{t}\\
 & =\Bigl(\Bigl(\bar{\boldsymbol{\boldsymbol{\boldsymbol{A}}}}_{i,j}(k)\Bigr)\boldsymbol{\boldsymbol{s}}^{t}\Bigr)r_{i,j}^{*}(k)+\Bigl(\Bigl(\bar{\boldsymbol{\boldsymbol{\boldsymbol{A}}}}_{i,j}(k)\Bigr)^{H}\boldsymbol{\boldsymbol{s}}^{t}\Bigr)r_{i,j}(k)\\
 & \hphantom{nnnn}-\Bigl(\lambda_{\text{max}}\Bigl(\boldsymbol{\boldsymbol{\boldsymbol{D}}}_{i,j}(k)\Bigr)\boldsymbol{\boldsymbol{I}}_{ML}\Bigr)\boldsymbol{\boldsymbol{s}}^{t}-\Bigl(M-k\Bigr)\boldsymbol{\boldsymbol{s}}^{t}\\
p_{i,j}(k) & =-2\biggl|\Bigl(\boldsymbol{\boldsymbol{\boldsymbol{s}}}^{t}\Bigr)^{H}\Bigl(\bar{\boldsymbol{\boldsymbol{\boldsymbol{A}}}}_{i,j}(k)\Bigr)\boldsymbol{\boldsymbol{\boldsymbol{s}}}^{t}\biggr|^{2}\\
 & \hphantom{nnn}-2\Bigl(\Bigl(\boldsymbol{\boldsymbol{\boldsymbol{s}}}^{t}\Bigr)^{H}\Bigl(\boldsymbol{\boldsymbol{\boldsymbol{D}}}_{i,j}(k)\Bigr)\boldsymbol{\boldsymbol{\boldsymbol{s}}}^{t}\Bigr)+4\lambda_{\text{max}}\Bigl(\boldsymbol{\boldsymbol{\boldsymbol{D}}}_{i,j}(k)\Bigr)\\
 & \hphantom{nnn}+4\Bigl(M-k\Bigr)ML.\\
 & =-6\Bigl|r_{i,j}(k)\Bigr|^{2}+4\lambda_{\text{max}}\Bigl(\boldsymbol{\boldsymbol{\boldsymbol{D}}}_{i,j}(k)\Bigr)+4\Bigl(M-k\Bigr)ML.
\end{aligned}
\label{eq:dtterms}
\end{equation}

Problem in (\ref{eq:Majfinal}) is a nonconvex problem because of
the presence of equality constraint. However, the constraint set can
be relaxed and the optimal minimizer for the relaxed problem will
lie on the boundary set \cite{boyd}. The epigraph form of the relaxed
problem can be given as:

\begin{equation}
\begin{aligned} & \underset{\boldsymbol{\boldsymbol{\boldsymbol{s}}},\alpha}{\text{\text{min}}} &  & \alpha\\
 & \text{ \text{s.t}} &  & 4\text{Re}\Bigl(\boldsymbol{\boldsymbol{s}}{}^{H}\boldsymbol{\boldsymbol{\boldsymbol{d}}}_{i,j}(k)\Bigr)+p_{i,j}(k)\leq\alpha,i,j=1,.,L,\,\forall k\in a_{k}\\
 &  &  & \bigl|\boldsymbol{\boldsymbol{s}}(m)\bigr|\leq1,\,m=1,..,ML.
\end{aligned}
\label{eq:finalprob}
\end{equation}

The problem in (\ref{eq:finalprob}) is a convex problem and there
exist many off-the-shelf interior point solvers \cite{cvx} to solve
the problem in (\ref{eq:finalprob}). The pseudocode of the Interior
point solver based PSL minimizer is given in the Algorithm-1.

\begin{algorithm}[h]
\textbf{Require}: Number of sequences $\text{\textquoteleft}L\text{\textquoteright}$
and length of each sequence $\text{\textquoteleft}M\text{\textquoteright}$

{\small{}1: set $t=0$, initialize $\left\{ \boldsymbol{\boldsymbol{s}}_{i}^{0}\right\} _{i=1}^{L}$}{\small\par}

{\small{}2: form $\boldsymbol{\boldsymbol{\boldsymbol{s}}},\boldsymbol{\boldsymbol{\boldsymbol{B}}}_{s},\boldsymbol{\boldsymbol{\boldsymbol{A}}}_{k}\,,\forall s,k$
using (\ref{eq:matrices})}{\small\par}

{\small{}3: $\bar{\boldsymbol{\boldsymbol{\boldsymbol{A}}}}_{i,j}(k)=\bigl(\boldsymbol{\boldsymbol{\boldsymbol{B}}}_{i}\bigr)^{H}\boldsymbol{\boldsymbol{\boldsymbol{A}}}_{k}\bigl(\boldsymbol{\boldsymbol{\boldsymbol{B}}}_{j}\bigr)\,,\forall k\in a_{k},i,j.$}{\small\par}

4: $\boldsymbol{\boldsymbol{\boldsymbol{\Phi}}}_{i,j}(k)=\begin{array}[t]{c}
\!\!\!\!\!\!\!\!\!\!\!\!\!\!\!\!\!\!\!\!\!\!\!\!\!\!\!\!\!\!\!\!\!\!\!\!\!\!\!\!\!\!\!\text{vec}\Bigl(\bar{\boldsymbol{\boldsymbol{\boldsymbol{A}}}}_{i,j}(k)\Bigr)\text{vec}^{H}\Bigl(\Bigl(\bar{\boldsymbol{\boldsymbol{\boldsymbol{A}}}}_{i,j}(k)\Bigr)^{H}\Bigr)\\
\!\!\!\!\!+\text{vec}\Bigl(\Bigl(\bar{\boldsymbol{\boldsymbol{\boldsymbol{A}}}}_{i,j}(k)\Bigr)^{H}\Bigr)\text{vec}^{H}\Bigl(\bar{\boldsymbol{\boldsymbol{\boldsymbol{A}}}}_{i,j}(k)\Bigr),\forall k\in a_{k},i,j.
\end{array}$

{\small{}5:}\textbf{\small{} repeat}{\small\par}

6: $\boldsymbol{\boldsymbol{\boldsymbol{D}}}_{i,j}(k)=\begin{array}[t]{c}
\!\!\!\!\!\!\!\!\!\!\!\!\!\!\!\!\!\!\!\!\!\!\!\!\!\!\!\!\!\!\!\!\!\!\!\!\!\!\!\!\!\!\!\!\Bigl(\bar{\boldsymbol{\boldsymbol{\boldsymbol{A}}}}_{i,j}(k)\Bigr)\Bigl(\Bigl(\boldsymbol{\boldsymbol{\boldsymbol{s}}}^{t}\Bigr)^{H}\Bigl(\bar{\boldsymbol{\boldsymbol{\boldsymbol{A}}}}_{i,j}(k)\Bigr)^{H}\boldsymbol{\boldsymbol{\boldsymbol{s}}}^{t}\Bigr)\\
\!\!\!\!\!+\Bigl(\bar{\boldsymbol{\boldsymbol{\boldsymbol{A}}}}_{i,j}(k)\Bigr)^{H}\Bigl(\Bigl(\boldsymbol{\boldsymbol{\boldsymbol{s}}}^{t}\Bigr)^{H}\Bigl(\bar{\boldsymbol{\boldsymbol{\boldsymbol{A}}}}_{i,j}(k)\Bigr)\boldsymbol{\boldsymbol{\boldsymbol{s}}}^{t}\Bigr),\forall k\in a_{k},i,j.
\end{array}$

{\small{}7: $\boldsymbol{\bar{\boldsymbol{\boldsymbol{D}}}}_{i,j}(k)=\Bigl(\Bigl(\boldsymbol{\boldsymbol{\boldsymbol{D}}}_{i,j}(k)\Bigr)-\Bigl(\lambda_{\text{max}}\Bigl(\boldsymbol{\boldsymbol{\boldsymbol{D}}}_{i,j}(k)\Bigr)\boldsymbol{\boldsymbol{I}}_{ML}\Bigr)\Bigr)\forall k\in a_{k},i,j.$}{\small\par}

{\small{}8: $\boldsymbol{\boldsymbol{\boldsymbol{d}}}_{i,j}(k)=\Bigl(\boldsymbol{\bar{\boldsymbol{\boldsymbol{D}}}}_{i,j}(k)\Bigr)\boldsymbol{\boldsymbol{\boldsymbol{s}}}^{t}-\Bigl(M-k\Bigr)\boldsymbol{\boldsymbol{s}}^{t}\,,\forall k\in a_{k},i,j.$}{\small\par}

{\small{}9: $\begin{array}[t]{c}
p_{i,j}(k)=-2\biggl|\Bigl(\boldsymbol{\boldsymbol{\boldsymbol{s}}}^{t}\Bigr)^{H}\Bigl(\bar{\boldsymbol{\boldsymbol{\boldsymbol{A}}}}_{i,j}(k)\Bigr)\boldsymbol{\boldsymbol{\boldsymbol{s}}}^{t}\biggr|^{2}-2\Bigl(\Bigl(\boldsymbol{\boldsymbol{\boldsymbol{s}}}^{t}\Bigr)^{H}\Bigl(\boldsymbol{\boldsymbol{\boldsymbol{D}}}_{i,j}(k)\Bigr)\boldsymbol{\boldsymbol{\boldsymbol{s}}}^{t}\Bigr)\\
\hphantom{nnnnnn}+4\lambda_{\text{max}}\Bigl(\boldsymbol{\boldsymbol{\boldsymbol{D}}}_{i,j}(k)\Bigr)+4\Bigl(M-k\Bigr)ML,\forall k\in a_{k},i,j.
\end{array}$ }{\small\par}

{\small{}10: get $\boldsymbol{\boldsymbol{s}}^{t+1}$ by solving the
problem in (\ref{eq:finalprob}).}{\small\par}

{\small{}11: $t$$\leftarrow$$t+1$}{\small\par}

{\small{}12:}\textbf{\small{} until }{\small{}convergence}\caption{:Interior point solver based PSL minimizer}
\end{algorithm}

However, when dimension of the problem ($L$ and $M$) increases,
off-the-shelf solvers will become computationally expensive. To overcome
this issue, in the following we present an efficient way to compute
the solution of (\ref{eq:finalprob}) further as follows. The problem
in (\ref{eq:Majfinal}) is a function of complex variable $\boldsymbol{s}$
and we first convert in terms of real variables as follows:

\begin{equation}
\begin{aligned} & \underset{\boldsymbol{\boldsymbol{\boldsymbol{x}}}}{\text{\text{min}}} &  & \text{\ensuremath{\underset{k\in a_{k}}{\text{\text{max}}}}}\hphantom{nn}4\boldsymbol{\boldsymbol{x}}^{T}\boldsymbol{\boldsymbol{d}}_{k}+p_{k}\\
 & \text{s.t} &  & \left|\boldsymbol{\boldsymbol{\boldsymbol{x}}}(i)\right|^{2}+\left|\boldsymbol{\boldsymbol{\boldsymbol{x}}}(i+ML)\right|^{2}\leq1,\,i=1,..,ML,
\end{aligned}
\label{eq:c2r}
\end{equation}

where

$\boldsymbol{\boldsymbol{s}}_{R}=Re(\boldsymbol{\boldsymbol{s}}),\boldsymbol{\boldsymbol{s}}_{I}=Im(\boldsymbol{\boldsymbol{s}}),\boldsymbol{\boldsymbol{x}}=[\boldsymbol{\boldsymbol{s}}_{R}^{T},\boldsymbol{\boldsymbol{s}}_{I}^{T}]^{T},$

$\boldsymbol{\boldsymbol{d}}_{Rk}=Re\Bigl(\boldsymbol{\boldsymbol{\boldsymbol{d}}}_{i,j}(k)\Bigr),\boldsymbol{\boldsymbol{d}}_{Ik}=Im\Bigl(\boldsymbol{\boldsymbol{\boldsymbol{d}}}_{i,j}(k)\Bigr),$

$\boldsymbol{\boldsymbol{d}}_{k}=[\boldsymbol{\boldsymbol{d}}_{Rk}^{T},\boldsymbol{\boldsymbol{d}}_{Ik}^{T}]^{T},p_{k}=p_{i,j}(k).$

By introducing a simplex variable $\boldsymbol{\boldsymbol{q}}$,
we can rewrite the discrete inner maximum problem as follows:

\begin{equation}
\begin{aligned} & \underset{\boldsymbol{\boldsymbol{q}}\geq0,\boldsymbol{\boldsymbol{1}}^{T}\boldsymbol{\boldsymbol{q}}=1}{\text{\text{max}}} &  & \underset{k\in a_{k}}{\sum}\Bigl[q_{k}(4\boldsymbol{\boldsymbol{x}}^{T}\boldsymbol{\boldsymbol{d}}_{k}+p_{k})\Bigr]\\
 & \underset{\boldsymbol{\boldsymbol{q}}\geq0,\boldsymbol{\boldsymbol{1}}^{T}\boldsymbol{\boldsymbol{q}}=1}{\text{\text{max}}} &  & 4\boldsymbol{\boldsymbol{x}}^{T}\tilde{\boldsymbol{\boldsymbol{\boldsymbol{D}}}}\boldsymbol{\boldsymbol{q}}+\boldsymbol{\boldsymbol{q}}^{T}\boldsymbol{\boldsymbol{p}}
\end{aligned}
\label{eq:simplex}
\end{equation}

where

$\tilde{\boldsymbol{\boldsymbol{\boldsymbol{D}}}}=\Bigl[\boldsymbol{\boldsymbol{d}}_{1},\boldsymbol{\boldsymbol{d}}_{2},...,\boldsymbol{\boldsymbol{d}}_{|a_{k}|}\Bigr],\boldsymbol{\boldsymbol{q}}=\Bigl[q_{1},q_{2},...,q_{|a_{k}|}\Bigr]^{T},$

$\boldsymbol{\boldsymbol{p}}=\Bigl[p_{1},p_{2},..,p_{|a_{k}|}\Bigr]^{T}$,
and $|a_{k}|$ denotes the total number of elements in the set $a_{k}$.

By using (\ref{eq:simplex}) and (\ref{eq:c2r}), the problem in (\ref{eq:Majfinal})
can be rewritten as:

\begin{equation}
\begin{aligned} & \underset{\boldsymbol{\boldsymbol{\boldsymbol{x}}}}{\text{\text{min}}} &  & \underset{\boldsymbol{\boldsymbol{q}}\geq0,\boldsymbol{\boldsymbol{1}}^{T}\boldsymbol{\boldsymbol{q}}=1}{\text{\text{max}}}\hphantom{nn}4\boldsymbol{\boldsymbol{x}}^{T}\tilde{\boldsymbol{\boldsymbol{\boldsymbol{D}}}}\boldsymbol{\boldsymbol{q}}+\boldsymbol{\boldsymbol{q}}^{T}\boldsymbol{\boldsymbol{p}}\\
 & \text{s.t} &  & \left|\boldsymbol{\boldsymbol{\boldsymbol{x}}}(i)\right|^{2}+\left|\boldsymbol{\boldsymbol{\boldsymbol{x}}}(i+ML)\right|^{2}\leq1,\,i=1,..,ML.
\end{aligned}
\label{eq:pfin}
\end{equation}

Problem in (\ref{eq:pfin}) is bilinear in the variables $\boldsymbol{\boldsymbol{x}}$
and $\boldsymbol{\boldsymbol{q}}$. By using the minmax theorem \cite{MINMAX},
without altering the solution, we can swap minmax to maxmin as follows:

\begin{equation}
\begin{aligned} & \text{\ensuremath{\underset{\boldsymbol{\boldsymbol{q}}\geq0,\boldsymbol{\boldsymbol{1}}^{T}\boldsymbol{\boldsymbol{q}}=1}{\text{\text{max}}}}} &  & \text{\ensuremath{\underset{\boldsymbol{\boldsymbol{\boldsymbol{x}}}}{\text{\text{min}}}}}\;4\boldsymbol{\boldsymbol{x}}^{T}\tilde{\boldsymbol{\boldsymbol{\boldsymbol{D}}}}\boldsymbol{\boldsymbol{q}}+\boldsymbol{\boldsymbol{q}}^{T}\boldsymbol{\boldsymbol{p}}\\
 & \text{ s.t} &  & \left|\boldsymbol{\boldsymbol{\boldsymbol{x}}}(i)\right|^{2}+\left|\boldsymbol{\boldsymbol{\boldsymbol{x}}}(i+ML)\right|^{2}\leq1,\,i=1,..,ML,
\end{aligned}
\label{eq:minmaxthe}
\end{equation}

Problem in (\ref{eq:minmaxthe}) can be rewritten as:

\begin{equation}
\begin{aligned} & \text{\ensuremath{\underset{\boldsymbol{\boldsymbol{q}}\geq0,\boldsymbol{\boldsymbol{1}}^{T}\boldsymbol{\boldsymbol{q}}=1}{\text{\text{max}}}}} & g(\boldsymbol{\boldsymbol{q}}) & \text{}\end{aligned}
\label{eq:hpprob}
\end{equation}

where

\begin{equation}
\begin{aligned}g(\boldsymbol{\boldsymbol{q}})=\; & \text{\ensuremath{\underset{\boldsymbol{\boldsymbol{\boldsymbol{x}}}}{\text{\text{min}}}}} &  & 4\boldsymbol{\boldsymbol{x}}^{T}\tilde{\boldsymbol{\boldsymbol{\boldsymbol{D}}}}\boldsymbol{\boldsymbol{q}}+\boldsymbol{\boldsymbol{q}}^{T}\boldsymbol{\boldsymbol{p}}\\
 & \text{s.t} &  & \left|\boldsymbol{\boldsymbol{\boldsymbol{x}}}(i)\right|^{2}+\left|\boldsymbol{\boldsymbol{\boldsymbol{x}}}(i+ML)\right|^{2}\leq1,\,i=1,..,ML,
\end{aligned}
\label{eq:minxprob}
\end{equation}

The problem in (\ref{eq:hpprob}) can be solved iteratively via the
Mirror Descent Algorithm (MDA), which is a very established algorithm
to solve minimization/maximization problems with non-differentiable
objective. Without getting into details of the MDA algorithm (we refer
the interested reader to \cite{MDA}), the iterative steps of MDA
for the problem in (\ref{eq:hpprob}) can be given as:

\noindent\fbox{\begin{minipage}[t]{1\columnwidth - 2\fboxsep - 2\fboxrule}%
step-1: Get subgradient of the objective $g(\boldsymbol{\boldsymbol{q}}),$
which is equal to $4\tilde{\boldsymbol{\boldsymbol{\boldsymbol{D}}}}^{T}\boldsymbol{\boldsymbol{z}}^{m}+\boldsymbol{\boldsymbol{p}}$,
where $\boldsymbol{z}^{m}$ denote a sequence like variables (similar
to $\boldsymbol{s}$) whose elements will have unit modulus.\\
step-2: Update the simplex variable as $\boldsymbol{\boldsymbol{q}}^{m+1}=\frac{\boldsymbol{\boldsymbol{q}}^{m}\odot e^{\gamma_{m}4\:\tilde{\boldsymbol{\boldsymbol{\boldsymbol{D}}}}^{T}\boldsymbol{\boldsymbol{z}}^{m}+\boldsymbol{\boldsymbol{p}}}}{\boldsymbol{\boldsymbol{1}}^{T}(\boldsymbol{\boldsymbol{q}}^{m}\odot e^{\gamma_{m}4\:\tilde{\boldsymbol{\boldsymbol{\boldsymbol{D}}}}^{T}\boldsymbol{\boldsymbol{z}}^{m}+\boldsymbol{\boldsymbol{p}}})}$,
where $\gamma_{m}$ is a suitable step size.

step-3: $m=m+1$ and go to step-1 unless convergence is achieved.%
\end{minipage}}

\noindent $\vphantom{nn}$

\noindent Once we get the optimal $\boldsymbol{\boldsymbol{q}}^{*}$,
the update for the variables $\boldsymbol{\boldsymbol{\boldsymbol{x}}}$
can be obtained as explained below:

\begin{equation}
\bigl[\boldsymbol{x}(i),\boldsymbol{x}(i+ML)\bigr]^{T}=\frac{v_{i}}{||v_{i}||_{2}}\label{eq:xupdate}
\end{equation}

where $v_{i}=\bigl[\boldsymbol{\tilde{c}}(i),\boldsymbol{\tilde{c}}({i+ML})\bigr]^{T},\,i=1,.,ML,$
and $\tilde{\boldsymbol{\boldsymbol{c}}}=-\tilde{\boldsymbol{\boldsymbol{\boldsymbol{D}}}}\boldsymbol{\boldsymbol{q}}^{*}$.

From the real variables $\boldsymbol{x}$, the complex sequence set
variable $\boldsymbol{s}$ can be recovered. The pseudocode of the
MDA based PSL minimization is given in the Algorithm-2.

\begin{algorithm}[h]
\textbf{Require}: Number of sequences $\text{\textquoteleft}L\text{\textquoteright}$
and length of each sequence $\text{\textquoteleft}M\text{\textquoteright}$

{\small{}1: set $t=0$, initialize $\left\{ \boldsymbol{\boldsymbol{s}}_{i}^{0}\right\} _{i=1}^{L}$}{\small\par}

{\small{}2: form $\boldsymbol{\boldsymbol{\boldsymbol{s}}},\boldsymbol{\boldsymbol{\boldsymbol{B}}}_{s},\boldsymbol{\boldsymbol{\boldsymbol{A}}}_{k}\,,\forall s,k$
using (\ref{eq:matrices})}{\small\par}

{\small{}3: $\bar{\boldsymbol{\boldsymbol{\boldsymbol{A}}}}_{i,j}(k)=\bigl(\boldsymbol{\boldsymbol{\boldsymbol{B}}}_{i}\bigr)^{H}\boldsymbol{\boldsymbol{\boldsymbol{A}}}_{k}\bigl(\boldsymbol{\boldsymbol{\boldsymbol{B}}}_{j}\bigr)\,,\forall k\in a_{k},i,j.$}{\small\par}

4: $\boldsymbol{\boldsymbol{\boldsymbol{\Phi}}}_{i,j}(k)=\begin{array}[t]{c}
\!\!\!\!\!\!\!\!\!\!\!\!\!\!\!\!\!\!\!\!\!\!\!\!\!\!\!\!\!\!\!\!\!\!\!\!\!\!\!\!\!\!\!\text{vec}\Bigl(\bar{\boldsymbol{\boldsymbol{\boldsymbol{A}}}}_{i,j}(k)\Bigr)\text{vec}^{H}\Bigl(\Bigl(\bar{\boldsymbol{\boldsymbol{\boldsymbol{A}}}}_{i,j}(k)\Bigr)^{H}\Bigr)\\
\!\!\!\!\!+\text{vec}\Bigl(\Bigl(\bar{\boldsymbol{\boldsymbol{\boldsymbol{A}}}}_{i,j}(k)\Bigr)^{H}\Bigr)\text{vec}^{H}\Bigl(\bar{\boldsymbol{\boldsymbol{\boldsymbol{A}}}}_{i,j}(k)\Bigr),\forall k\in a_{k},i,j.
\end{array}$

{\small{}5:}\textbf{\small{} repeat}{\small\par}

6: $\boldsymbol{\boldsymbol{\boldsymbol{D}}}_{i,j}(k)=\begin{array}[t]{c}
\!\!\!\!\!\!\!\!\!\!\!\!\!\!\!\!\!\!\!\!\!\!\!\!\!\!\!\!\!\!\!\!\!\!\!\!\!\!\!\!\!\!\!\!\Bigl(\bar{\boldsymbol{\boldsymbol{\boldsymbol{A}}}}_{i,j}(k)\Bigr)\Bigl(\Bigl(\boldsymbol{\boldsymbol{\boldsymbol{s}}}^{t}\Bigr)^{H}\Bigl(\bar{\boldsymbol{\boldsymbol{\boldsymbol{A}}}}_{i,j}(k)\Bigr)^{H}\boldsymbol{\boldsymbol{\boldsymbol{s}}}^{t}\Bigr)\\
\!\!\!\!\!+\Bigl(\bar{\boldsymbol{\boldsymbol{\boldsymbol{A}}}}_{i,j}(k)\Bigr)^{H}\Bigl(\Bigl(\boldsymbol{\boldsymbol{\boldsymbol{s}}}^{t}\Bigr)^{H}\Bigl(\bar{\boldsymbol{\boldsymbol{\boldsymbol{A}}}}_{i,j}(k)\Bigr)\boldsymbol{\boldsymbol{\boldsymbol{s}}}^{t}\Bigr),\forall k\in a_{k},i,j.
\end{array}$

{\small{}7: $\boldsymbol{\bar{\boldsymbol{\boldsymbol{D}}}}_{i,j}(k)=\Bigl(\Bigl(\boldsymbol{\boldsymbol{\boldsymbol{D}}}_{i,j}(k)\Bigr)-\Bigl(\lambda_{\text{max}}\Bigl(\boldsymbol{\boldsymbol{\boldsymbol{D}}}_{i,j}(k)\Bigr)\boldsymbol{\boldsymbol{I}}_{ML}\Bigr)\Bigr)\forall k\in a_{k},i,j.$}{\small\par}

{\small{}8: $\boldsymbol{\boldsymbol{\boldsymbol{d}}}_{i,j}(k)=\Bigl(\boldsymbol{\bar{\boldsymbol{\boldsymbol{D}}}}_{i,j}(k)\Bigr)\boldsymbol{\boldsymbol{\boldsymbol{s}}}^{t}-\Bigl(M-k\Bigr)\boldsymbol{\boldsymbol{s}}^{t}\,,\forall k\in a_{k},i,j.$}{\small\par}

{\small{}9: $\begin{array}[t]{c}
p_{i,j}(k)=-2\biggl|\Bigl(\boldsymbol{\boldsymbol{\boldsymbol{s}}}^{t}\Bigr)^{H}\Bigl(\bar{\boldsymbol{\boldsymbol{\boldsymbol{A}}}}_{i,j}(k)\Bigr)\boldsymbol{\boldsymbol{\boldsymbol{s}}}^{t}\biggr|^{2}-2\Bigl(\Bigl(\boldsymbol{\boldsymbol{\boldsymbol{s}}}^{t}\Bigr)^{H}\Bigl(\boldsymbol{\boldsymbol{\boldsymbol{D}}}_{i,j}(k)\Bigr)\boldsymbol{\boldsymbol{\boldsymbol{s}}}^{t}\Bigr)\\
\hphantom{nnnnnn}+4\lambda_{\text{max}}\Bigl(\boldsymbol{\boldsymbol{\boldsymbol{D}}}_{i,j}(k)\Bigr)+4\Bigl(M-k\Bigr)ML,\forall k\in a_{k},i,j.
\end{array}$ }{\small\par}

{\small{}10: evaluate $\boldsymbol{\boldsymbol{q}}^{*}$ using Mirror
Descent Algorithm}{\small\par}

{\small{}11: $\tilde{\boldsymbol{\boldsymbol{c}}}=-\tilde{\boldsymbol{\boldsymbol{\boldsymbol{D}}}}\boldsymbol{\boldsymbol{q}}^{*}$}{\small\par}

{\small{}12: $v_{i}=\bigl[\boldsymbol{\tilde{c}}(i),\boldsymbol{\tilde{c}}({i+ML})\bigr]^{T},\,i=1,..,ML.$}{\small\par}

{\small{}13: $\bigl[\boldsymbol{x}(i),\boldsymbol{x}(i+ML)\bigr]^{T}=\frac{v_{i}}{||v_{i}||_{2}}$.}{\small\par}

{\small{}14: Recover $\boldsymbol{\boldsymbol{s}}^{t+1}\text{ from }\boldsymbol{\boldsymbol{x}}^{t+1}$
and get required sequence set from it.}{\small\par}

{\small{}15: $t$$\leftarrow$$t+1$}{\small\par}

{\small{}16:}\textbf{\small{} until }{\small{}convergence}\caption{:MDA based PSL minimizer}
\end{algorithm}

\subsection{Convergence of the algorithm}

As the proposed algorithm is based on the MM technique, the descent
property in (\ref{eq:MM-1}) would be applicable here i.e.,

\begin{equation}
f(\boldsymbol{\boldsymbol{\boldsymbol{x}}}^{t+1})\leq f(\boldsymbol{\boldsymbol{\boldsymbol{x}}}^{t})\label{eq:objless}
\end{equation}

We minimize the PSL function which is bounded below by zero and the
sequence of iterates which decrease the objective function at every
iteration will sure converge to a finite value. Now, we will discuss
the convergence of iterates $\{\boldsymbol{\boldsymbol{\boldsymbol{x}}}^{t}\}$
to a stationary point, which is defined as:

\textbf{Proposition} 1 \cite{bertsekas_convex}: Let $g:\mathbb{\mathbb{\mathbb{R}}}^{N}\rightarrow\mathbb{R}$
be any smooth function with $\boldsymbol{\boldsymbol{\boldsymbol{z}}}^{*}$
as a local minimum, then

\begin{equation}
\nabla g(\boldsymbol{\boldsymbol{\boldsymbol{z}}}^{*})\boldsymbol{\boldsymbol{\boldsymbol{v}}}\geq0,\:\forall\boldsymbol{\boldsymbol{\boldsymbol{v}}}\in T_{\boldsymbol{\boldsymbol{\boldsymbol{\chi}}}}(\boldsymbol{\boldsymbol{\boldsymbol{z}}}^{*})\label{eq:sp}
\end{equation}

where $T_{\boldsymbol{\boldsymbol{\boldsymbol{\chi}}}}(\boldsymbol{\boldsymbol{\boldsymbol{z}}}^{*})$
represents the tangent cone of $\boldsymbol{\boldsymbol{\boldsymbol{\chi}}}$
at $\boldsymbol{\boldsymbol{\boldsymbol{z}}}^{*}$.

Assume that there exists a converging subsequence $\boldsymbol{\boldsymbol{\boldsymbol{x}}}^{a_{i}}\rightarrow\boldsymbol{\boldsymbol{x}}^{*}$,
then the MM method confirms that,

\[
\begin{array}{c}
u(\boldsymbol{\boldsymbol{x}}^{(a_{i+1})}|\boldsymbol{\boldsymbol{x}}^{(a_{i+1})})=f(\boldsymbol{\boldsymbol{x}}^{(a_{i+1})})\leq f(\boldsymbol{\boldsymbol{x}}^{(a_{i}+1)})\\
\leq u(\boldsymbol{\boldsymbol{x}}^{(a_{i}+1)}|\boldsymbol{\boldsymbol{x}}^{(a_{i})})\leq u(\boldsymbol{\boldsymbol{\boldsymbol{x}}}|\boldsymbol{\boldsymbol{x}}^{(a_{i})})
\end{array}
\]

\[
u(\boldsymbol{\boldsymbol{x}}^{(a_{i+1})}|\boldsymbol{\boldsymbol{x}}^{(a_{i+1})})\leq u(\boldsymbol{\boldsymbol{\boldsymbol{x}}}|\boldsymbol{\boldsymbol{x}}^{(a_{i})})
\]

Letting $i\rightarrow+\infty$, we obtain

\begin{equation}
u(\boldsymbol{\boldsymbol{x}}^{\infty}|\boldsymbol{\boldsymbol{x}}^{\infty})\leq u(\boldsymbol{\boldsymbol{\boldsymbol{x}}}|\boldsymbol{\boldsymbol{x}}^{\infty})\label{eq:sp2}
\end{equation}

Replacing $\boldsymbol{\boldsymbol{x}}^{\infty}\text{ with }\boldsymbol{\boldsymbol{\boldsymbol{x}}}^{*}$,
we have

\begin{equation}
u(\boldsymbol{\boldsymbol{x}}^{*}|\boldsymbol{\boldsymbol{\boldsymbol{x}}}^{*})\leq u(\boldsymbol{\boldsymbol{\boldsymbol{x}}}|\boldsymbol{\boldsymbol{\boldsymbol{x}}}^{*})\label{eq:sp3}
\end{equation}

So, (\ref{eq:sp3}) conveys that $\boldsymbol{\boldsymbol{\boldsymbol{x}}}^{*}$
is a stationary point and also a global minimizer of $u(.)$.

From the majorization step, we know that the first-order behavior
of majorized function $u(\boldsymbol{\boldsymbol{\boldsymbol{x}}}|\boldsymbol{\boldsymbol{\boldsymbol{x}}}^{t})$
is equal to the original cost function $f(\boldsymbol{\boldsymbol{\boldsymbol{x}}})$.
So, we can show

\begin{equation}
u(\boldsymbol{\boldsymbol{\boldsymbol{x}}}^{*}|\boldsymbol{\boldsymbol{\boldsymbol{x}}}^{*})\leq u(\boldsymbol{\boldsymbol{\boldsymbol{x}}}|\boldsymbol{\boldsymbol{\boldsymbol{x}}}^{*})\Leftrightarrow f(\boldsymbol{\boldsymbol{\boldsymbol{x}}}^{*})\leq f(\boldsymbol{\boldsymbol{\boldsymbol{x}}})\label{eq:spg2}
\end{equation}

So, the set of points generated by the proposed algorithm are stationary
points and $\boldsymbol{\boldsymbol{\boldsymbol{x}}}^{*}$ is the
minimizer of $f(\boldsymbol{\boldsymbol{\boldsymbol{x}}})$. This
concludes the proof.$\hphantom{nnnnnnnnnnnnnnnnnnnnnnnnnnn}$$\blacksquare$

\subsection{Computational and space complexity of the proposed algorithm}

Our proposed algorithm consists of two loops, in which the inner loop
calculates $\boldsymbol{\boldsymbol{q}}^{*}$ using MDA and the outer
loop will update the elements of the sequence set. As shown in the
Algorithm-2, per iteration computational complexity of the outer loop
is dominated in the calculation of $\boldsymbol{\boldsymbol{\boldsymbol{D}}}_{i,j}(k),\boldsymbol{\bar{\boldsymbol{\boldsymbol{D}}}}_{i,j}(k),\boldsymbol{\boldsymbol{\boldsymbol{d}}}_{i,j}(k),p_{i,j}(k),\tilde{\boldsymbol{\boldsymbol{c}}}$.
The quantity $\Bigl(\boldsymbol{\boldsymbol{\boldsymbol{s}}}^{t}\Bigr)^{H}\Bigl(\bar{\boldsymbol{\boldsymbol{\boldsymbol{A}}}}_{i,j}(k)\Bigr)\boldsymbol{\boldsymbol{\boldsymbol{s}}}^{t}$
which appears in the some of the constants of the algorithm is nothing
but $r_{i,j}(k)$ which can be calculated using FFT and IFFT operations,
one can implement the above quantities very efficiently. Once the
optimal $\boldsymbol{\boldsymbol{q}}^{*}$ is obtained using MDA which
will be sparse, then the quantity $\tilde{\boldsymbol{\boldsymbol{c}}}$
can be calculated efficiently using a sparse matrix-vector multiplication.
The per iteration computational complexity of the inner loop (i.e.
MDA) is dominated by the calculation of subgradient, which can also
be efficiently calculated via FFT operations. So, the per iteration
computational complexity of the proposed algorithm is dominated by
two matrix-vector multiplications, $L(L+1)/2$ FFT (of length $M$)
and $L(L+1)/4$ IFFT (of length $M$) operations, so the total number
of flops would be around $\mathcal{O}(ML|a_{k}|)+\mathcal{O}(M\,log\,M)$.
The space complexity of proposed algorithm is dominated by two $(ML\times ML)$
matrices, one $(ML\times|a_{k}|)$ matrix, two $(|a_{k}|\times1)$
vectors and one $(ML\times1)$ vector and hence, total space complexity
is around $\mathcal{O}(ML|a_{k}|)$.

\section{NUMERICAL SIMULATIONS AND MIMO RADAR IMAGING EXPERIMENT}

\subsection{NUMERICAL SIMULATIONS}

To highlight the strength of the proposed algorithm, we conduct numerical
experiments for different dimensions of sequence set: $(L,M)=(2,100)$,
$(L,M)=(2,200)$, $(L,M)=(3,150)$ and $(L,M)=(4,256)$ using the
random initialization sequence. The random initialization sequence
is chosen as $\bigl\{ e^{j2\pi\theta_{a,b}}\bigr\},a=1,..,M;b=1,..,L$,
where $\bigl\{\theta_{a,b}\bigr\}$ are drawn randomly from the uniform
distribution $\left[0,1\right]$. All the numerical experiments were
performed in MATLAB R2018a on a PC with i7 processor, 12GB RAM. The
proposed algorithm is implemented and compared with the Multi-CAN,
MM-Corr, ISL-NEW, BiST (which is a PSL minimization algorithm but
the sequences designed will only take values from the finite unimodular
alphabets, in the simulations we have simulated BiST method taking
values from set with 8 alphabets) algorithms in terms of obtained
PSL value. In numerical experiments, for a fair comparison, all algorithms
are initialized using the same initial sequence set and stopped using
the same convergence criterion of either $500$ iterations (or) the
convergence threshold of $\epsilon\leq10^{-6}$ where 
\begin{equation}
\epsilon=\frac{\left|\text{PSL}(t)-\text{PSL}(t-1)\right|}{\text{PSL}(t-1)}\label{eq:stopcriteria}
\end{equation}

where $\text{PSL}(t)$ is the PSL value at $t^{th}$ iteration.

\subsubsection*{(i) PSL vs Iteration}

\begin{figure*}[tp]
\subfloat[Sequence set of dimension $(L,M)=(2,100)$]{\includegraphics[scale=0.55]{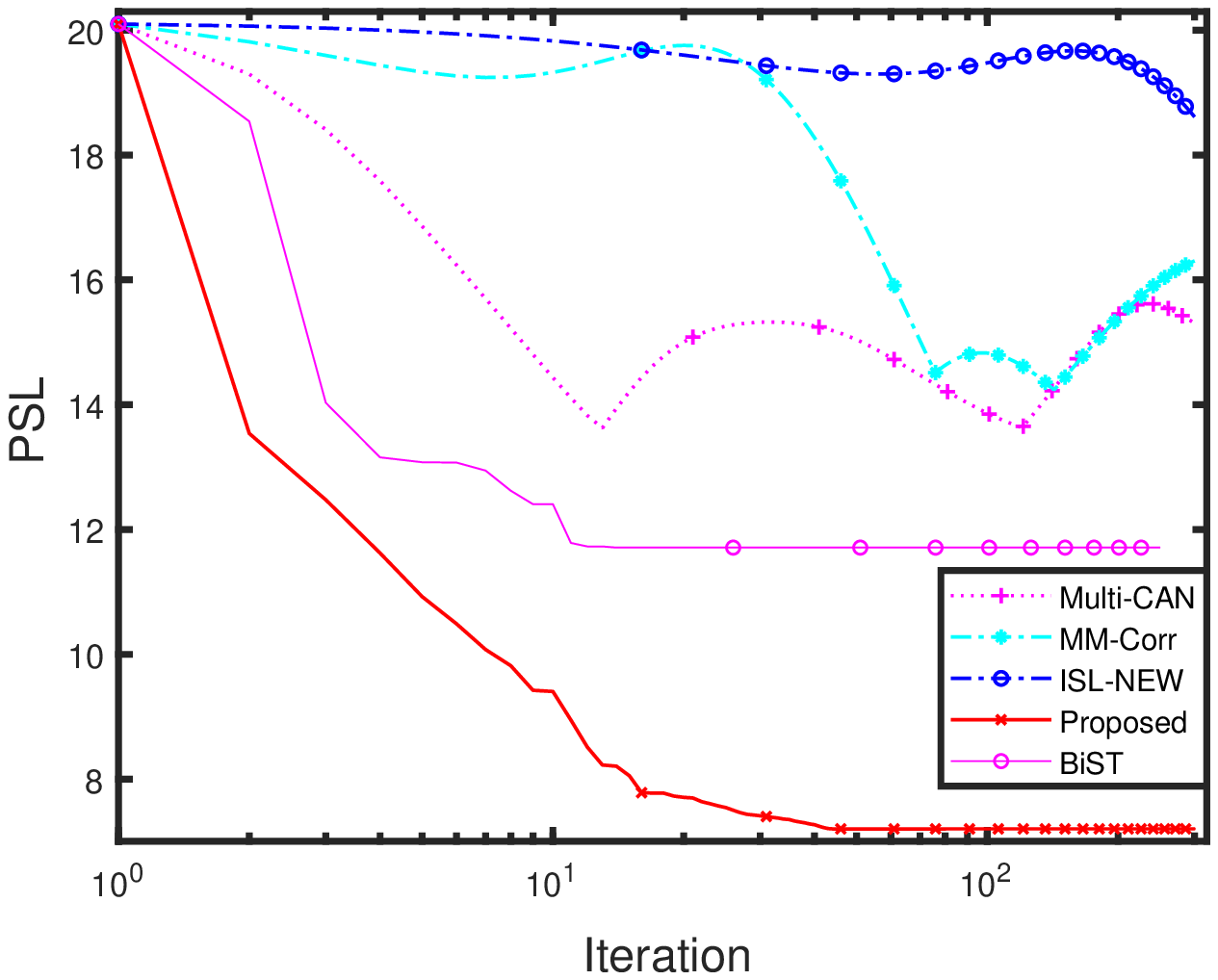}

}\subfloat[Sequence set of dimension $(L,M)=(2,200)$]{\includegraphics[scale=0.55]{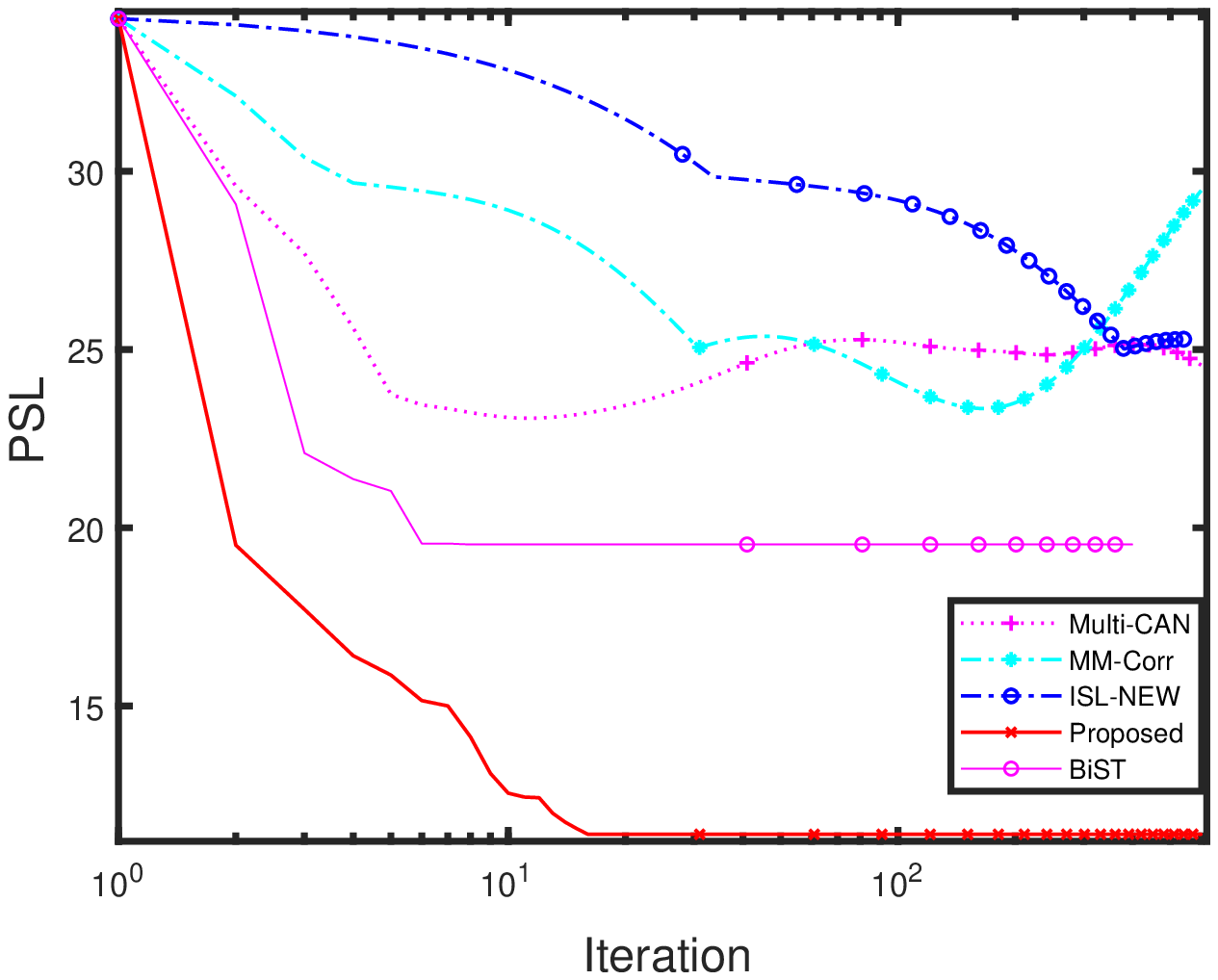}

}

\subfloat[Sequence set of dimension $(L,M)=(3,150)$]{\includegraphics[scale=0.4]{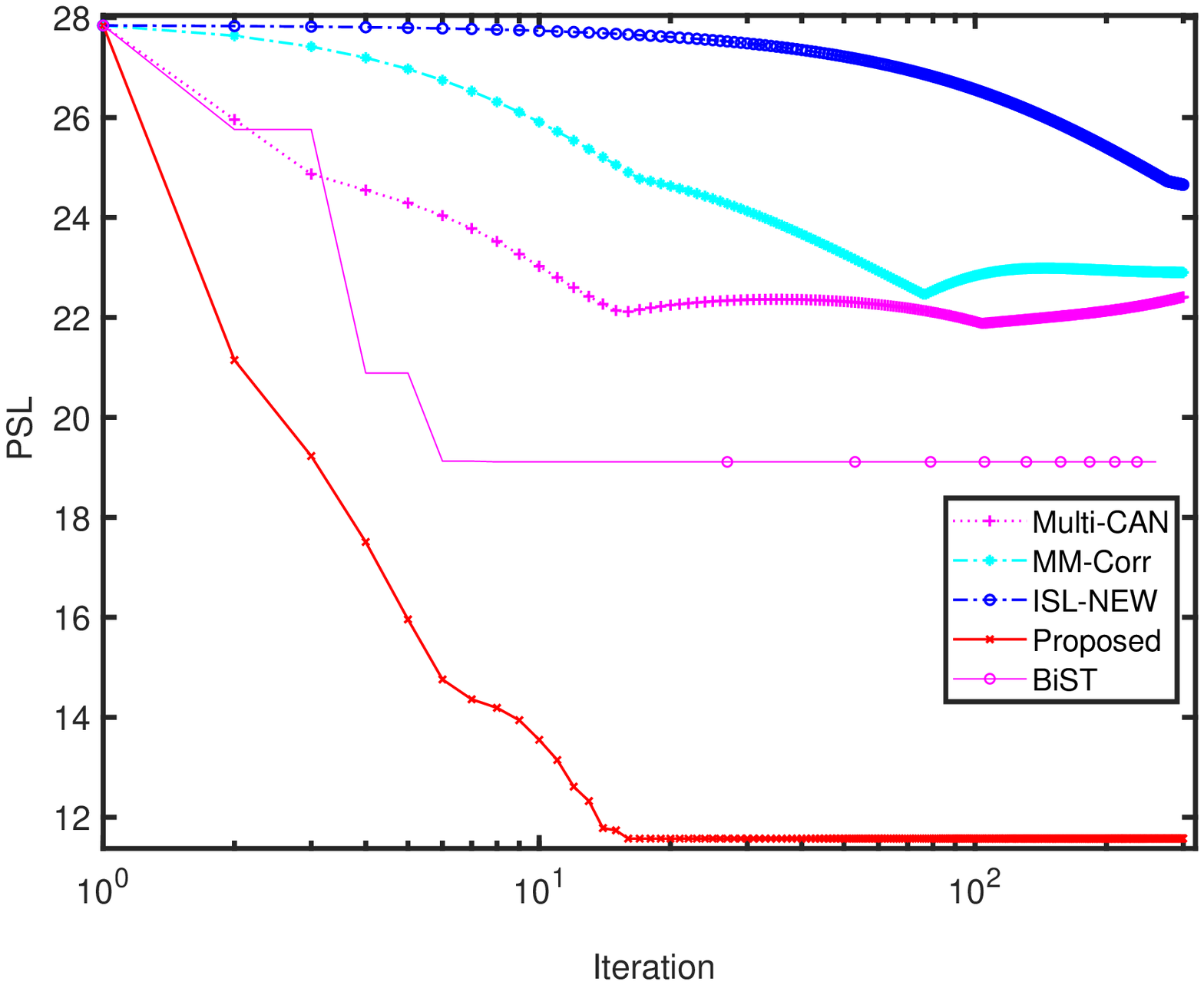}

}\subfloat[Sequence set of dimension $(L,M)=(4,256)$]{\includegraphics[scale=0.55]{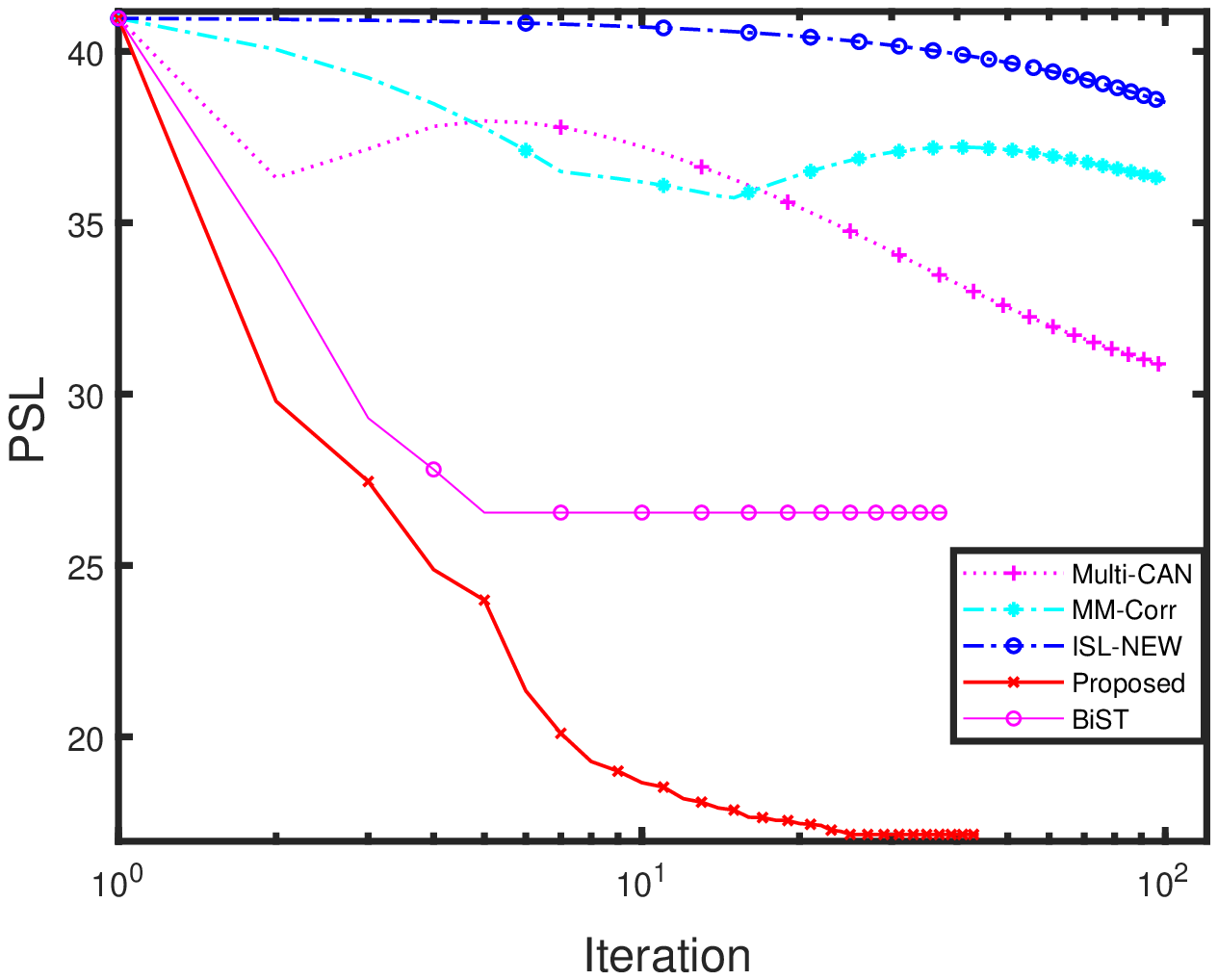}

}

\caption{PSL vs Iteration}
\end{figure*}

Figure. 1 show the plots of PSL value with respect to iterations for
different dimensions of sequence sets $(L,M)=(2,100)$, $(L,M)=(2,200)$,
$(L,M)=(3,150)$ and $(L,M)=(4,256)$. From the simulation plots,
it can be seen that though all the algorithms have started at the
same PSL value, they all converged to different PSL values and the
proposed algorithm has reached the PSL value which is better than
the state-of-the-art algorithms. For instance, from figure-1(b), one
can observe that for a sequence set dimension of $(L,M)=(2,200)$,
the proposed algorithm has converged to a PSL value $11$, while the
state-of-the-art methods converged to $24$ (which is roughly two
times more than that of the proposed method). Hence, we conclude that
irrespective of sequence set dimension, our proposed algorithm exhibits
better performance than the state-of-the-art algorithms in terms of
PSL value.

\subsubsection*{(ii) Aperiodic correlations vs Lag}

\begin{figure*}[tp]
\subfloat[$|r_{1,1}(k)|$ vs. $k$]{\includegraphics[scale=0.55]{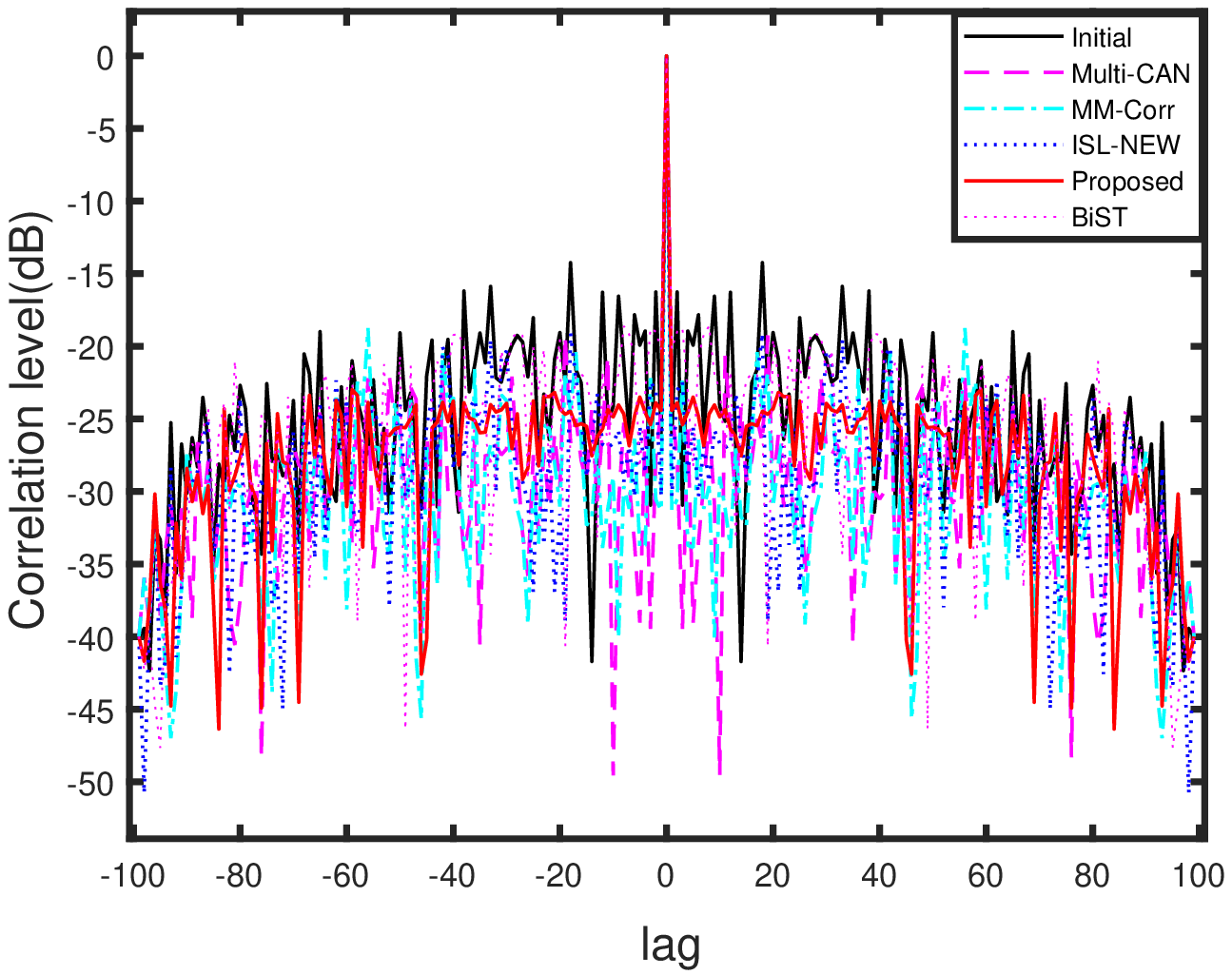}

}\subfloat[$|r_{1,2}(k)|$ vs. $k$]{\includegraphics[scale=0.55]{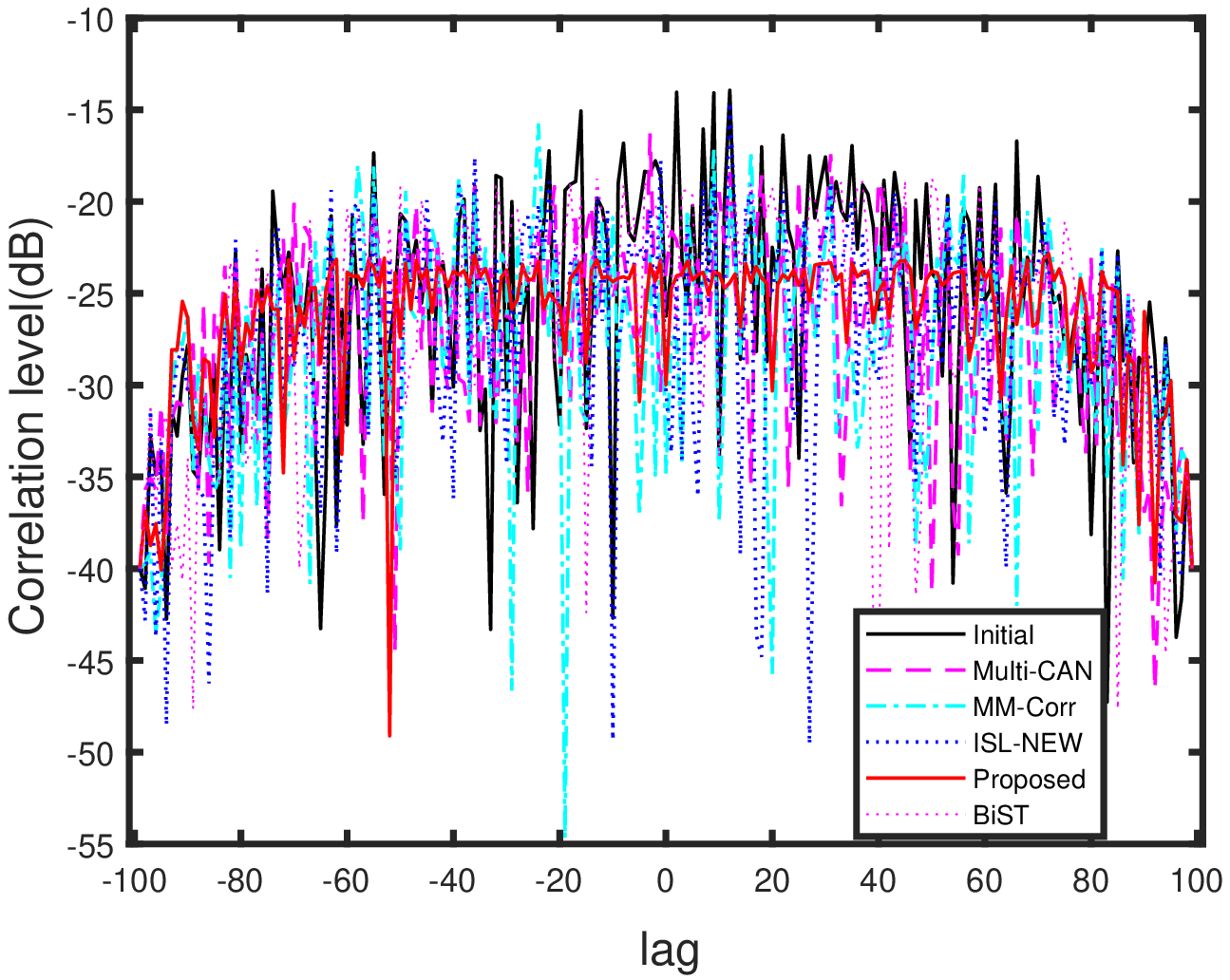}

}

\subfloat[$|r_{2,1}(k)|$ vs. $k$]{\includegraphics[scale=0.55]{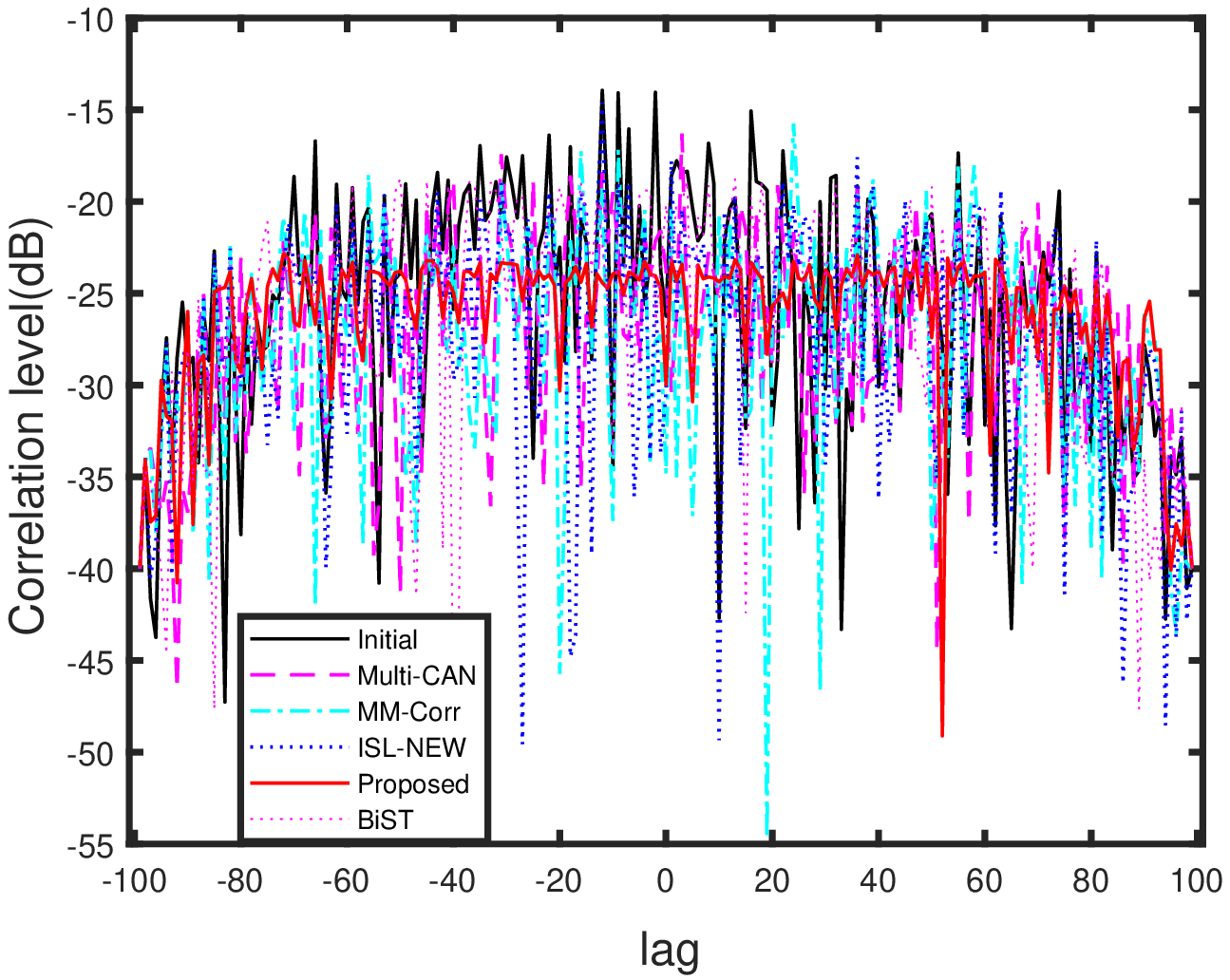}

}\subfloat[$|r_{2,2}(k)|$ vs. $k$]{\includegraphics[scale=0.55]{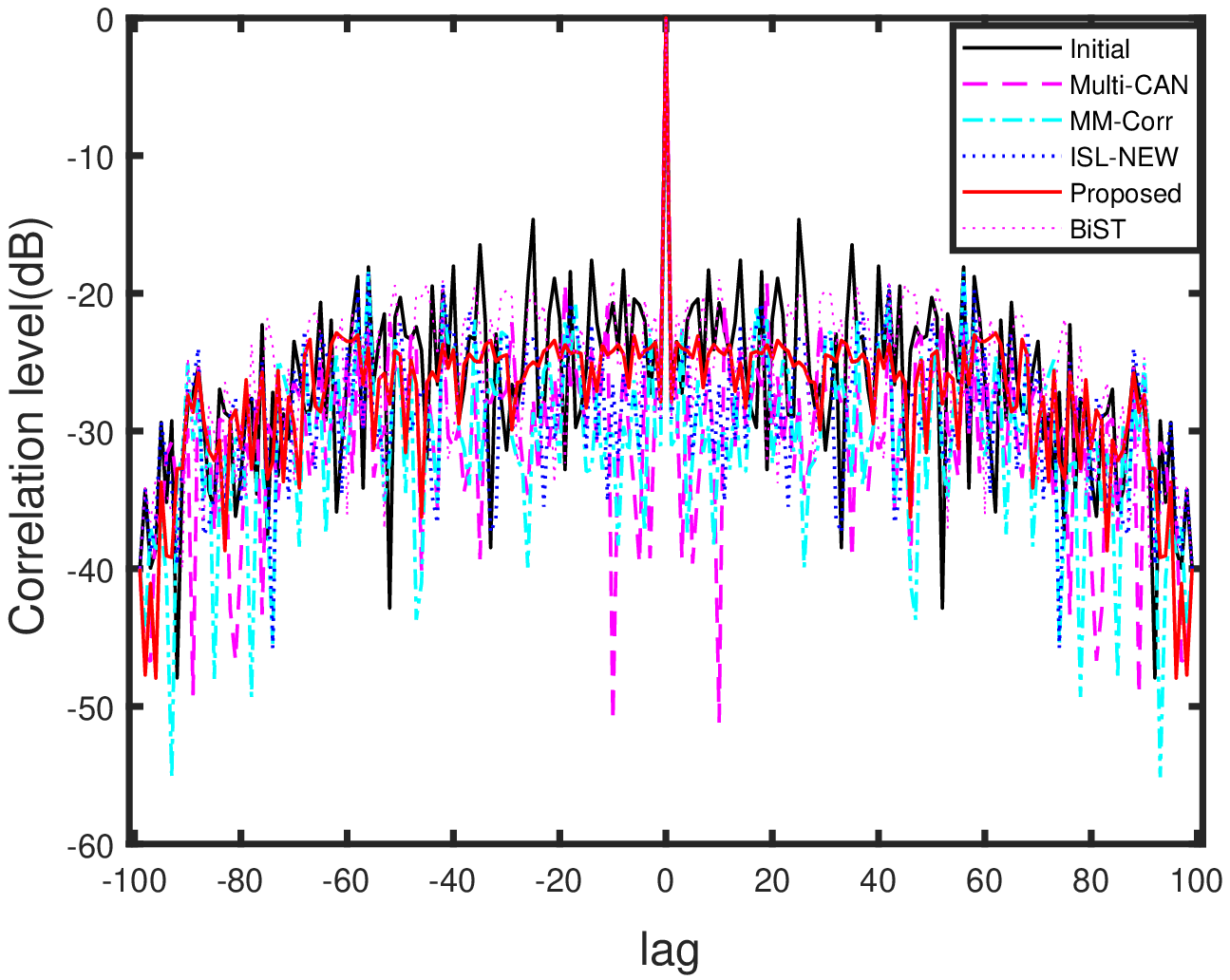}

}

\caption{Correlations plots for sequence set design for dimensions $(L,M)=(2,100)$,
please note that plots of $r_{1,2}(k)$ and $r_{2,1}(k)$ are mirror
images of each other. }
\end{figure*}

\begin{figure*}[tp]
\subfloat[$|r_{1,1}(k)|$ vs. $k$]{\includegraphics[scale=0.55]{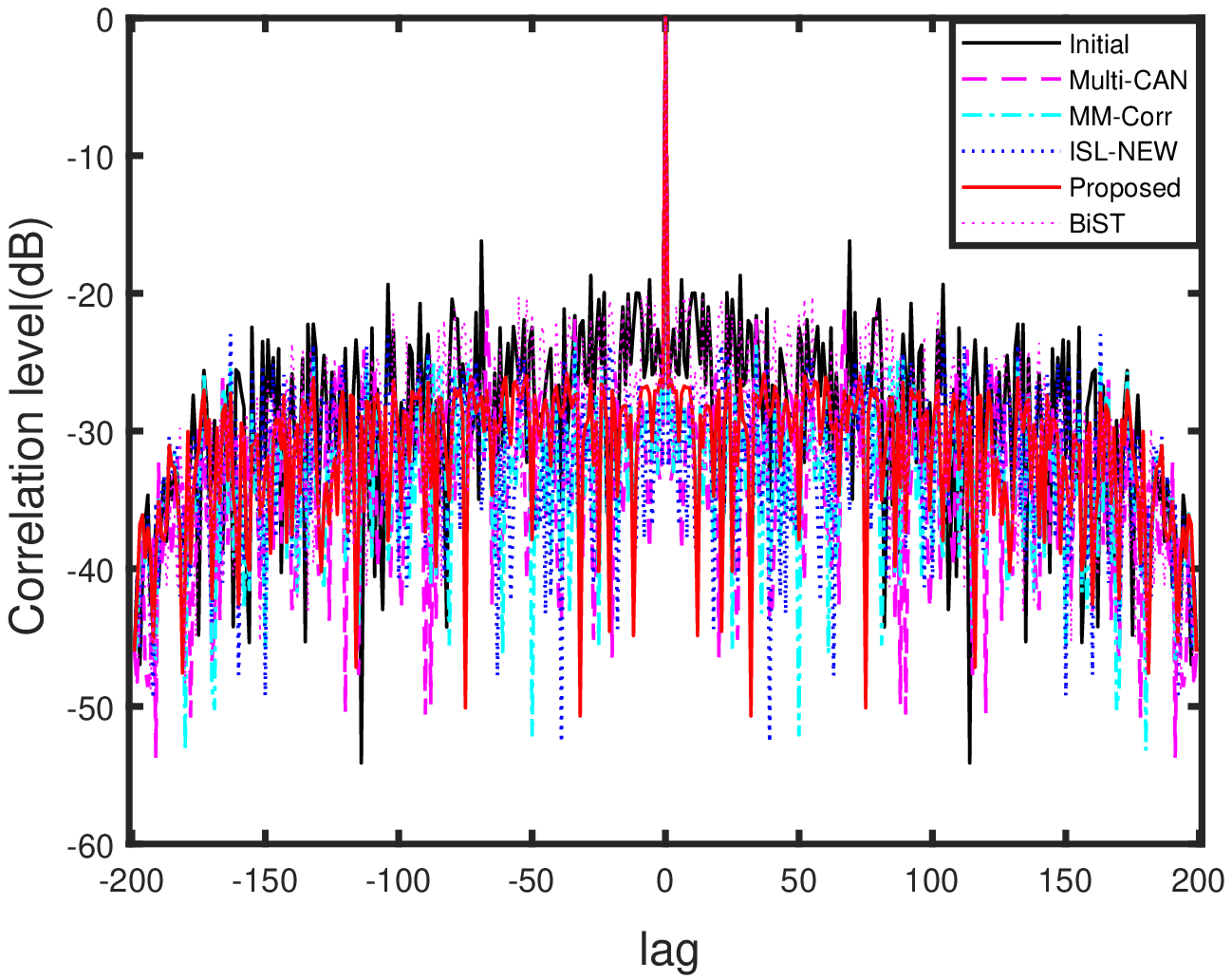}

}\subfloat[$|r_{1,2}(k)|$ vs. $k$]{\includegraphics[scale=0.55]{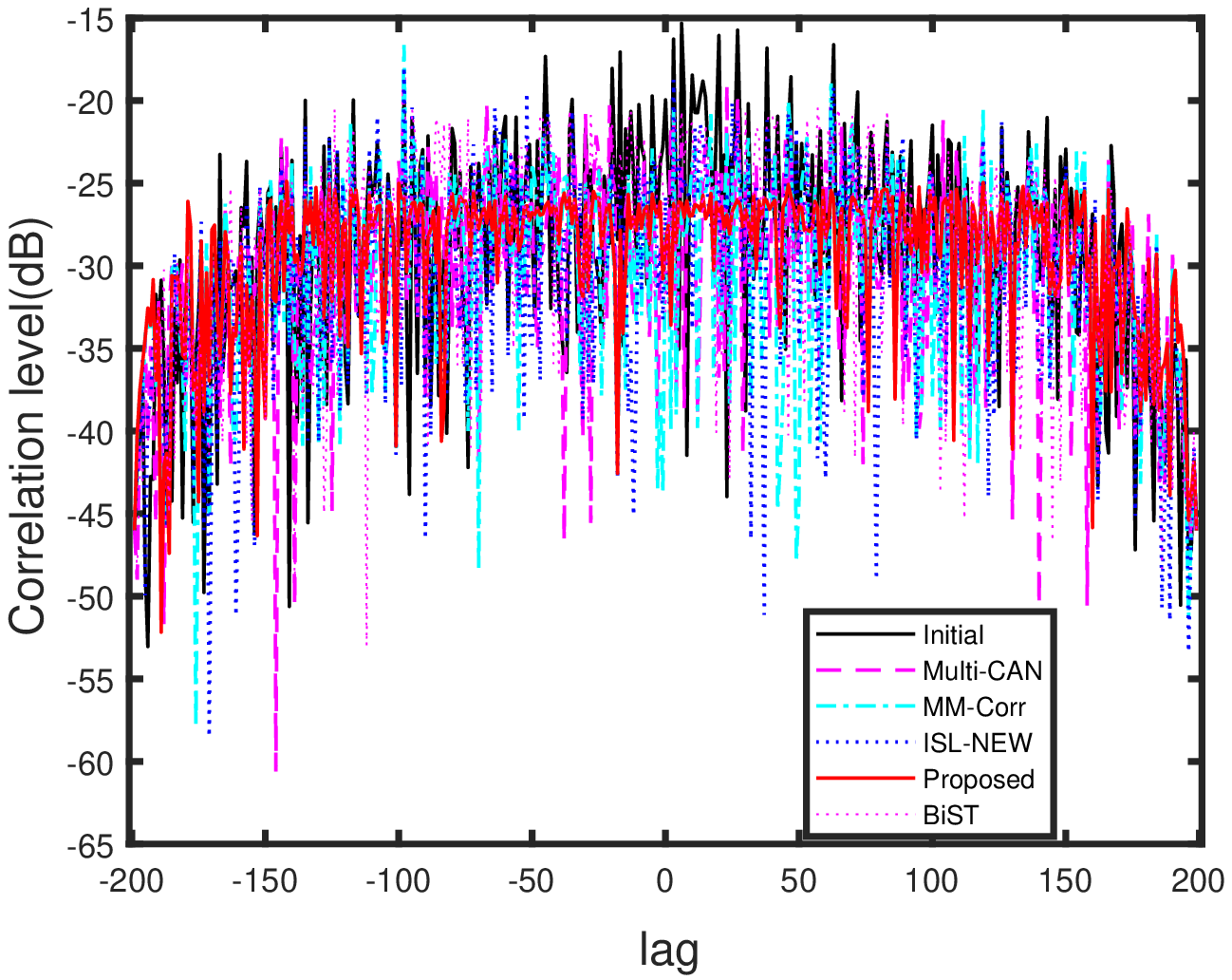}

}

\subfloat[$|r_{2,1}(k)|$ vs. $k$]{\includegraphics[scale=0.55]{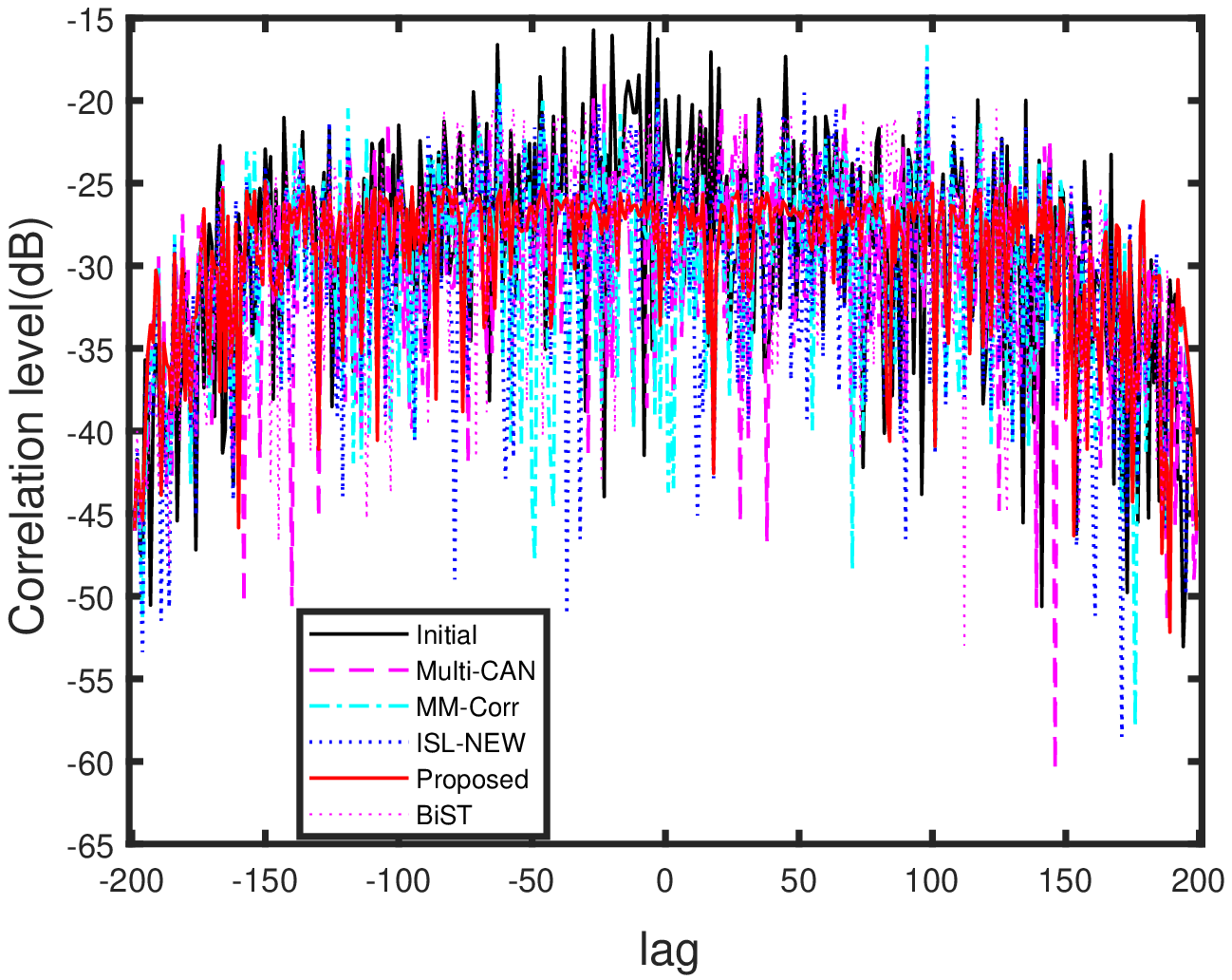}

}\subfloat[$|r_{2,2}(k)|$ vs. $k$]{\includegraphics[scale=0.55]{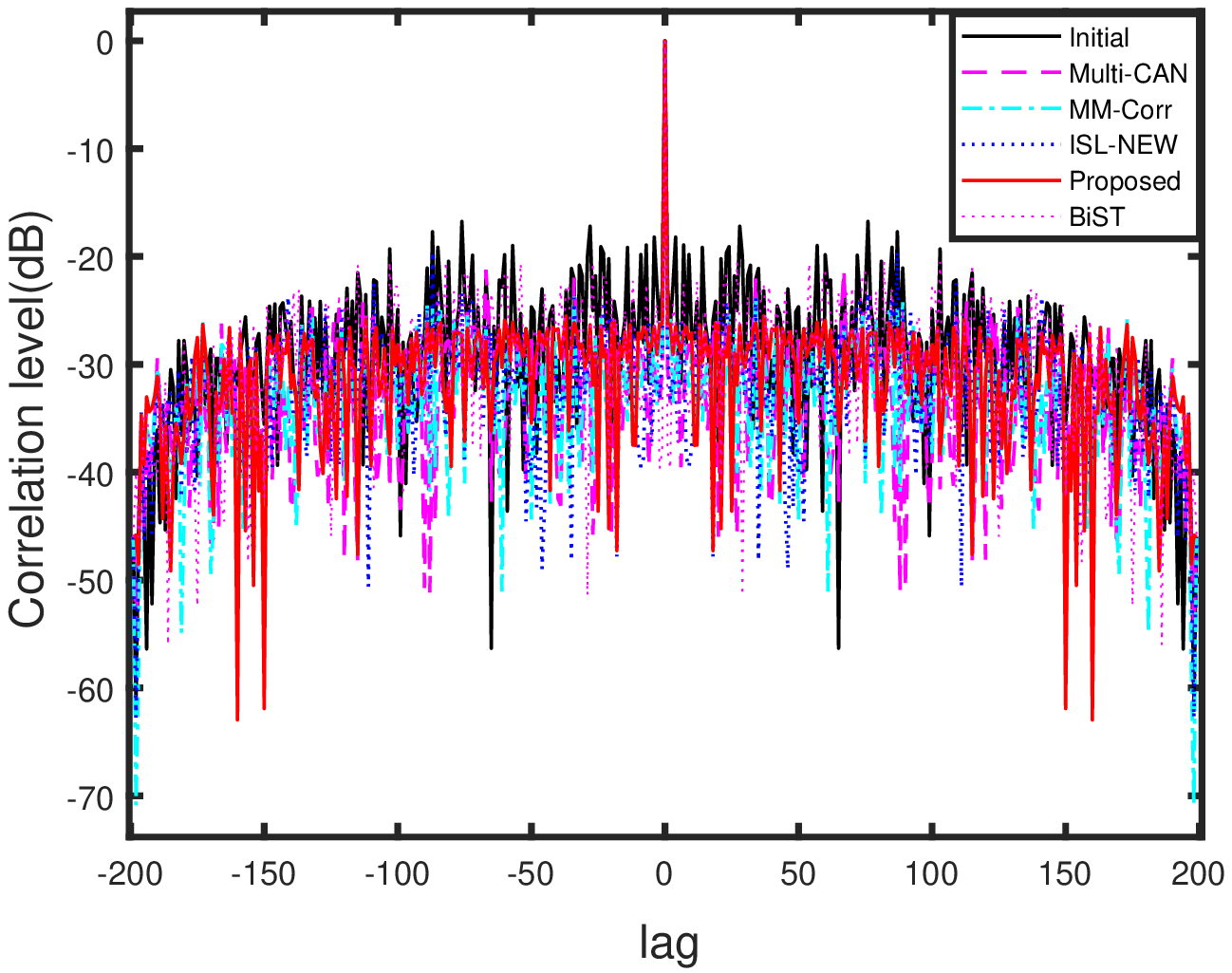}

}

\caption{Correlations plots for sequence set design for dimensions $(L,M)=(2,200)$,
please note that plots of $r_{1,2}(k)$ and $r_{2,1}(k)$ are mirror
images of each other.}
\end{figure*}

\begin{figure*}[tp]
\subfloat[$|r_{1,1}(k)|$ vs $k$]{\includegraphics[scale=0.4]{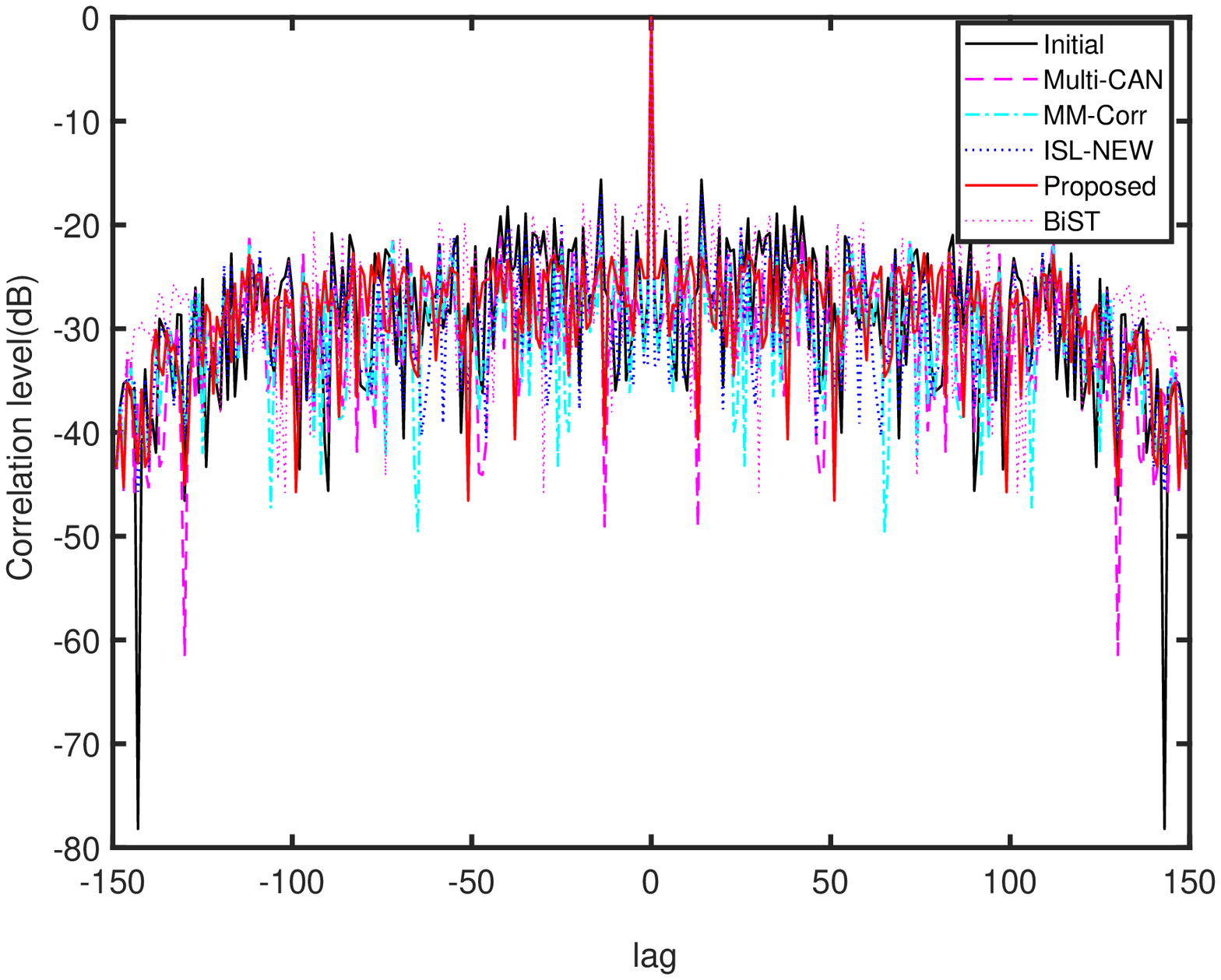}

}\subfloat[$|r_{2,2}(k)|$ vs $k$]{\includegraphics[scale=0.4]{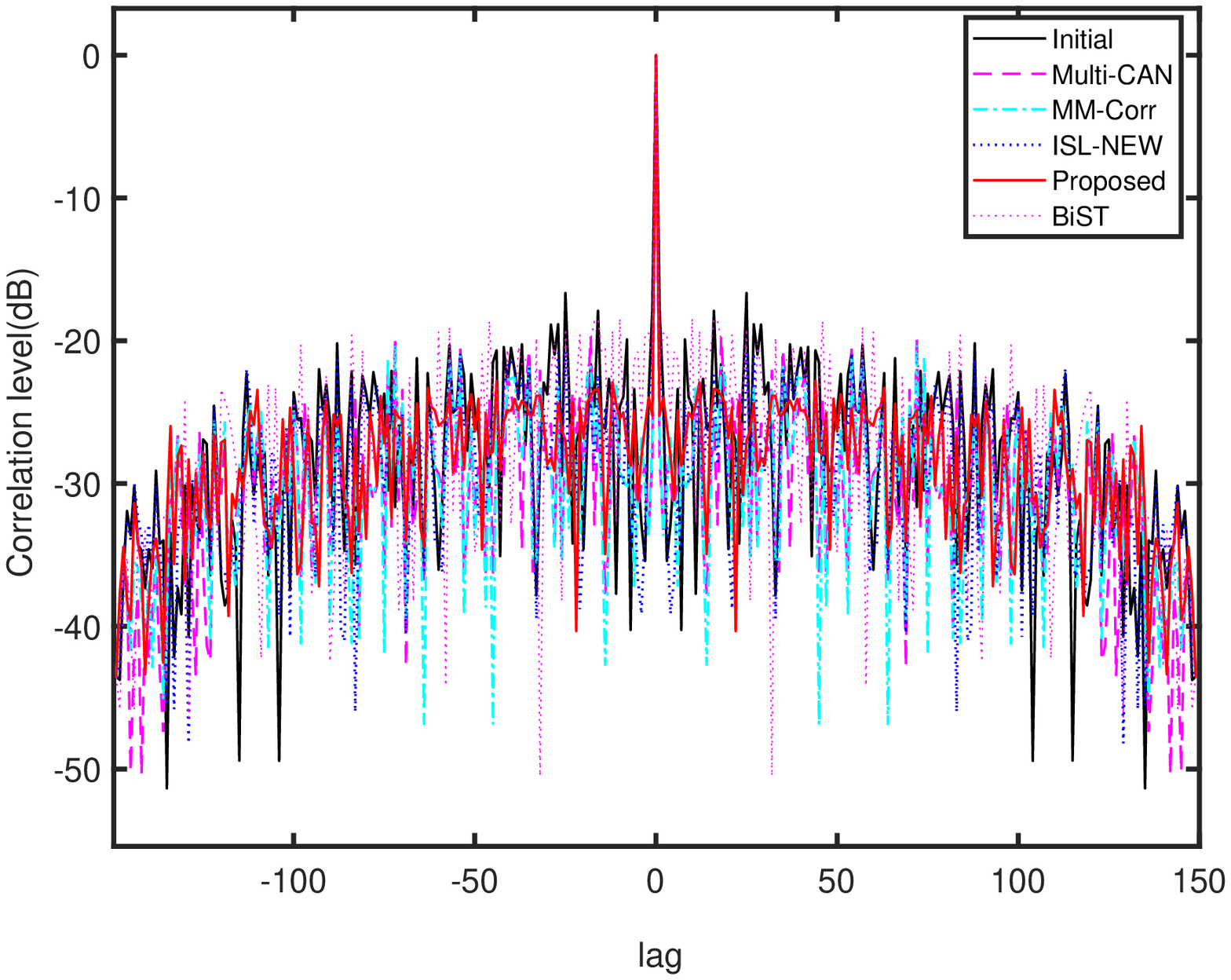}

}

\subfloat[$|r_{3,3}(k)|$ vs $k$]{\includegraphics[scale=0.4]{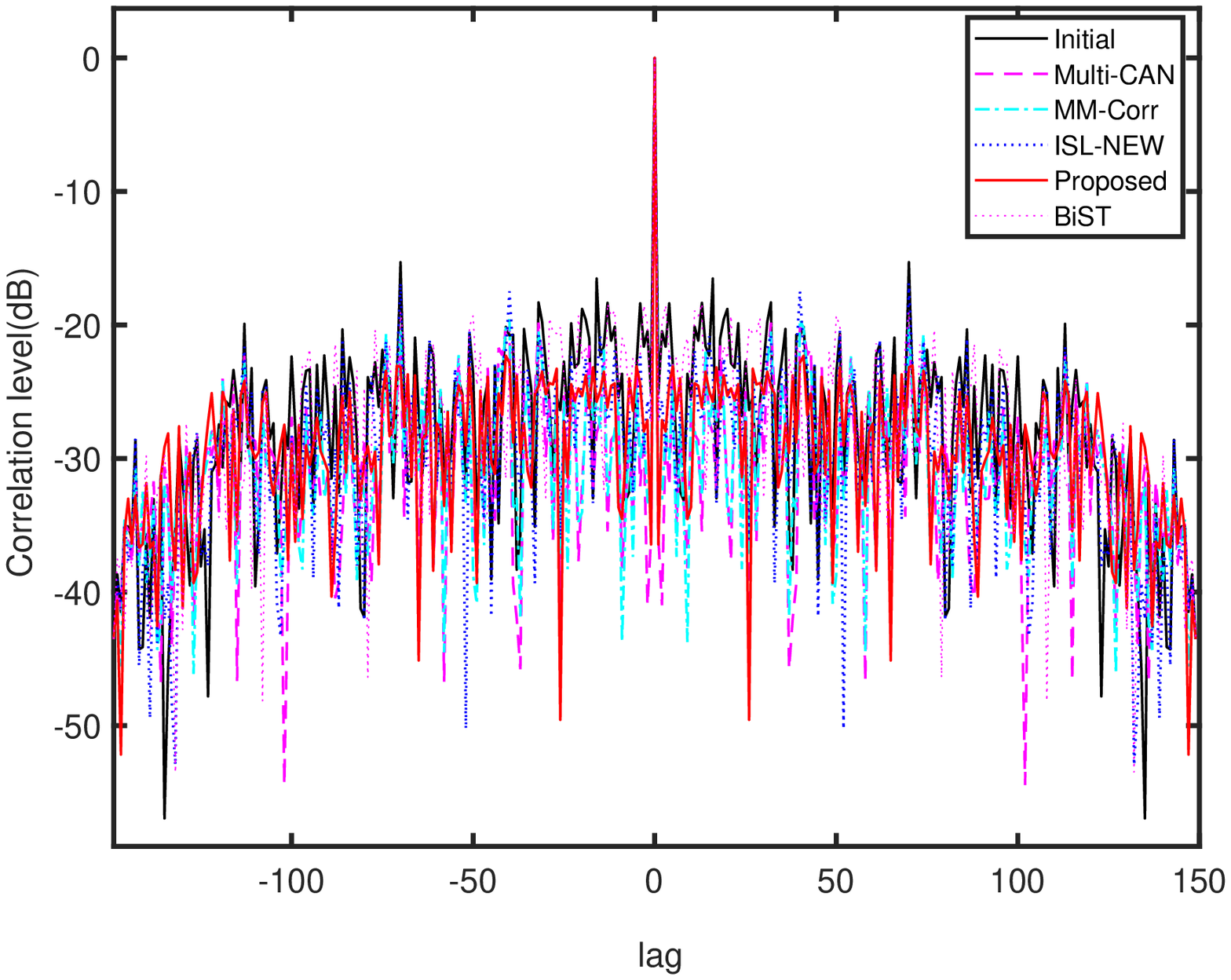}

}\subfloat[$|r_{1,2}(k)|$ vs $k$]{\includegraphics[scale=0.4]{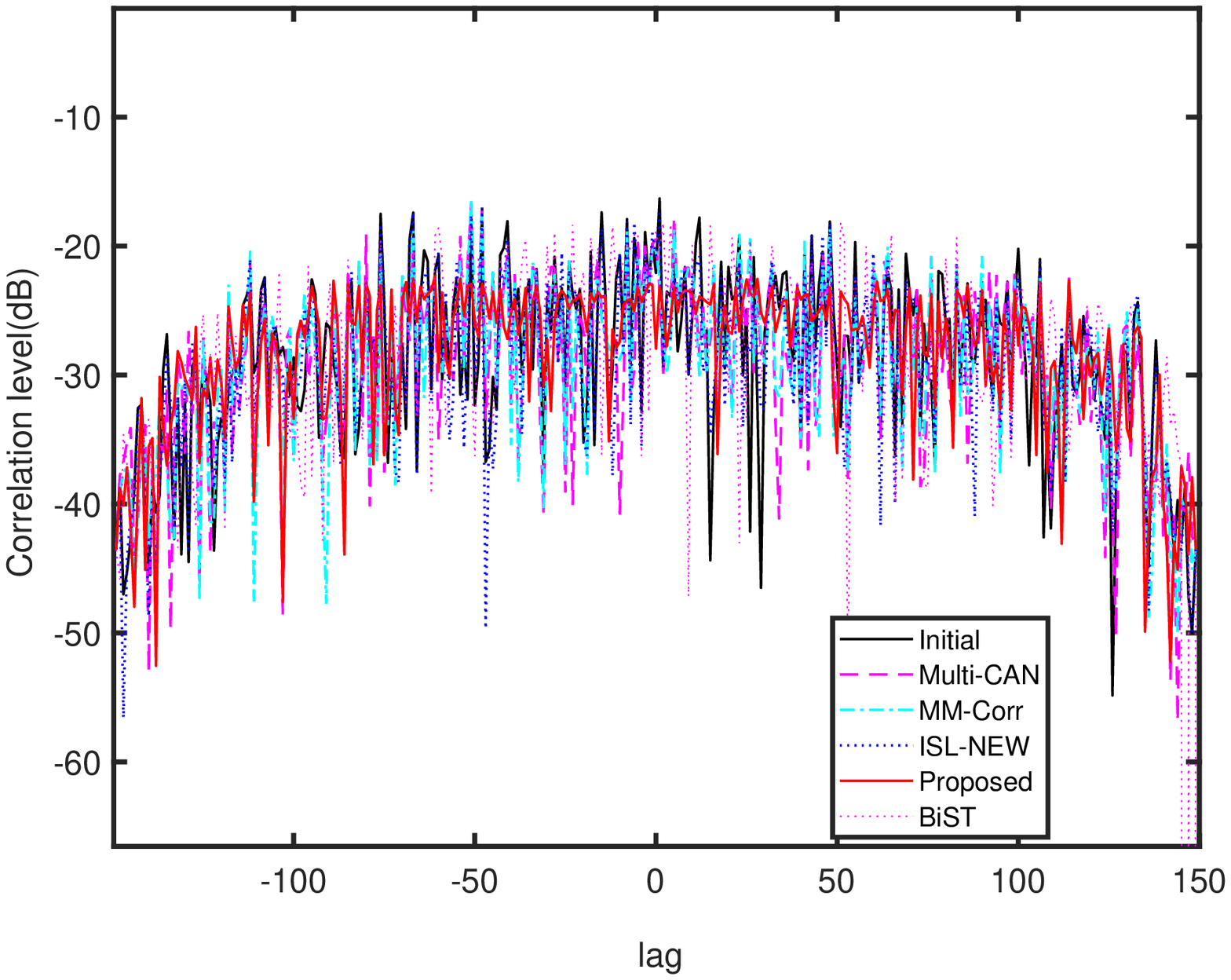}

}

\subfloat[$|r_{1,3}(k)|$ vs $k$]{\includegraphics[scale=0.4]{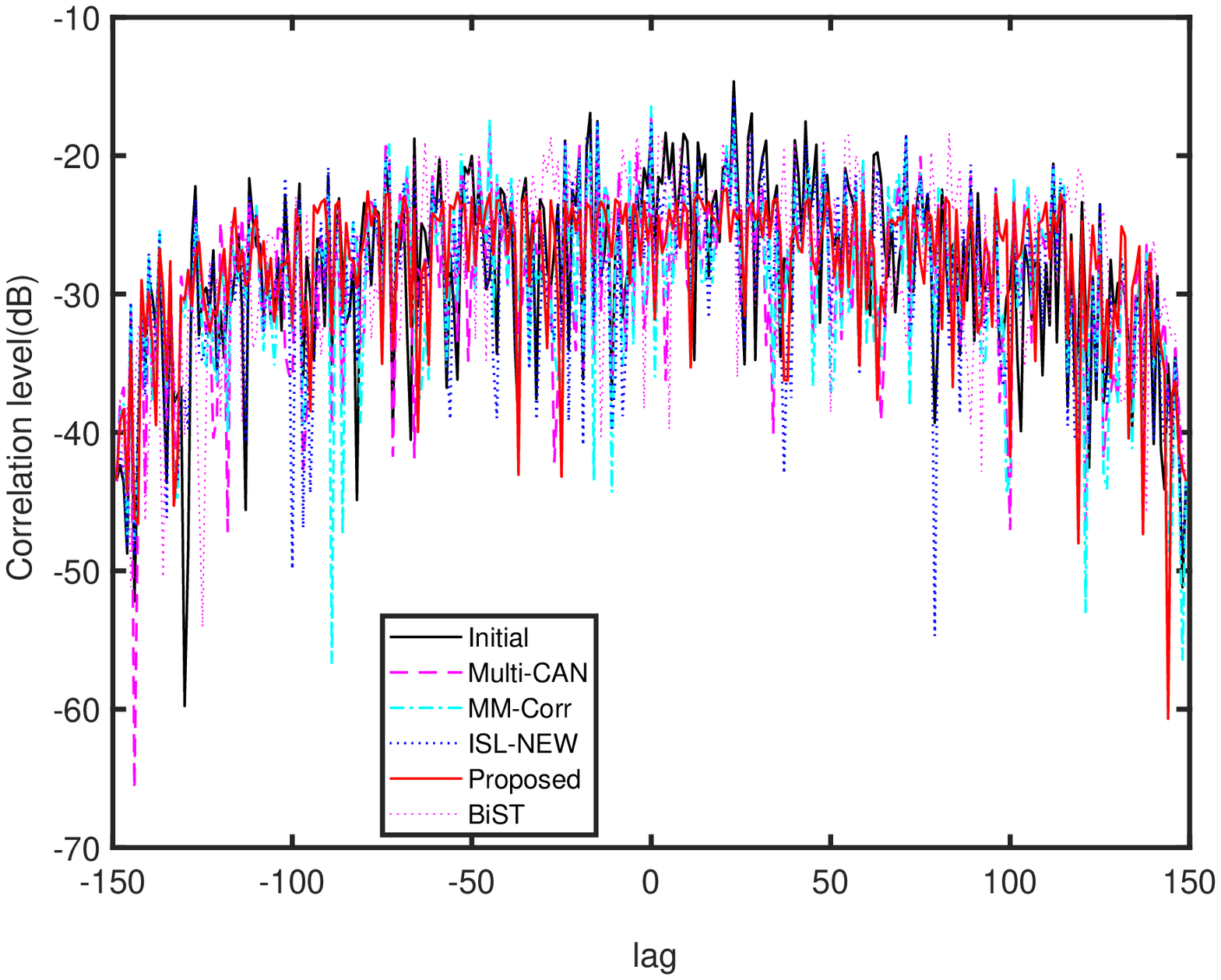}

}\subfloat[$|r_{2,3}(k)|$ vs $k$]{\includegraphics[scale=0.4]{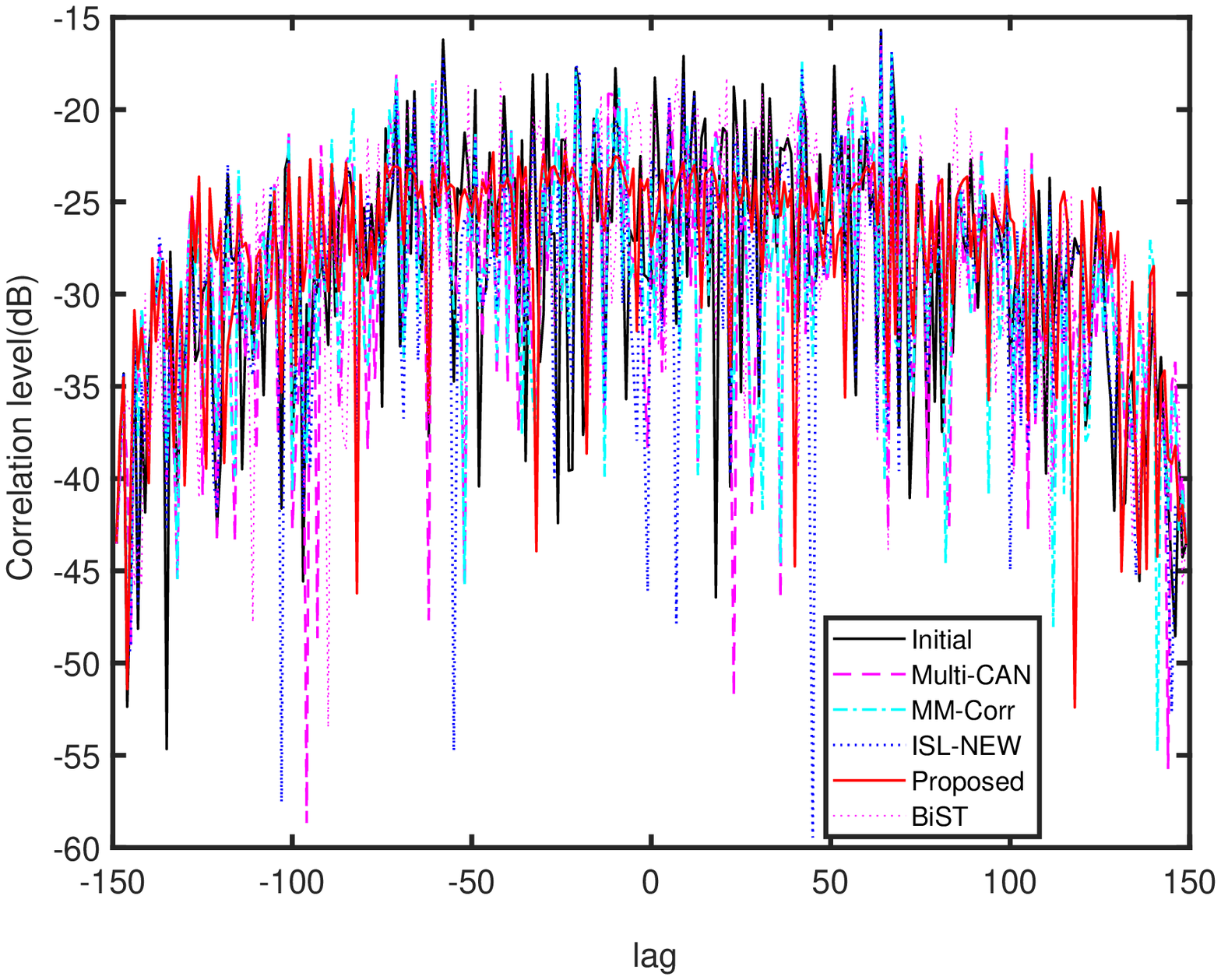}

}

\caption{Correlation plots vs lag for sequence set design for dimensions $(L,M)=(3,150)$}
\end{figure*}

\begin{figure*}[tp]
\subfloat[$|r_{1,1}(k)|$ vs. $k$]{\includegraphics[scale=0.55]{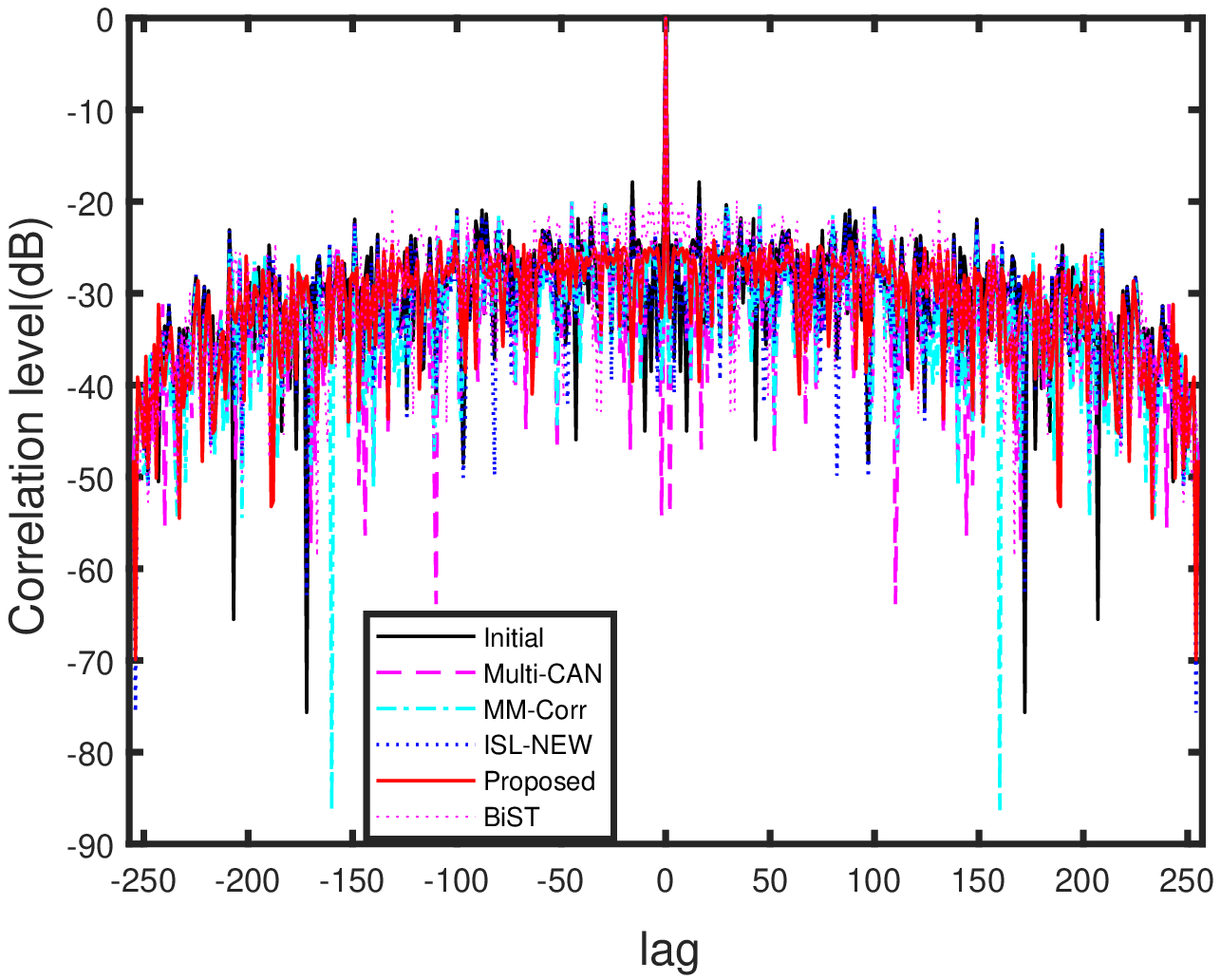}

}\subfloat[$|r_{2,2}(k)|$ vs. $k$]{\includegraphics[scale=0.55]{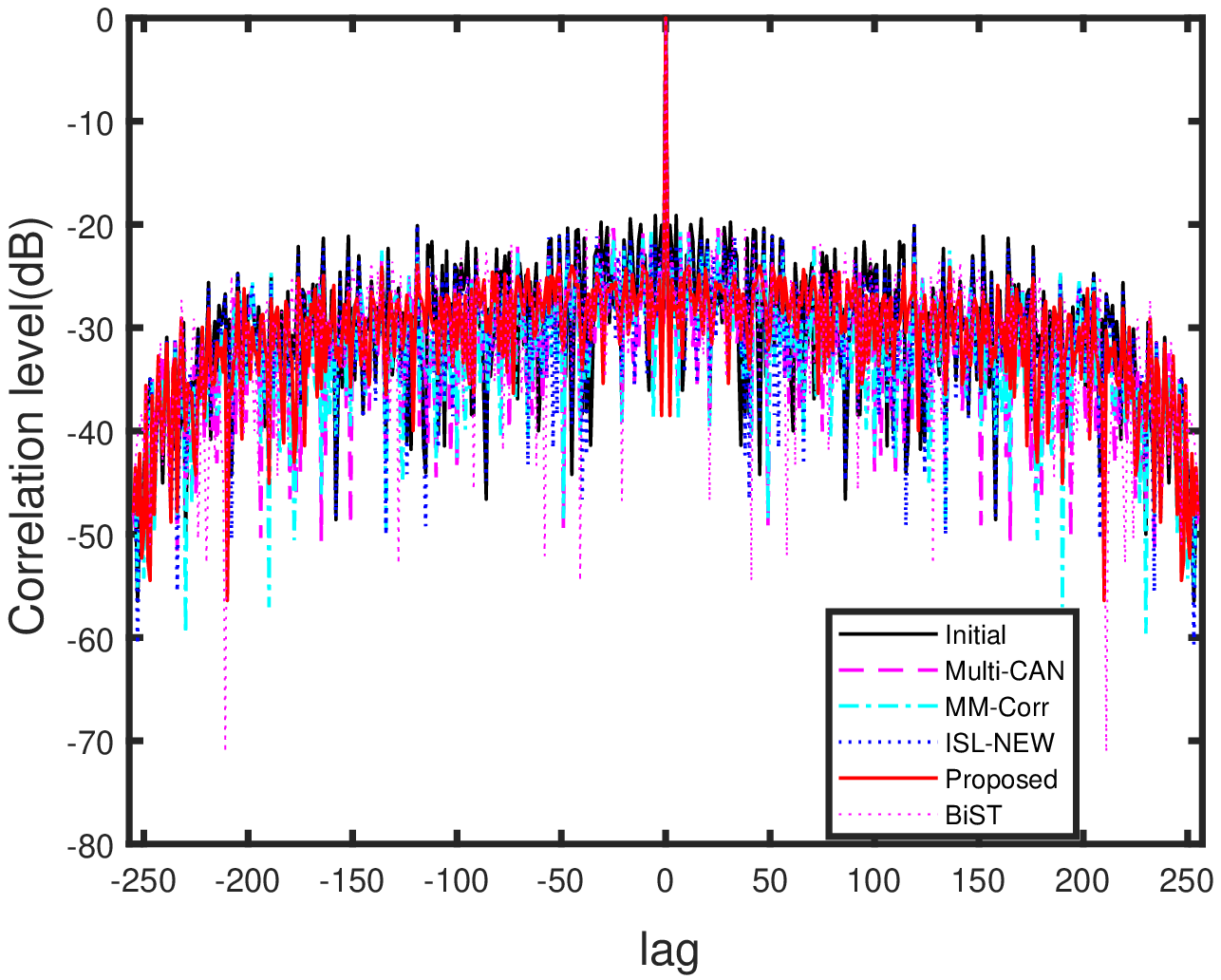}

}

\subfloat[$|r_{3,3}(k)|$ vs. $k$]{\includegraphics[scale=0.55]{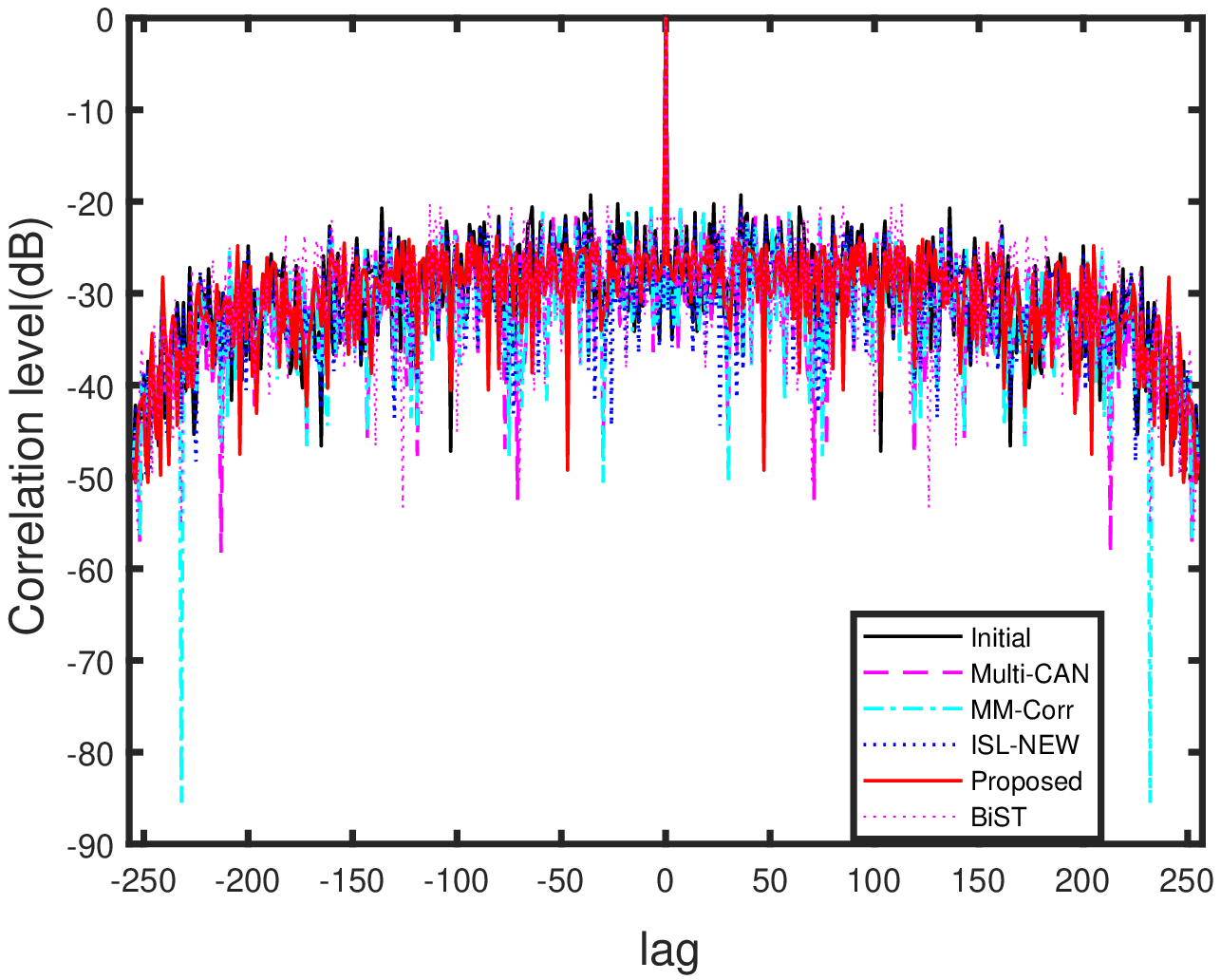}

}\subfloat[$|r_{4,4}(k)|$ vs. $k$]{\includegraphics[scale=0.55]{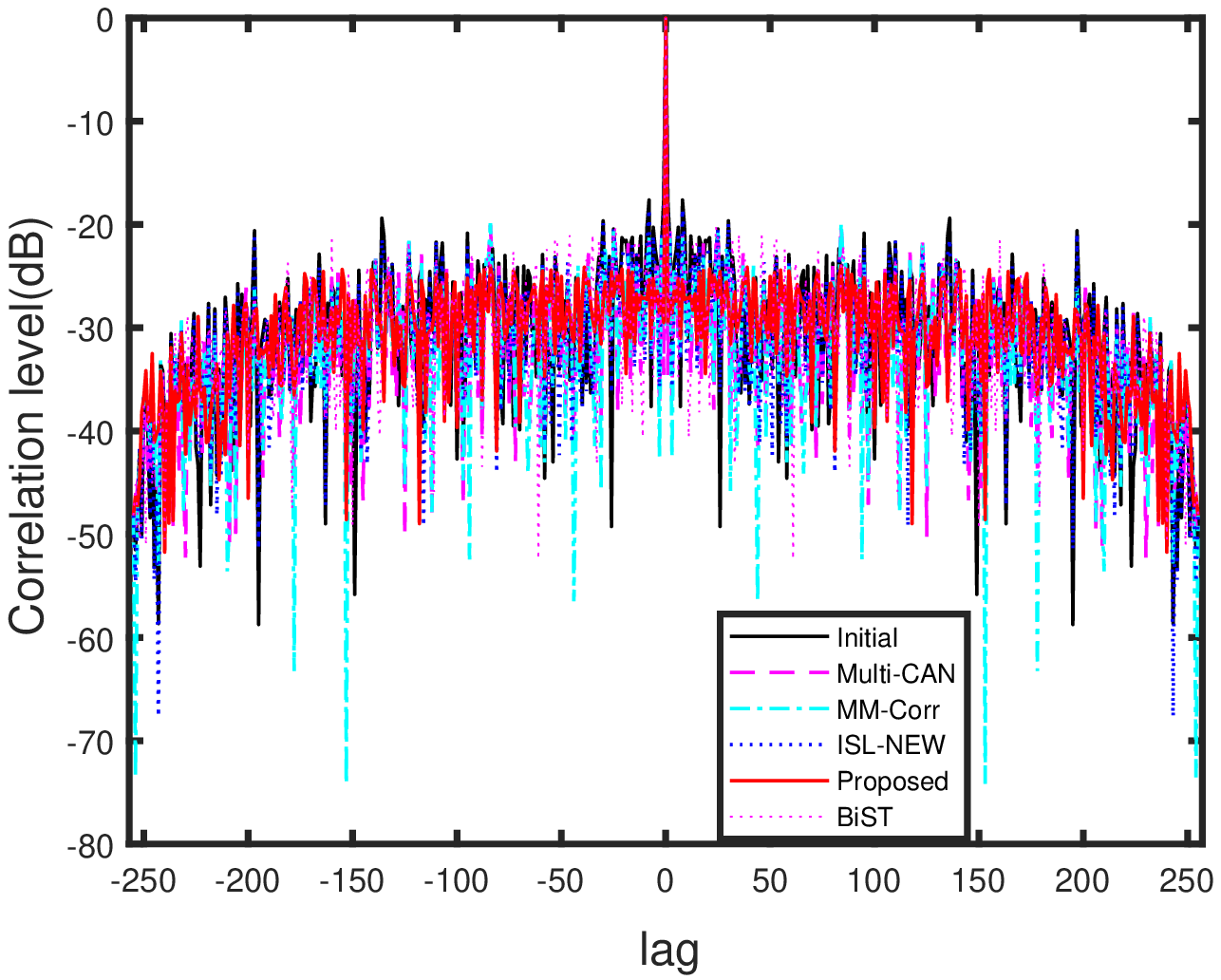}

}

\subfloat[$|r_{1,2}(k)|$ vs. $k$]{\includegraphics[scale=0.55]{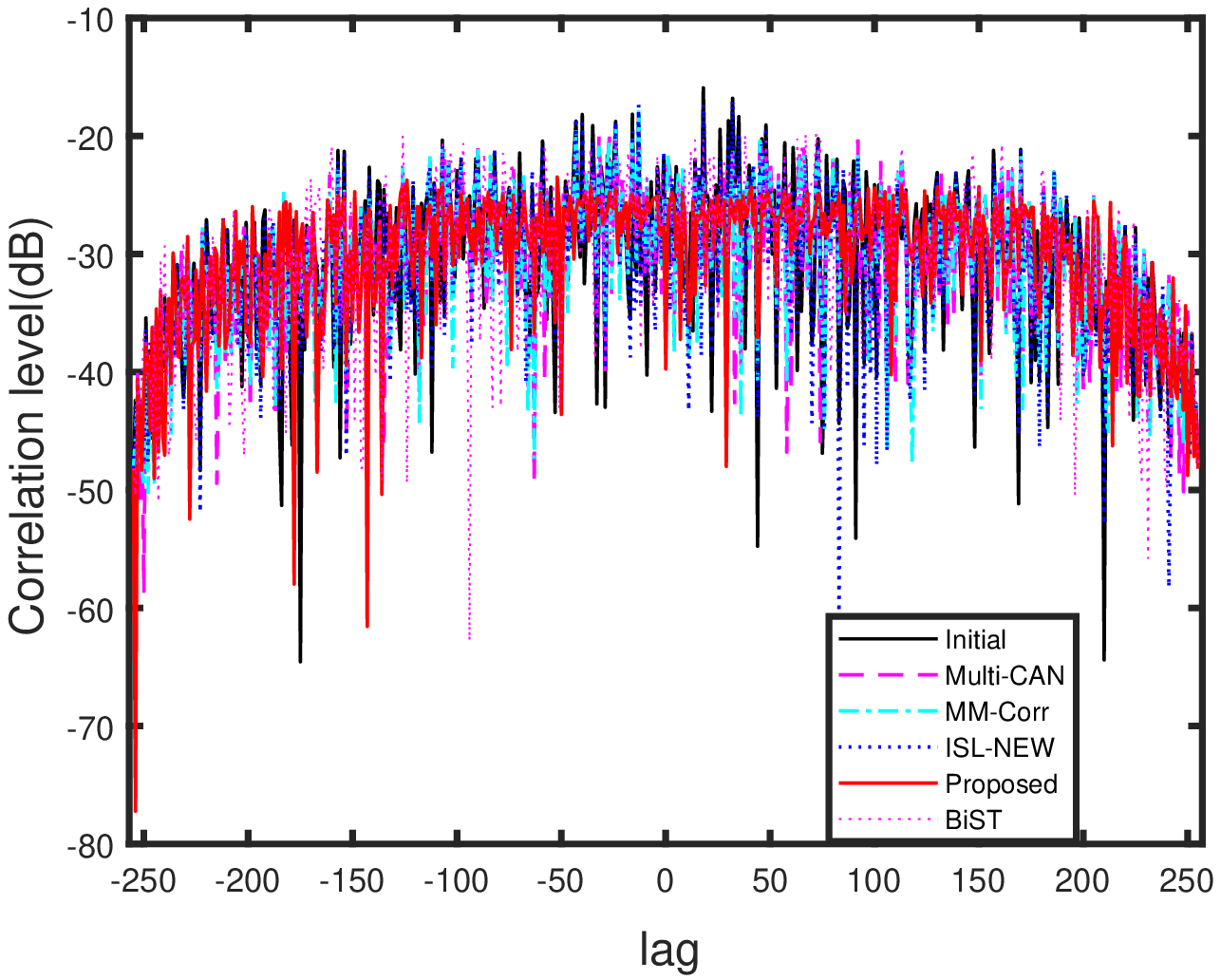}

}\subfloat[$|r_{3,4}(k)|$ vs. $k$]{\includegraphics[scale=0.55]{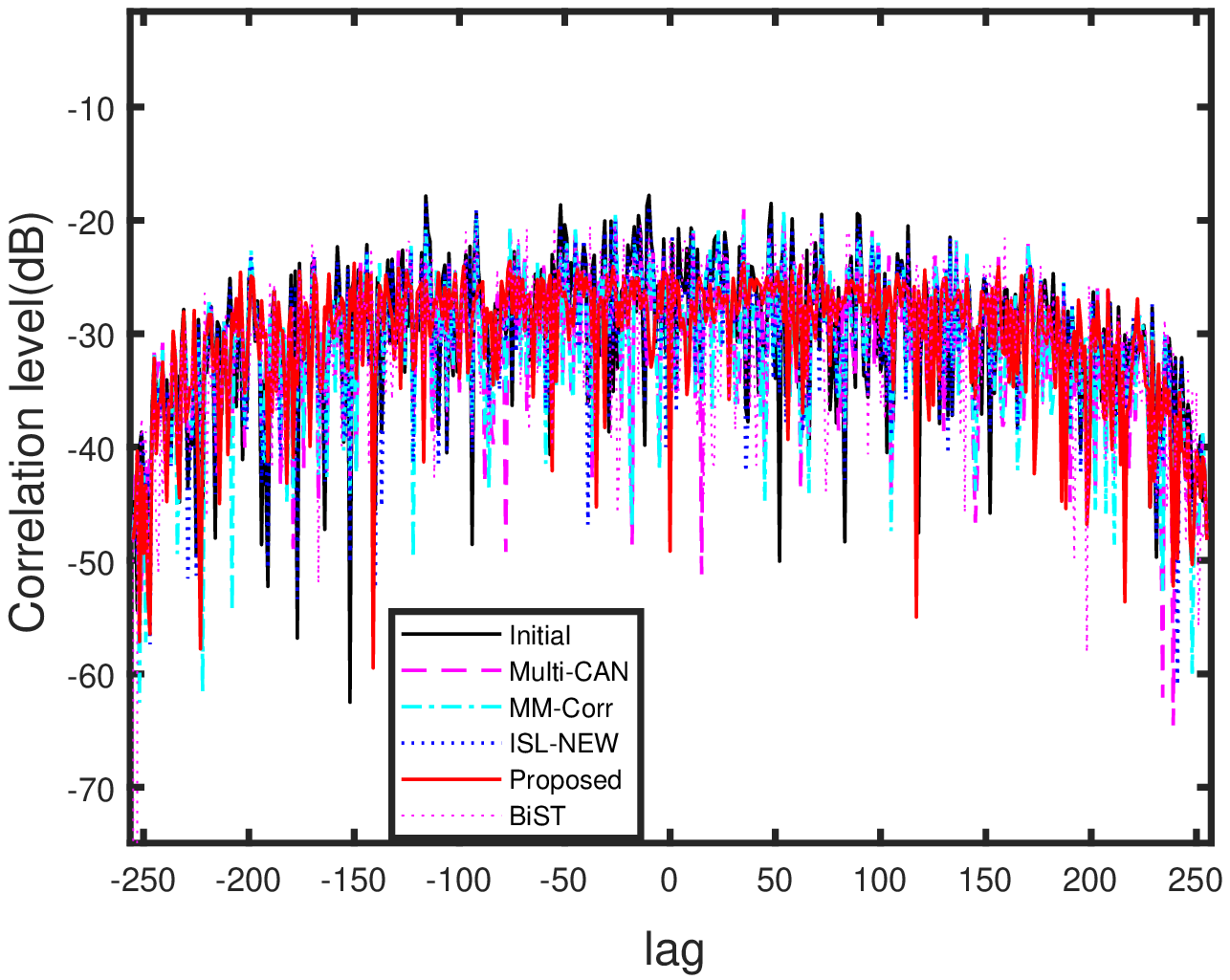}

}

\caption{Correlations plots for sequence set design for dimensions $(L,M)=(4,256)$,
please note that the plots of $r_{1,3}(k)$, $r_{1,4}(k)$, $r_{2,3}(k)$,
$r_{2,4}(k)$ are not included here.}
\end{figure*}

Figure. 2 to Figure. 5 show the plots of correlation values (auto-correlations
as well as cross-correlations) with respect to lag for different dimensions
of sequence sets $(L,M)=(2,100)$, $(L,M)=(2,200)$, $(L,M)=(3,150)$
and $(L,M)=(4,256)$ respectively. From the simulation plots, we observe
that, in comparison to the state-of-the-art algorithms, our proposed
algorithm is performing well in terms of the PSL value and more importantly
the sequences obtained via our approach have almost equi-sidelobe
level (both autocorrelation and cross correlation), which is one of
the main goals of our approach.

\subsection{MIMO RADAR SAR IMAGING EXPERIMENT }

In MIMO RADAR high-resolution imaging application \cite{R_Imaging_Li_PS,Multi_CAN,sd3},
probing sequences with lower peak side-lobe level are usually preferred.
So, to highlight the strength of the proposed algorithm (generated
sequence set) we conduct a MIMO RADAR angle-ranging (with negligible
doppler effect) experiment. Consider the colocated Uniform Linear
Array (ULA) at both the transmitting and receiving ends with $4$
transmitters and $4$ receivers with inter-element spacing between
the elements equal to $2\lambda$ and $\frac{\lambda}{2}$ ($\lambda$
is wavelength) respectively. If we assume that the targets are in
the far-field with a simulated pattern \textbf{LT} occupying $Q=60$
range bins (vertical) and $P=81\,(-40^{\circ}\text{ to }40^{\circ})$
number of scanning angles (horizontal). Let $\boldsymbol{\boldsymbol{S}}$
denote the probing signal matrix, then the received data can be modeled
as given as:

\begin{equation}
\boldsymbol{\boldsymbol{B}}^{H}=\stackrel[r=0]{Q-1}{\sum}\stackrel[p=1]{P}{\sum}\beta_{rp}\boldsymbol{\boldsymbol{c}}_{p}\boldsymbol{\boldsymbol{d}}_{p}^{T}\boldsymbol{\hat{\boldsymbol{S}}}^{H}\boldsymbol{\boldsymbol{J}}_{p}+\boldsymbol{\boldsymbol{N}}^{H}\label{eq:Recei}
\end{equation}

where $\left\{ \beta_{rp}\right\} _{r=0,p=1}^{Q-1,P}$ denote the
radar cross sections (rcs) of the target and $\boldsymbol{\boldsymbol{d}}_{p}$,
$\boldsymbol{\boldsymbol{c}}_{p}$ are the transmitting and receiving
steering vectors, respectively. $\boldsymbol{\boldsymbol{J}}_{p}$
is the $p^{th}$ lag shifting matrix ( like $\boldsymbol{\boldsymbol{\boldsymbol{A}}}_{k}$
in (\ref{eq:matrices})) but with dimension $\bigl((M+Q-1)\times(M+Q-1)\bigr)$,
$\boldsymbol{\boldsymbol{N}}^{H}$ is the noise matrix and $\hat{\boldsymbol{\boldsymbol{S}}}$
is the zero padded matrix which is given as $\hat{\boldsymbol{\boldsymbol{S}}}=[\boldsymbol{\boldsymbol{S}},\boldsymbol{\boldsymbol{0}}]^{T}$
( $\boldsymbol{\boldsymbol{S}}$ is $M\times L$ dimension and $\hat{\boldsymbol{\boldsymbol{S}}}$
is $(M+Q-1)\times L$ dimension). For the experiment the target strengths
$\left\{ \beta_{rp}\right\} _{r=0,p=1}^{Q-1,P}$ are selected as i.i.d
complex Gaussian random variables with mean $0$ and variance $1$.
The steering vectors are given by:

\begin{equation}
\begin{array}{c}
\boldsymbol{\boldsymbol{d}}_{p}=[1,e^{-j(4)\pi sin(\theta_{p})},e^{-j(8)\pi sin(\theta_{p})},e^{-j(12)\pi sin(\theta_{p})}]^{T}\end{array}\label{eq:rb}
\end{equation}

\begin{equation}
\begin{array}{c}
\boldsymbol{\boldsymbol{c}}_{p}=[1,e^{-j\pi sin(\theta_{p})},e^{-j\pi2sin(\theta_{p})},e^{-j\pi3sin(\theta_{p})}]^{T}\end{array}\label{eq:Tbp}
\end{equation}

where $\theta_{p}$ denote the scanning angle. In the simulation,
noise statistics is chosen to be i.i.d Gaussian with zero mean and
variance $\sigma^{2}$. The SNR in the experiment is taken to be $30$dB
($\sigma^{2}=0.001$). To form an high resolution image, goal is to
estimate $\left\{ \beta_{rp}\right\} _{r=0,p=1}^{Q-1,P}$, which is
done as follows. First, the matched filter $\boldsymbol{\boldsymbol{S}}_{q}^{MF}$
is applied on the received data $\boldsymbol{\boldsymbol{B}}^{H}$
to do the range compression on $q^{th}$ range bin , with the expression
for filter given by:

\begin{equation}
\boldsymbol{\boldsymbol{S}}_{q}^{MF}=\boldsymbol{\boldsymbol{J}}_{p}^{H}\hat{\boldsymbol{\boldsymbol{S}}}(\hat{\boldsymbol{\boldsymbol{S}}}^{H}\hat{\boldsymbol{\boldsymbol{S}}})^{-1}\label{eq:MF}
\end{equation}
Then the filter output is given by: 
\begin{equation}
\tilde{\boldsymbol{\boldsymbol{B}}}_{q}^{H}=\Biggl(\stackrel[r=0]{Q-1}{\sum}\stackrel[p=1]{P}{\sum}\beta_{rp}\boldsymbol{\boldsymbol{c}}_{p}\boldsymbol{\boldsymbol{d}}_{p}^{T}\boldsymbol{\hat{\boldsymbol{S}}}^{H}\boldsymbol{\boldsymbol{J}}_{p}+\boldsymbol{\boldsymbol{N}}^{H}\Biggr)\boldsymbol{\boldsymbol{S}}_{q}^{MF}\label{eq:app}
\end{equation}

\begin{equation}
\begin{array}[t]{c}
\tilde{\boldsymbol{\boldsymbol{B}}}_{q}^{H}=\Biggl(\stackrel[p=1]{P}{\sum}\beta_{qp}\boldsymbol{\boldsymbol{c}}_{p}\boldsymbol{\boldsymbol{d}}_{p}^{T}+\stackrel[r=0,r\neq q]{Q-1}{\sum}\stackrel[p=1]{P}{\sum}\beta_{rp}\boldsymbol{\boldsymbol{c}}_{p}\boldsymbol{\boldsymbol{d}}_{p}^{T}\boldsymbol{\hat{\boldsymbol{S}}}^{H}\boldsymbol{\boldsymbol{J}}_{p}\boldsymbol{\boldsymbol{S}}_{q}^{MF}\\
+\boldsymbol{\boldsymbol{N}}^{H}\boldsymbol{\boldsymbol{S}}_{q}^{MF}\Biggr)
\end{array}\label{eq:expan}
\end{equation}

The parameter of interest $\beta_{qp}$ can then be estimated in two
different ways:

(a) The Least Squares Estimator:

\begin{equation}
\hat{\beta}_{qp}^{LS}=\frac{\boldsymbol{\boldsymbol{c}}_{p}^{H}\tilde{\boldsymbol{\boldsymbol{B}}}_{q}^{H}{\boldsymbol{\boldsymbol{d}}}_{p}}{\bigl\Vert\boldsymbol{\boldsymbol{c}}_{p}\bigr\Vert^{2}\bigl\Vert{\boldsymbol{\boldsymbol{d}}}_{p}\bigr\Vert^{2}},\,p=1,..,P,\,q=0,..,Q-1\label{eq:LSEst}
\end{equation}

(b) The CAPON Estimator:

\begin{equation}
\hat{\beta}_{qp}^{C}=\frac{\boldsymbol{\boldsymbol{c}}_{p}^{H}\boldsymbol{\boldsymbol{V}}_{q}^{-1}\tilde{\boldsymbol{\boldsymbol{B}}}_{q}^{H}{\boldsymbol{d}}_{p}}{\boldsymbol{\boldsymbol{c}}_{p}^{H}\boldsymbol{\boldsymbol{V}}_{q}^{-1}\boldsymbol{\boldsymbol{c}}_{p}\bigl\Vert{\boldsymbol{\boldsymbol{d}}}_{p}\bigr\Vert^{2}},\,p=1,..,P,\,q=0,..,Q-1\label{eq:CAEst}
\end{equation}

where $\boldsymbol{\boldsymbol{V}}_{q}^{-1}=\tilde{\boldsymbol{\boldsymbol{B}}}_{q}^{H}\tilde{\boldsymbol{\boldsymbol{B}}}_{q}$
is the covariance matrix of compressed received data.

The estimated $\left\{ \beta_{rp}\right\} _{r=0,p=1}^{Q-1,P}$ using
different probing sequences (Multi-CAN, MM-Corr, ISL-NEW, BiST ( with
8 alphabets), and the proposed algorithm) of length ($M=256$) are
shown in the figures 6-7. It can be seen from the plots, for both
approaches to estimate the target strengths, the sequence set generated
by the proposed algorithm gives a better resolution image when compared
with the images obtained by employing the sequence sets generated
by other competing methods.

\begin{figure*}
\subfloat[True Target]{\includegraphics[scale=0.55]{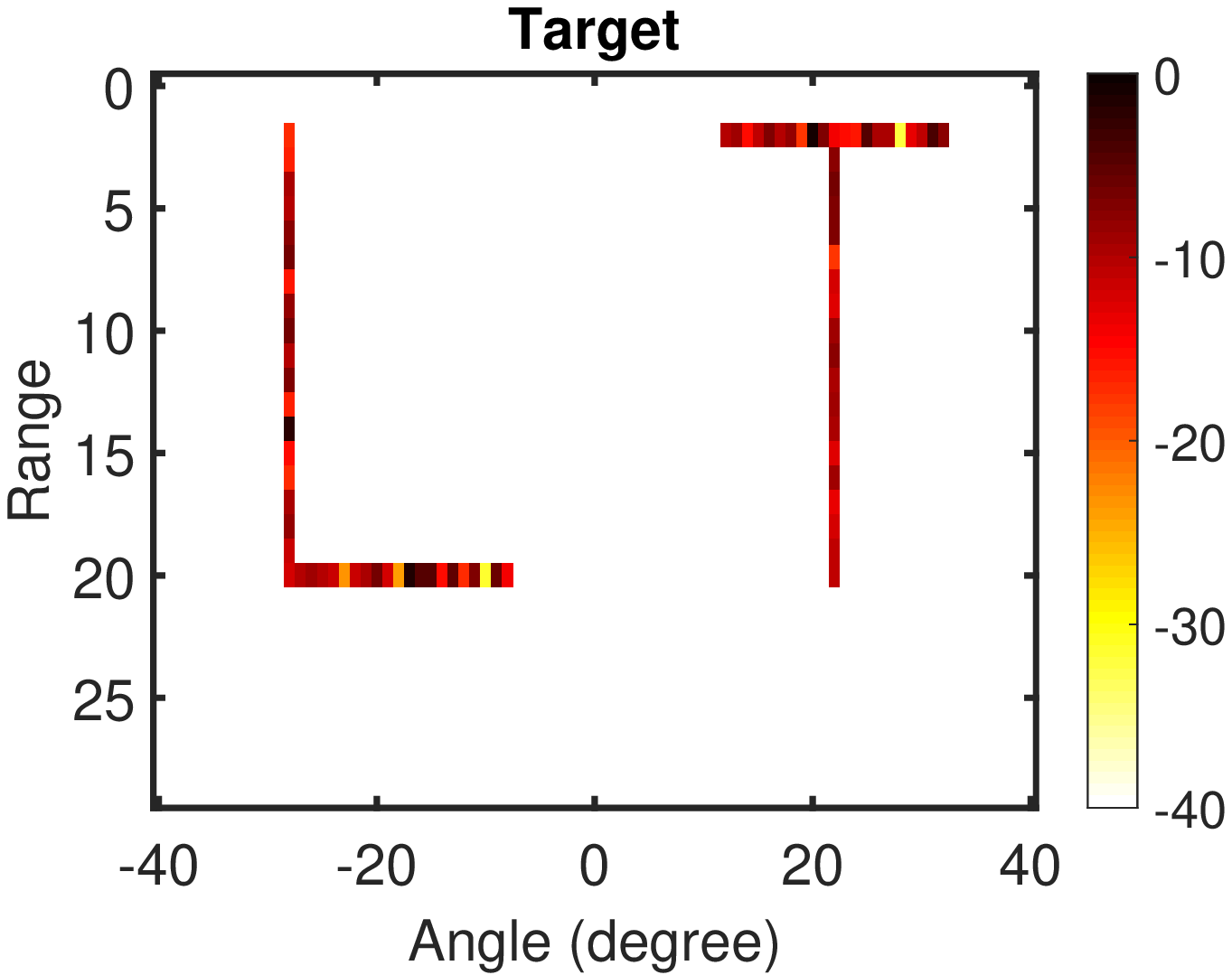}

}\subfloat[Multi-CAN sequence set]{\includegraphics[scale=0.55]{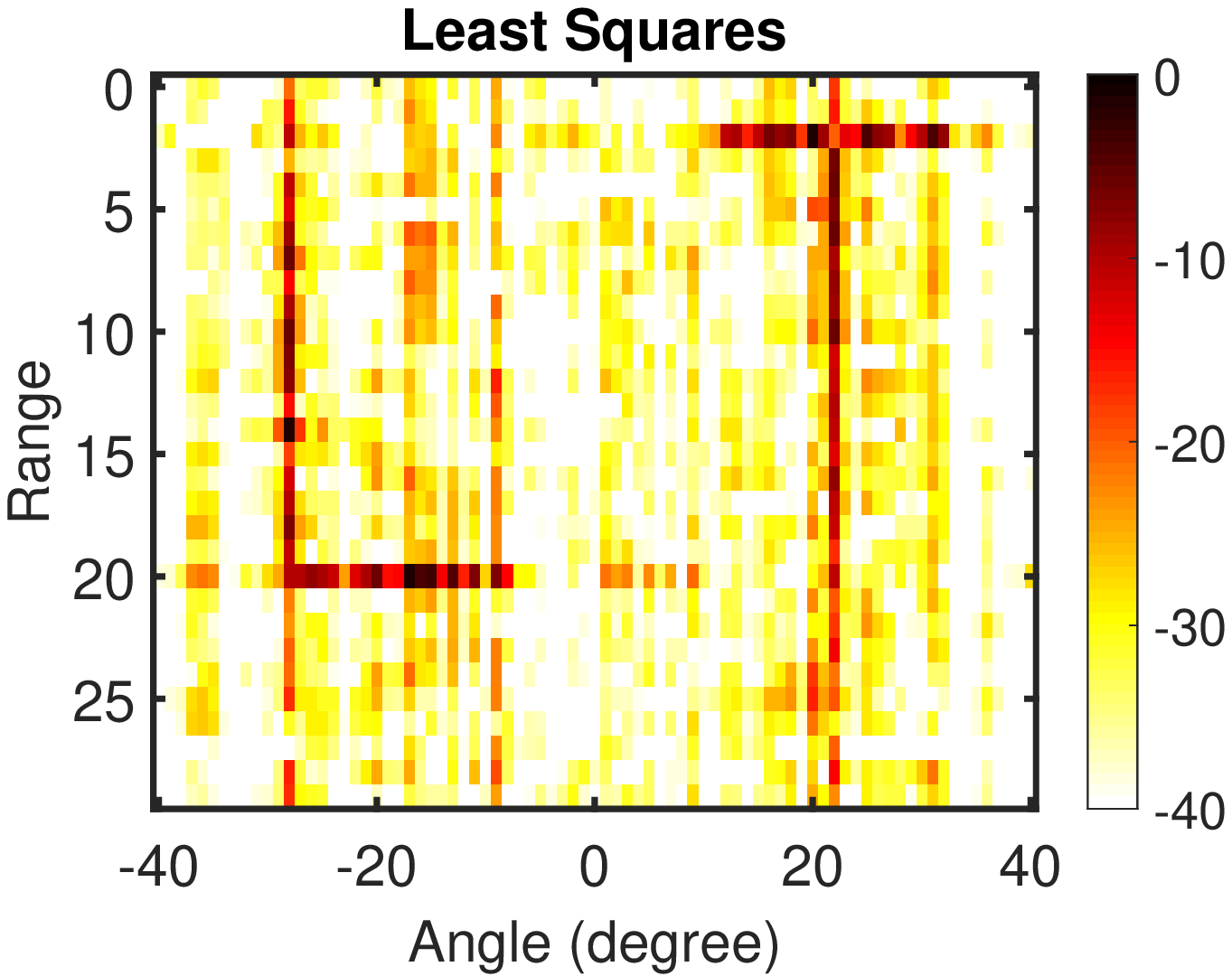}

}

\subfloat[MM-Corr sequence set]{\includegraphics[scale=0.55]{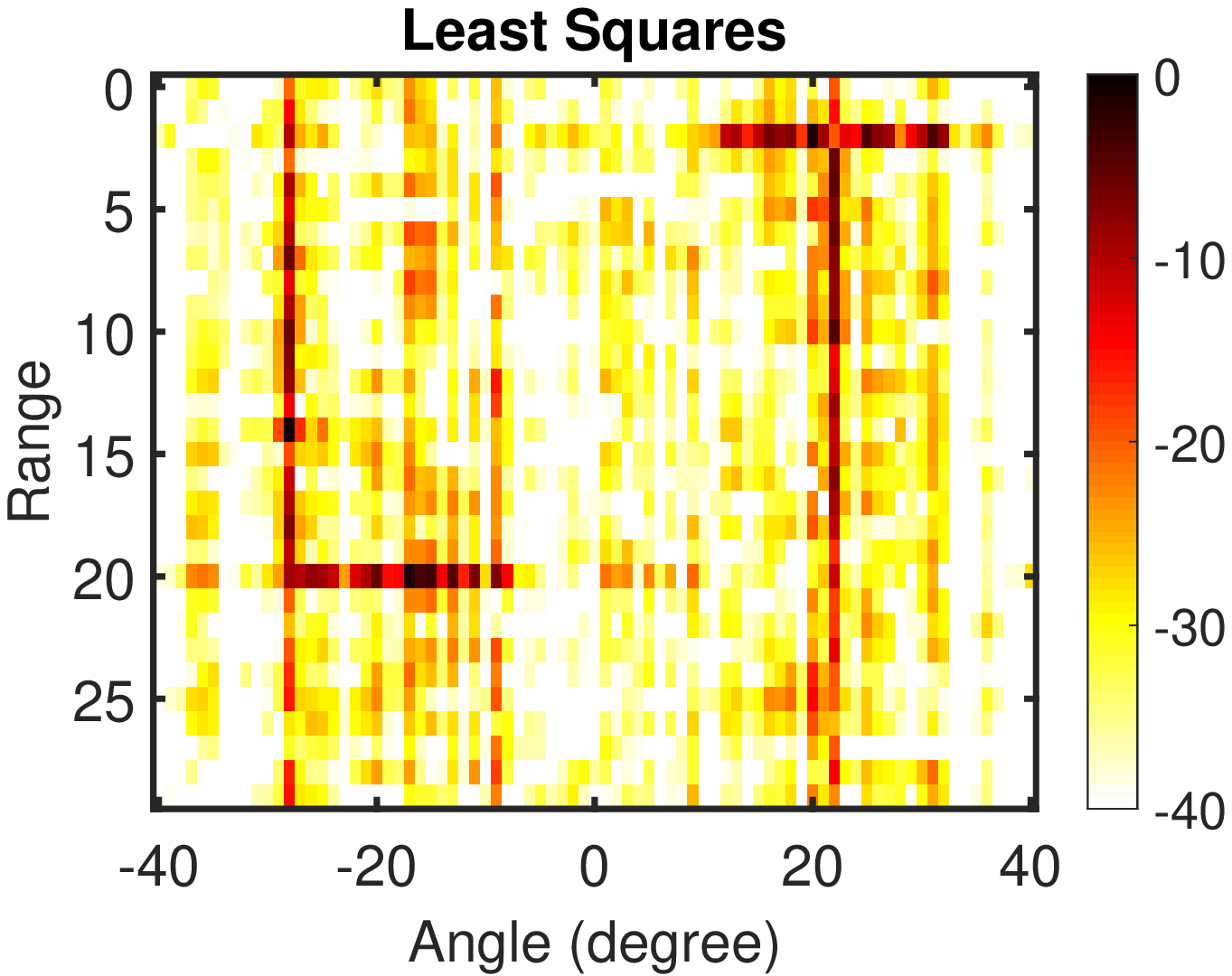}

}\subfloat[ISL-NEW sequence set]{\includegraphics[scale=0.55]{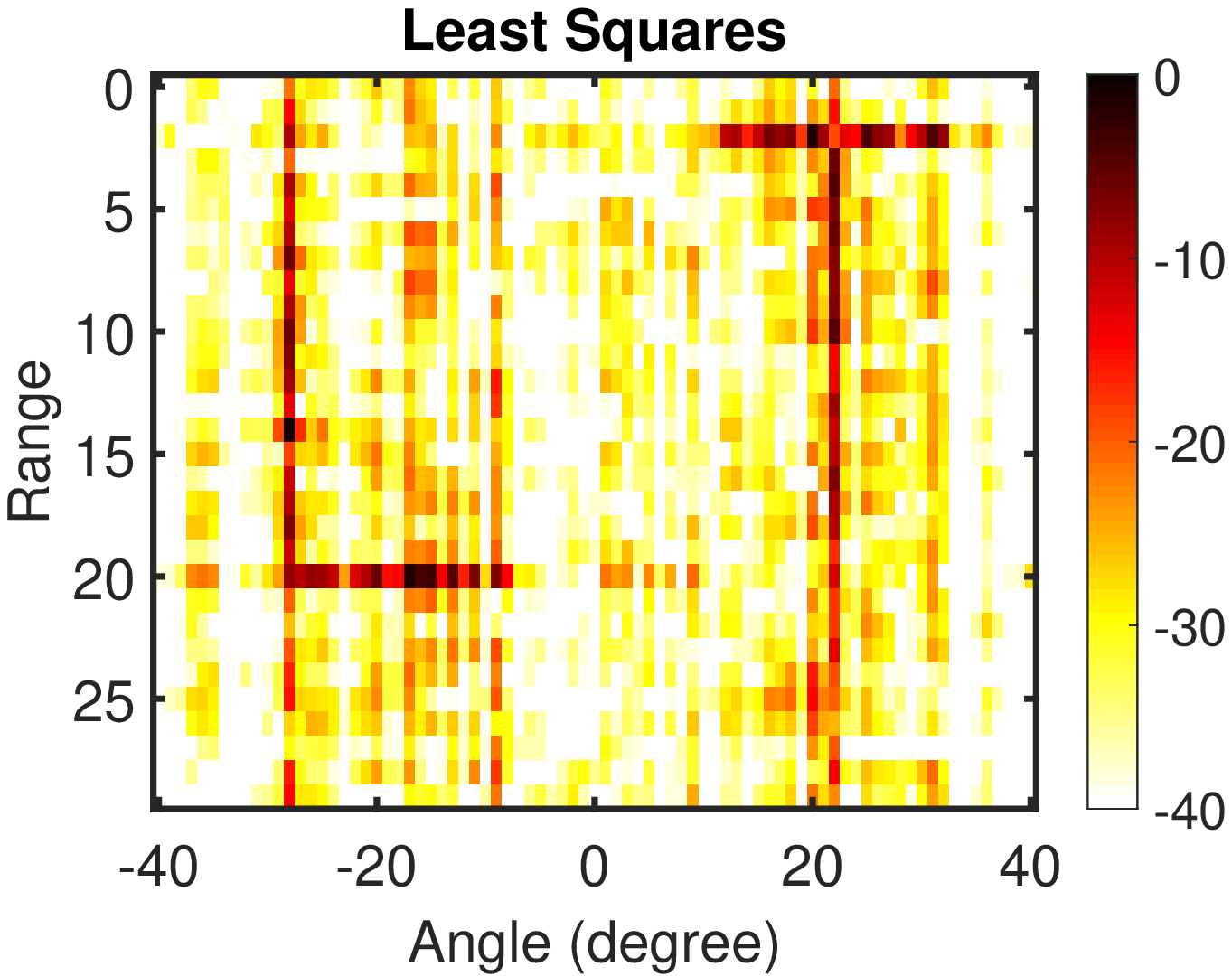}

}

\subfloat[BiST sequence set]{\includegraphics[scale=0.55]{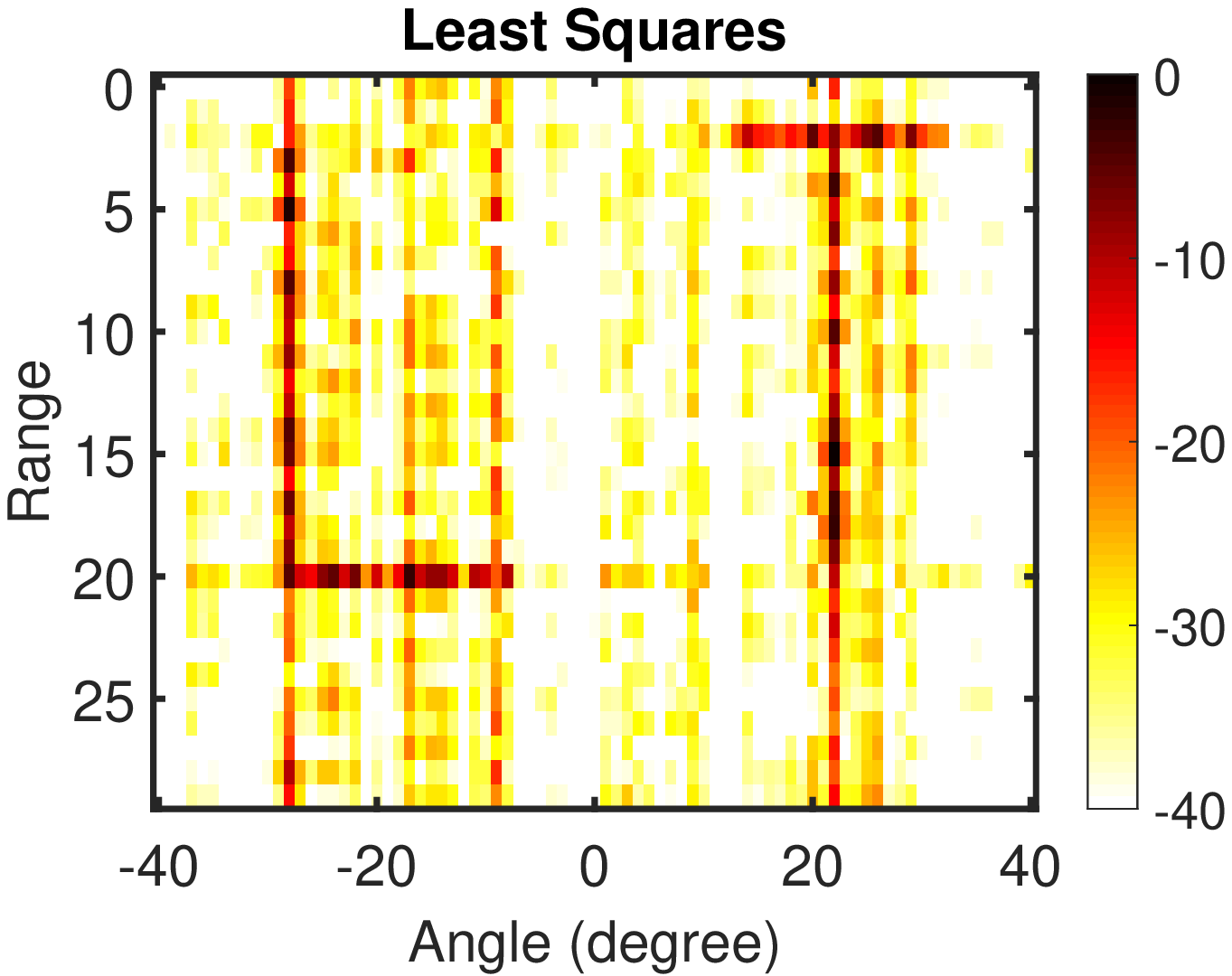}

}\subfloat[Proposed sequence set]{\includegraphics[scale=0.55]{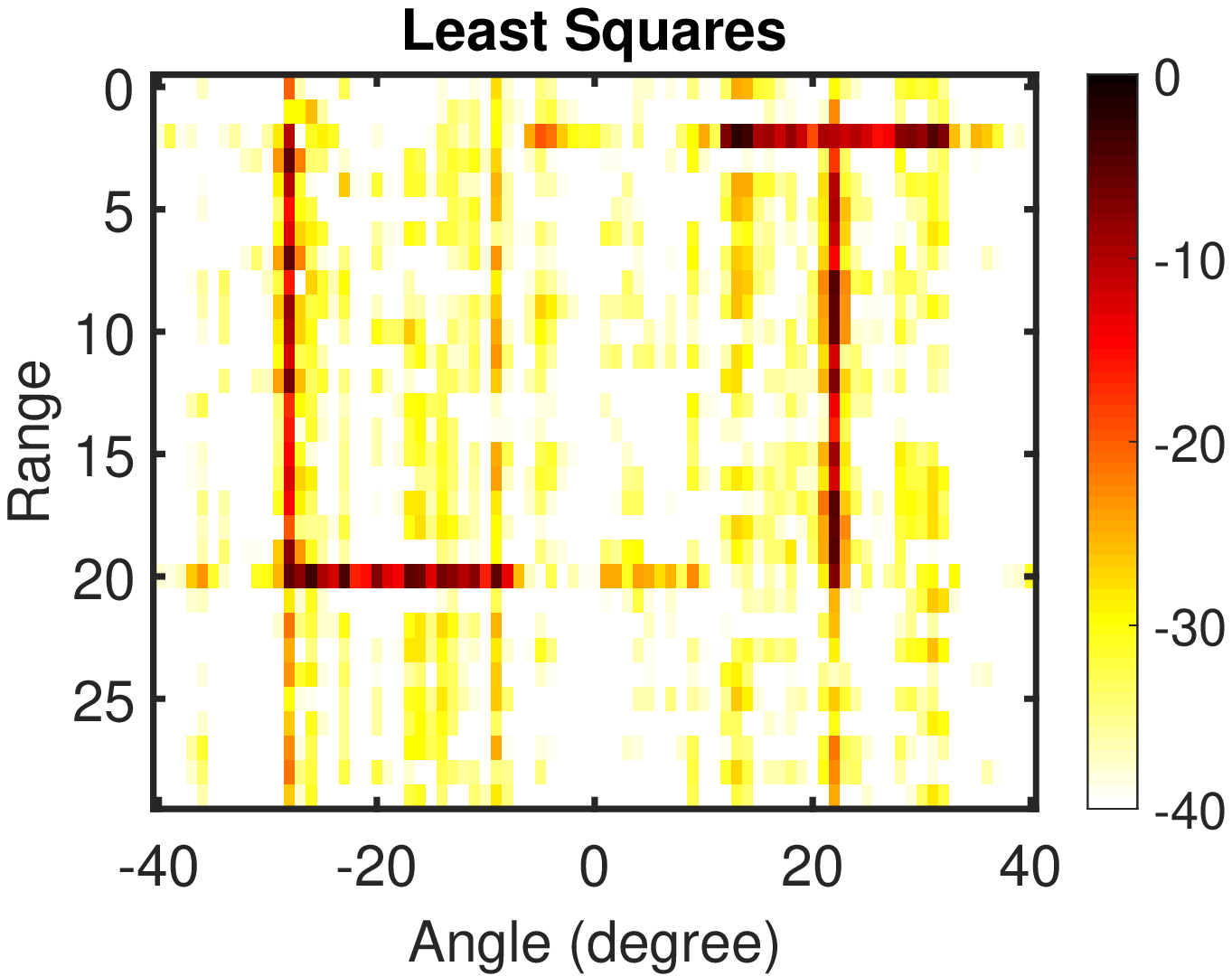}

}

\caption{MIMO RADAR target image reconstruction via the Least Squares Estimation
method for problem dimensions $(L,M)=(4,256)$ }
\end{figure*}

\begin{figure*}
\subfloat[True Target]{\includegraphics[scale=0.55]{R_Target}

}\subfloat[Multi-CAN sequence set]{\includegraphics[scale=0.55]{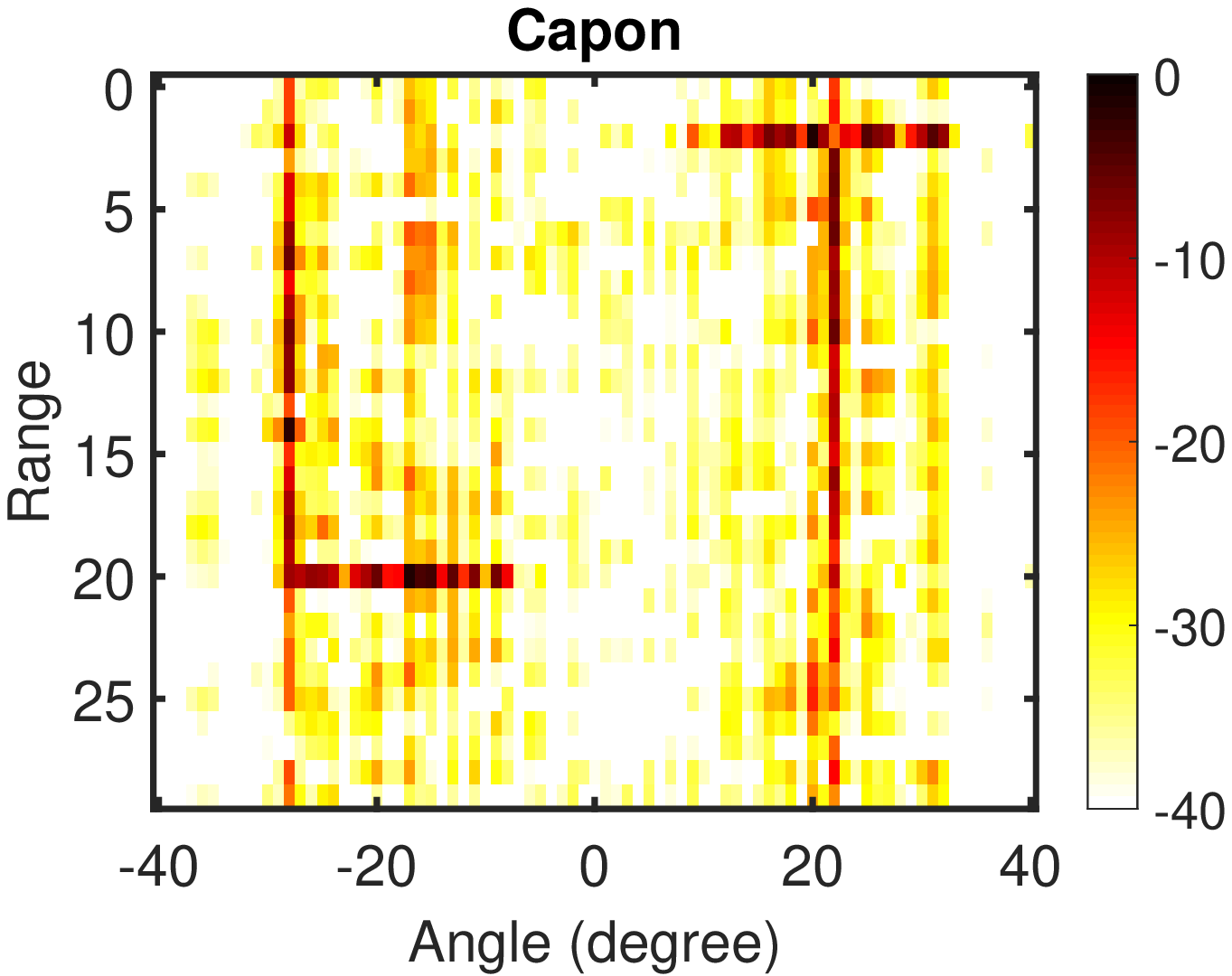}

}

\subfloat[MM-Corr sequence set]{\includegraphics[scale=0.55]{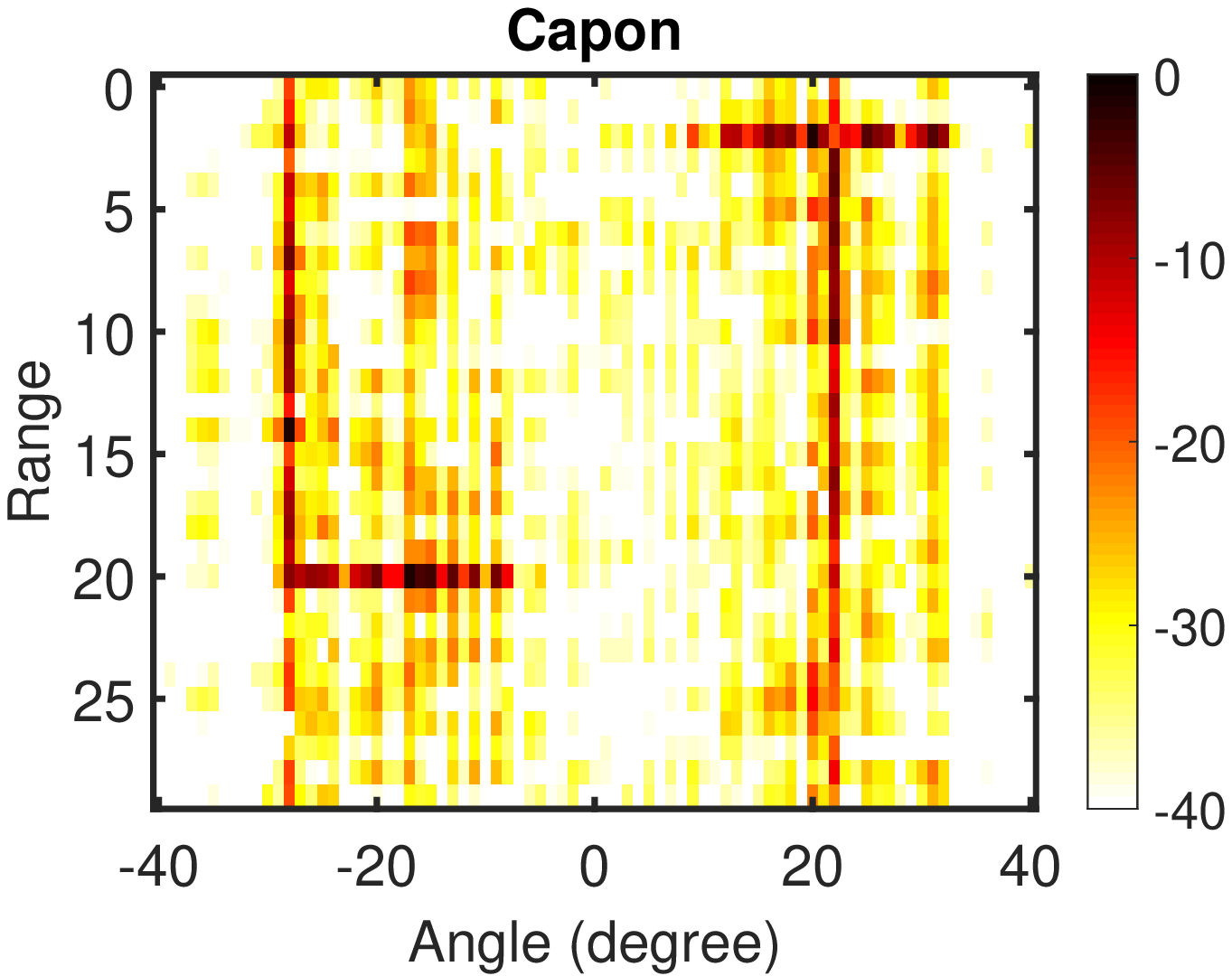}

}\subfloat[ISL-NEW sequence set]{\includegraphics[scale=0.55]{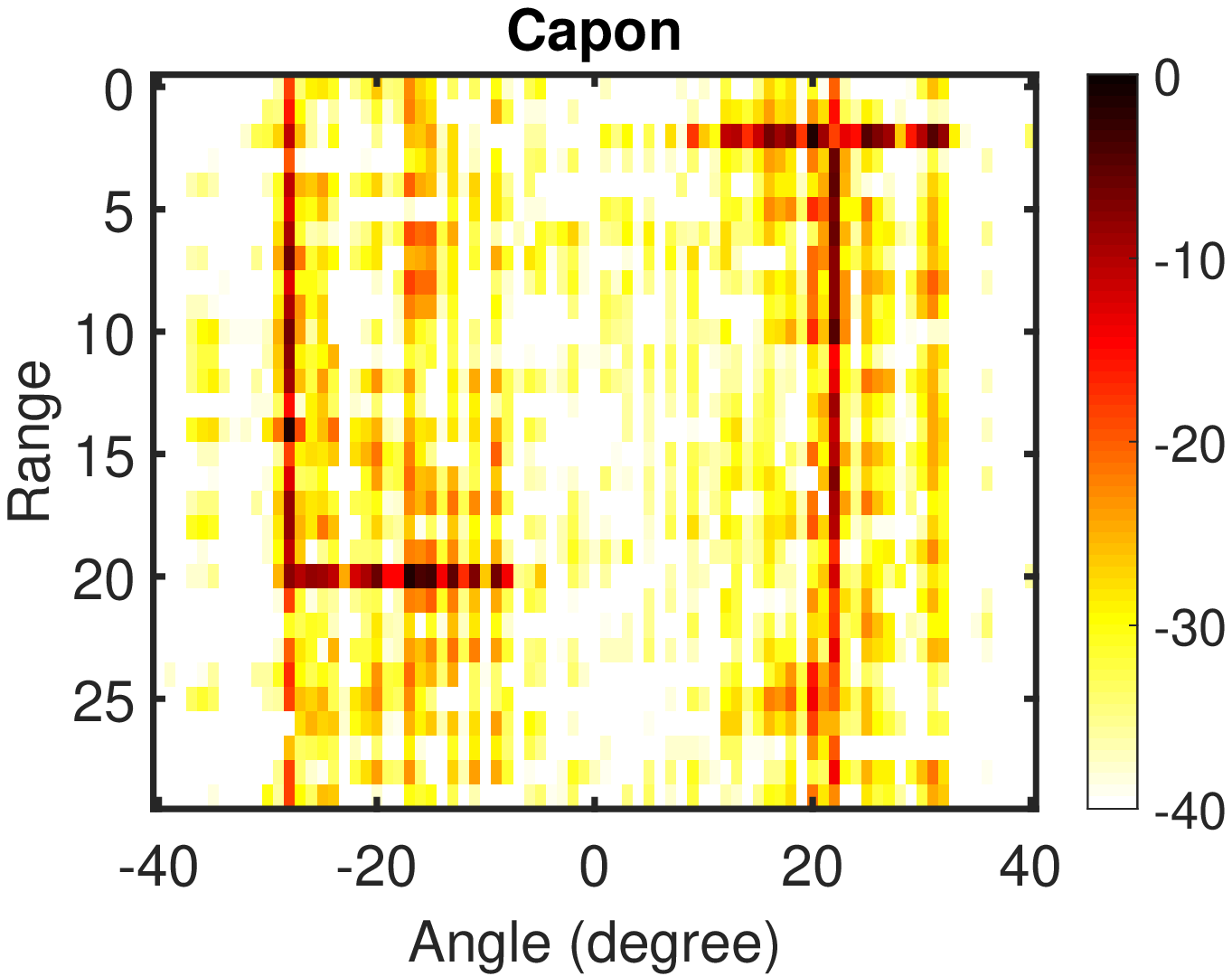}

}

\subfloat[BiST sequence set]{\includegraphics[scale=0.55]{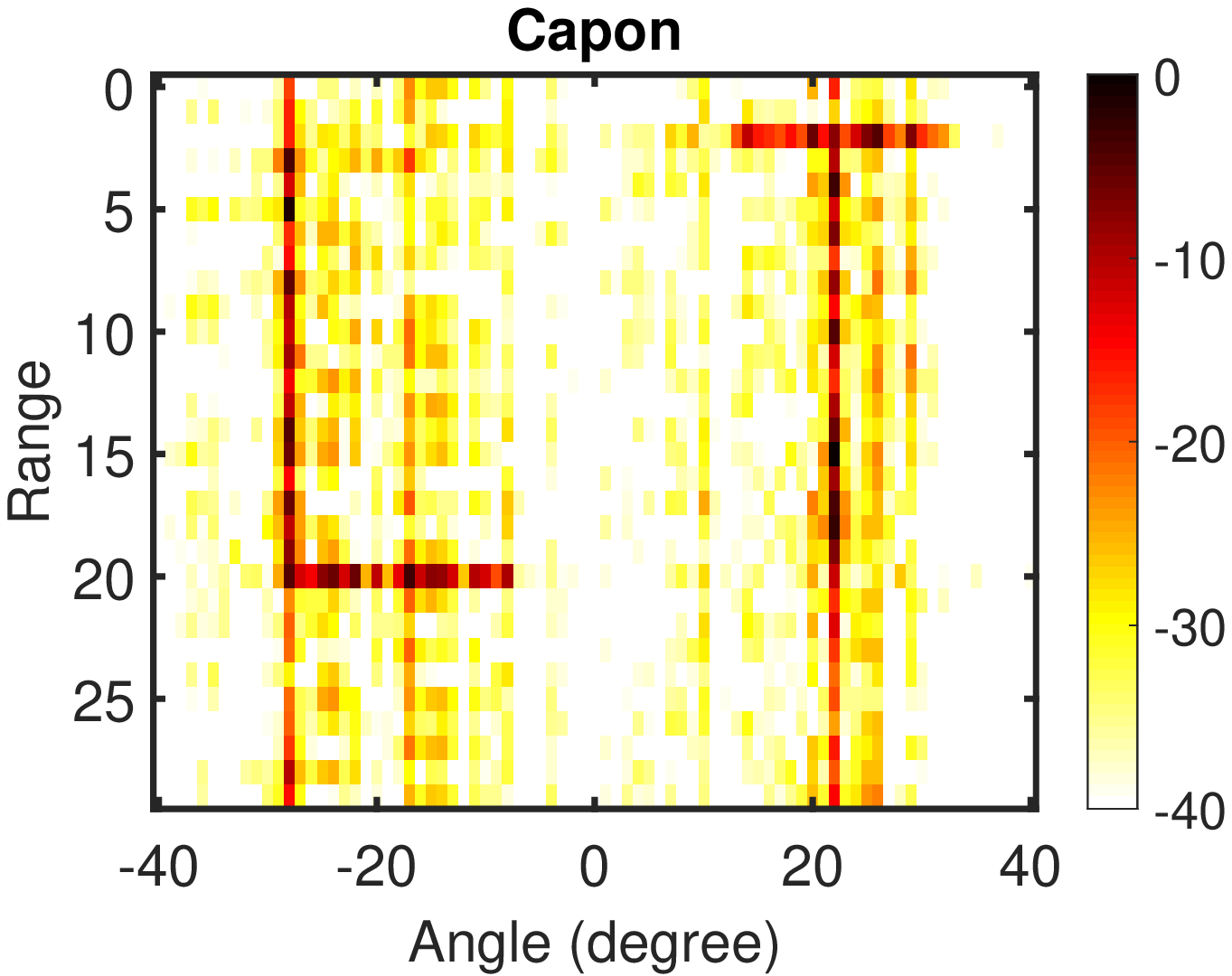}

}\subfloat[Proposed sequence set]{\includegraphics[scale=0.55]{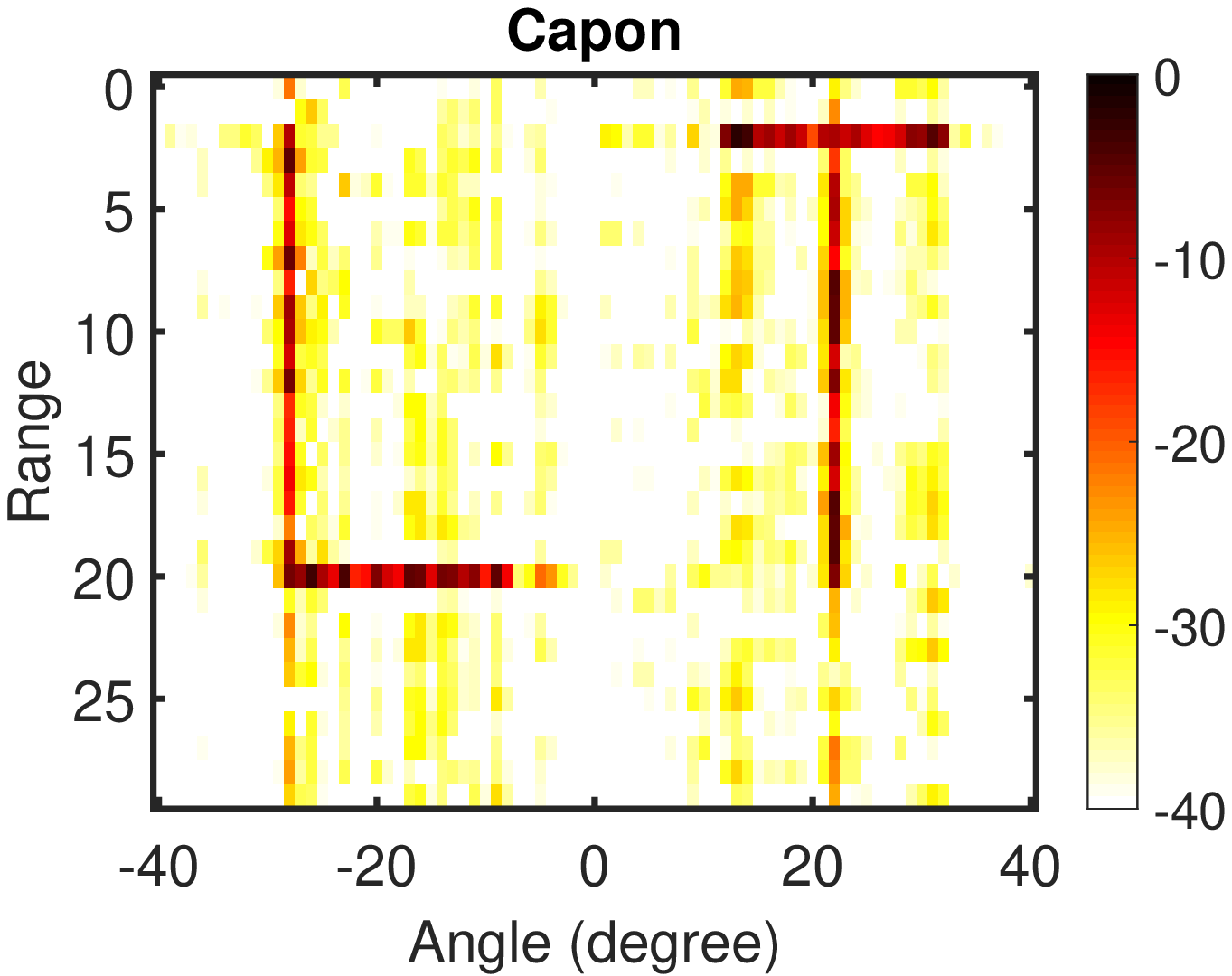}

}

\caption{MIMO RADAR target image reconstruction via the CAPON method for problem
dimensions $(L,M)=(4,256)$ }
\end{figure*}

\section*{\centerline{V.CONCLUSION}}

In this paper, we addressed the problem of designing sequence set
by directly minimizing the peak side-lobe level and proposed a Majorization-Minimization
technique based algorithm, which can be efficiently implemented using
the FFT and IFFT operations. To evaluate the performance of the proposed
algorithm, we conducted numerical simulations and compared with the
state-of-the-art algorithms, and observed that the proposed algorithm
is able to generate a sequence set with better PSL values. To highlight
the strength of the generated sequence set, we also conduct a MIMO
RADAR angle-ranging imaging experiment and showed that the sequence
set designed via the proposed algorithm produces very high-resolution
images when compared with the competing methods.

\bibliographystyle{IEEEtran}
\bibliography{PSL_SEQ_SET}

\end{document}